\documentclass[fleqn,usenatbib,times]{mnras}

\usepackage{array} 
\usepackage{framed}
\usepackage{graphicx,epsfig,psfig,multicol}
\usepackage[dvipsnames]{xcolor}
\usepackage[]{inputenc,amssymb}
\usepackage{amsmath}
\usepackage{color}
\usepackage{epsfig}
\usepackage{epstopdf}
\usepackage{textcomp}
\usepackage{xcolor,cancel}
\usepackage{comment}
\usepackage{subcaption}

\usepackage[export]{adjustbox}
\usepackage{adjustbox}
\usepackage{placeins}

\usepackage{float}

\usepackage{lipsum}
\usepackage{mwe}
\usepackage{arydshln} 

\definecolor{webgreen}{rgb}{0,.5,0}
\definecolor{webbrown}{rgb}{.6,0,0}

\usepackage[toc,page]{appendix}

\def \beq{\begin{equation}}
\def \eeq{\end{equation}}
\def \bea{\begin{eqnarray}}
\def \eea{\end{eqnarray}}

\newcommand{\OmgK}{\mbox{${\Omega}_{\rm K}$}}

\newcommand{\p}{\partial}
\newcommand{\rmd}{\mbox{${\rm d}$}}

\newcommand{\Msun}{\mbox{$ M_\odot $  }}

\newcommand{\Pmag}{\mbox{$P_{\rm m}$}} 
\newcommand{\Prad}{\mbox{$P_{\rm r}$}}
\newcommand{\Pgas}{\mbox{$P_{\rm g}$}}
\newcommand{\kes}{\mbox{$\kappa_{\rm es}$}}

\newcommand{\mP}{\mbox{$m_{\rm p}$}}
\newcommand{\vK}{\mbox{$v_{\rm K}$}}
\newcommand{\mathF}{\mbox{$\mathcal{F}$}}
\newcommand{\sigSB}{\mbox{$\sigma_{\rm SB}$}}
\newcommand{\kB}{\mbox{$k_{\rm B}$}} 
\newcommand{\qadv}{\mbox{$q_{\rm adv}^{-}$}} 
\newcommand{\qrad}{\mbox{$q_{\rm rad}^{-}$}}
\newcommand{\lamP}{\mbox{$\lambda_{P} $}}
\newcommand{\Crad}{\mbox{$\mathcal{C}_{\rm rad}$}}

\newcommand{\MdotEdd}{\mbox{$\dot{M}_{\rm Edd}$}}

\newcommand{\alT}{\mbox{$\alpha_{T} $}}

\newcommand{\alrho}{\mbox{$\alpha_{\rho} $}}

\newcommand{\Alf}{Alfv$\acute{\text{e}}$n } 

\newcommand{\Sth}{\mbox{$\mathcal{S}_{\rm th}$}}

\newcommand{\Pj}{\mbox{$\mathcal{P}_{\rm j}$}}

\newcommand{\PhiH}{\mbox{$\Phi_{\rm H}$}}

\newcommand{\OmgH}{\mbox{$\Omega_{\rm H}$}}
\newcommand{\kj}{\mbox{$k_{\rm j}$}}
\newcommand{\rH}{\mbox{$r_{\rm H}$}}
\newcommand{\rg}{\mbox{$r_{g}$}}
\newcommand{\Rtil}{\mbox{$\widetilde{R}$} } 
\newcommand{\Rd}{\mbox{$R_{\rm d}$}}
\newcommand{\Rdtil}{\mbox{$\widetilde{R}_{\rm d}$}}
\newcommand{\Bpd}{\mbox{$B_{\rm p d}$}}
\newcommand{\Btd}{\mbox{$B_{\rm t d}$}}
\newcommand{\rHtil}{\mbox{$\widetilde{r}_{\rm H}$}}
\newcommand{\Ej}{\mbox{$E_{\rm j}$}} 
\newcommand{\Mdotfb}{\mbox{$\dot{M}_{\rm fb}$}}
\newcommand{\Mdotpk}{\mbox{$\dot{M}_{\rm pk}$}}

\newcommand{\tfb}{\mbox{$t_{\rm fb}$}}

\usepackage[T1]{fontenc}

\DeclareRobustCommand{\VAN}[3]{#2}
\let\VANthebibliography\thebibliography
\def\thebibliography{\DeclareRobustCommand{\VAN}[3]{##3}\VANthebibliography}

\title[Magnetic Disks]{Magnetically Dominated Disks in Tidal Disruption Events and Quasi-Periodic Eruptions}

\author[K. Kaur et al.]{
Karamveer Kaur,\thanks{E-mail: karamveer.kaur@mail.huji.ac.il}
Nicholas C. Stone, Shmuel Gilbaum
\\
Racah Institute of Physics, The Hebrew University, 91904, Jerusalem, Israel}

\date{Accepted XXX. Received YYY; in original form ZZZ}

\pubyear{2021}

\begin{document}

\label{firstpage}
\pagerange{\pageref{firstpage}--\pageref{lastpage}}
\maketitle


\begin{abstract}

The classical radiation pressure instability has been a persistent theoretical feature of thin, radiatively efficient accretion disks with accretion rates $\sim 1-100\%$ of the Eddington rate. But there is only limited evidence of its occurrence in nature: rapid heartbeat oscillations of a few X-ray binaries and now, perhaps, the new class of hourly X-ray transients called quasi-periodic eruptions (QPEs). The accretion disks formed in tidal disruption events (TDEs) have been observed to peacefully trespass through the range of unstable accretion rates  without exhibiting any clear sign of the instability. We try to explain the occurrence or otherwise of this instability in these systems, by constructing steady state 1D models of thin magnetic accretion disks. The local magnetic pressure in the disk is assumed to be dominated by toroidal fields arising from a dynamo sourced by magneto-rotational instability (MRI). We choose a physically motivated criterion of MRI saturation, validated by recent magnetohydrodynamic simulations, to determine the strength of magnetic pressure in the disk. The resulting magnetic pressure support efficiently shrinks: (1) the parameter space of unstable mass accretion rates, explaining the absence of instability in systems such as TDEs and (2) the range of unstable radii in the inner accretion disk, which can shorten the quasi-periods of instability limit-cycles by more than three orders of magnitude, explaining the observed periods ($\sim $ a few hrs) of QPEs. In addition to examining stability properties of strongly magnetized disks, we predict other observational signatures such as spectral hardening factors and jet luminosities to test the compatibility of our disk models with observations of apparently stable TDE disks.

\end{abstract}

\begin{keywords}
accretion, accretion discs -- hydrodynamics -- instabilities -- magnetic fields
\end{keywords}



\section{Introduction}

 The question of stability in hot accretion flows is an old problem that is only beginning to be answered by modern simulations.  A thin, radiatively efficient $\alpha$-disk \citep{ShakuraSunyaev73} will often have inner regions dominated by radiation pressure, where electron scattering is the primary opacity source.  In a 1D (vertically averaged, azimuthally symmetric) disk model, these regions are both thermally \citep{Pringle+73, ShakuraSunyaev76} and viscously \citep{LightmanEardley74} unstable; both types of instability can also be analyzed together in a unified way \citep{Piran78}.  Subsequent analytic works found that the inner regions of hot accretion disks can stabilize for very low or very high accretion rates, when the disk becomes radiatively inefficient and advection dominates local cooling \citep{Abramowicz+88,narayan_yi1995}.  However, thermal and viscous instability are persistent features of semi-analytic black hole (BH) disk models when accretion rates fall roughly between $\sim 1\%$ and $\sim 100\%$ of the Eddington-limited rate, leading to global limit cycles of outburst and dormancy in time-dependent 1D \citep{Honma+91,SzuszkiewiczMiller98, Janiuk+02} and 2D \citep{Ohsuga06} calculations.

Observations of BH (and neutron star)
accretion disks in X-ray binaries (XRBs) have long delivered a mixed verdict on the picture of thermo-viscous instability described above.  While a wide range of accretion disks around stellar-mass compact objects show evidence for thermo-viscous instability limit cycles driven by hydrogen ionization in cold outer regions \citep{MeyerMeyerHofmeister81, Cannizzo93, Lasota01}, most of the observed XRBs are apparently stable to the radiation pressure driven thermo-viscous instability in spite of the anticipated dominance of the radiation pressure in the inner regions of these accretion disks \citep{Done+07}. Only a handful of these systems display the ``heartbeat oscillations'' on timescales $\sim 1-100$~s potentially associated with the radiation pressure instability \citep{Belloni+97,Taam1997,Massaro2010,Altamirano_2011_XRB_obs,Bagnoli2015,Maselli2018}.   

Recently, a new category of BH accretion disks have also exhibited significant evidence for thermal stability: tidal disruption events (TDEs).  TDEs occur in the centers of galaxies, when stars orbiting supermassive BHs are scattered onto radial orbits and are torn apart by tides \citep{Hills75, Rees88}.  The early stages of TDE disk formation are characterized by (usually) super-Eddington mass fallback rates \citep{EvansKochanek89, GuillochonRamirezRuiz13, Stone+13} and the initial formation of a globally eccentric accretion flow quite distinct from standard (axisymmetric) accretion disks \citep{Guillochon+14, Piran+15, Shiokawa+15}.  Although the subsequent circularization process is complex, and not fully understood (see \citealt{BonnerotStone21} for a recent review), it is likely that at late enough times, the disk has dissipated its excess energy and circularized into something like a standard accretion disk. 

Unless circularization is extremely slow, TDE disks should have sufficiently high accretion rates for radiation pressure to dominate gas pressure\footnote{Indeed, this theoretical expectation is confirmed by X-ray continuum fitting, which infers a range of near-Eddington accretion rates over the first several years of multiple TDE flares \citep{Wen+20, Wen+21}.}.  \citet{shen_matzner2014} applied the standard theory of time-dependent 1D accretion flows to TDE disks and found that, after a brief initial period of super-Eddington accretion, the dominance of radiation pressure will lead to global limit cycles and TDE disks will spend the vast majority of their lifetime in a cold and quiescent state.  This picture is strongly challenged by late-time (5-10 yr post-peak) {\it Hubble Space Telescope} observations of TDEs \citep{Gezari+15, vanVelzen+19}, which find persistently high UV luminosities. These observations are compatible with simple $\alpha$-disks in thermal equilibrium, but are far brighter than the cold branch of a thermo-viscous instability limit cycle \citep{shen_matzner2014}. 

More recently, a new class of X-ray transients exhibiting quasi-periodic eruptions (QPEs) on timescales $\sim$ a few hrs \citep{Miniutti_2019,Giustini_2020,Arcodia_2021,Chakraborty_2021} have been discovered.  These QPEs may potentially be associated with accretion disk instabilities \citep{Miniutti_2019,Raj_Nixon2021, Sniegowska2022, Pan_2022_viscosity_model}, although the standard radiation pressure instability predicts timescales longer by roughly 2-3 orders of magnitudes 
than observed QPE recurrence times \citep{Arcodia_2021}.
This timescale problem has led to a rich variety of proposed physical scenarios for QPE origins, often involving one or more donor stars episodically transferring mass to a massive BH \citep{King_2020,sukova2021,Xian2021,Ingram2021,King_2022_QPE_models,Metzger_Stone_2022,Wang2022,Nixon2022,Zhao2022,Krolik2022, LuQuataert22}. They usually involve accretion around a BH with mass $M \sim 10^{5-6}\Msun$ at Eddington ratios $\dot{M}/\MdotEdd \sim 0.1-0.5$.  

Various explanations for the apparent stability (or modified instability) of the inner zones of BH accretion disks have been proposed.  Four such explanations include (i) magnetic pressure support \citep{BegelmanPringle07, Oda+09}, (ii) stochastic variability due to turbulence \citep{JaniukMisra12}, (iii) intrinsic delays between stress and pressure fluctuations in magnetohydrodynamic disks \citep{Hirose+09} and (iv) the ``iron opacity bump'' \citep{Jiang+16}. Of these explanations, the first appears the most generally promising: if a strong magnetic field $B$ exists such that the associated magnetic pressure $\Pmag = B^2/8\pi \gtrsim \Prad$ (the disk radiation pressure), then the disk can become stable to thermal and viscous perturbations \citep{BegelmanPringle07}. Some evidence for this has emerged in self-consistent, global radiation magnetohydrodynamic (MHD) simulations \citep{Sadowski16, Jiang+19}.  Other semi-analytical and numerical studies \citep{Dexter2019,Pan2021,Sniegowska2022} of magnetic accretion disks also demonstrate the stabilizing tendency of magnetic pressure support, while significantly modifying the characteristics of instability (if it still exists). Notably, magnetic fields tend to shorten the instability limit-cycles and shrink the range of unstable mass accretion rates. Explanations (ii) and (iii) have been challenged by global radiation-magnetohydrodynamic simulations that find persistent thermal instability in weakly magnetized disks (\citealt{Sadowski16}; see also \citealt{Ross+17} for local simulations finding that stochastic variability alone does not prevent thermal instability). Even absent a dynamically important magnetic field, radiation-dominated disks may be thermally stabilized by the complicated opacity structure provided by bound-bound transitions in iron-group elements, as is seen in some local MHD simulations \citep{Jiang+16}; however, global 1D models with realistic opacities still exhibit viscous limit cycles \citep{Grzdzielski+17}.

In this paper, we aim to explain the occurrence of radiation pressure stability or instability in this wide variety of observed systems by constructing magnetically dominated steady-state 
accretion disk models. The local magnetic pressure in the disk is dominantly contributed by toroidal fields arising from magneto-rotational instability (MRI), that also provides the necessary viscosity for accretion to occur in the first place \citep{Balbus1991}. We assign the local disk magnetic pressure on the basis of physically motivated MRI saturation criterion suggested by \citet{pessah_psaltis2005, BegelmanPringle07}. The recent MHD simulations confirm the validity of this criterion for the disks accreting at both sub-Eddington and super-Eddington rates \citep{Jiang+19,Jiang2019superEdd}. More recent MHD simulations \citep{Mishra2022} further verify the necessity of dominant toroidal fields in the disk mid-plane for stability by exploring various initial seed field configurations. We find that the resultant magnetic pressure due to this saturation criterion is sufficiently strong to either stabilize the disks or significantly alter the characteristics of any putative instability. The timescales of instability limit cycles exhibited by these models are shorter by more than roughly three orders of magnitude, tentatively explaining the rapid hourly eruptions observed in QPEs (without invoking smaller TDE disks unlike previous magnetic disk models of \citealt{Sniegowska2022}). 

In \S~\ref{sec:disks}, we provide a detailed formalism for our steady-state 
disk models.  Specifically, we discuss how we model the magnetic pressure support for these disks and its impact on generic disk properties. In \S~\ref{sec:thermal}, we analyse the stability properties of magnetic disks, comparing closely with the standard radiation pressure instability characterizing non-magnetic disks. In \S~\ref{sec:observables} we make predictions for observational signatures of strong magnetic fields in our disk models and compare them against relevant observational constraints such as the short quasi-periods of QPEs (\S~\ref{sec_qpe}), the hardened X-ray blackbody emission and stability of TDEs, (\S~\ref{sec_TDE_hardBB}) and the apparent lack of high-energy jet emission in most TDEs (\S~\ref{sec_TDE_jets}). We present the summary and conclusions in \S~\ref{sec:conclusions}.

\section{ Accretion disk models}
\label{sec:disks}



We construct 1D, steady-state, slim/thin disk models for magnetized accretion disks around a central BH of mass $M$ in Newtonian gravity, starting from the classical Shakura-Sunyaev models \citep{ShakuraSunyaev73}.  We add to these models (a) magnetic pressure support arising from the saturation of MRI
; (b) advective cooling due to the inward flow of matter in thick disks (e.g. \citealt{frank_king_raine2002}); and (c) realistic, tabulated opacities \citep{Cloudy2013,Cloudy2017}. However, in the inner regions of interest to us, electron scattering opacity $\kes = 0.34$ (in cgs) dominates and therefore, we neglect absorption 
in computing disk structure\footnote{We have also confirmed the sub-dominant contribution of absorption opacity $\kappa_{\rm abs}$ to disk properties by evaluating Rosseland mean opacities using the photoionization code Cloudy \citep{Cloudy2013,Cloudy2017}. Within $R \lesssim 100 r_g$, the local disk properties $(\rho, T_{\rm c})$ differ less than a few percent for $\dot{M} \geq 0.01 \MdotEdd$. Realistic opacity matters only for local stability in the outer disk regions and can not stabilize the inner disk; check appendix~\ref{app_opacity_effects} for details.  }, although we include them while computing emission properties of the disks in \S~\ref{sec_TDE_hardBB}.

The mid-plane density $\rho$ and temperature $T_{\rm c}$ determine the local thermodynamic state of the disk. The surface density $\Sigma = \rho H$, where disk height $H = c_s/\OmgK$ is determined from vertical hydrostatic equilibrium. The sound speed $c_s = \sqrt{P/\rho}$ and total pressure $P(\rho,T_{\rm c}) =  \Pgas + \Prad + \Pmag$. The gas pressure $\Pgas = \rho \kB T_{\rm c}/ (\mu \mP) $ with $\mu = 0.6$ for Solar metallicity gas, and radiation pressure $\Prad = 4 \sigSB T_{\rm c}^4/(3 c)$. 

The magnetic pressure $\Pmag=B^2/8\pi$ is controlled by the mean mid-plane magnetic field strength $B$, the amplitude and geometry of which is uncertain.  We follow \citet{BegelmanPringle07} (hereafter BP07) in assuming that a predominantly toroidal field will result from the saturation of an MRI-driven dynamo and that this saturated value is such that the \Alf speed $v_{\rm A} = \sqrt{\Pmag / \rho}$ roughly equals the geometric mean of the local Keplerian velocity $\vK$ and the speed of sound in gas $c_{\rm s,g} = \sqrt{\Pgas / \rho}$. 
More explicitly: 
\beq 
 \Pmag = p_0 \vK \rho \sqrt{\frac{\kB T_{\rm c}}{\mu \mP}} ,
\label{Pmag}
\eeq 
where we have introduced a dimensionless scaling parameter $p_0$. In this section, we consider the three tentative values for  $p_0 = \{ 1,10 , 100 \}$ to demonstrate its impact on the nature of magnetic disks; note that BP07 considered $p_0 =1$ in their analysis.  This parameter decides the fractional contribution of $\Pmag$ to the total pressure, albeit in a nonlinear fashion. This MRI saturation criterion was first conjectured by BP07 on the basis of local linear stability studies of MRI in magnetically dominated disks by \citet{pessah_psaltis2005} (hereafter PP05). PP05 investigated the stability of axisymmetric modes of magnetized disks dominated by toroidal magnetic fields, with only a small non-zero value for poloidal field (necessary for the action of MRI). PP05 found that the growth rates of unstable MRI modes were highly suppressed if $v_{\rm A} \gtrsim  \sqrt{\vK c_{\rm s,g}}$. BP07 argued that the high magnetic tension prevents the growth or even emergence of MRI in the disks exceeding the magnetic pressure beyond this critical limit (see their section~3.1 for a detailed discussion). More recent radiation-magnetohydrodynamic simulations have found mid-plane magnetic field strengths roughly consistent with this conjecture.  For example, in \citet{Jiang+19}, simulations of a massive BH with $M = 5 \times 10^{8} \Msun$ accreting at $\dot{M}/\MdotEdd = 0.07-0.2$ find that the saturated magnetic pressure in the disk mid-plane corresponds to $p_0 \simeq 4-9$.

These steady state disks are characterized by a constant mass accretion rate $\dot{M}$ usually measured in terms of $\MdotEdd = L_{\rm Edd}/(\eta c^2) = 0.023 \Msun {\rm yr}^{-1} M_6/\eta_{-1}(a)$, the accretion rate for which accretion luminosity equals the Eddington luminosity $L_{\rm Edd}$. Here $\eta_{-1} = \eta(a)/0.1$ is the normalized radiative efficiency of a BH with spin parameter $a$, and $M_6 = M/(10^6 \Msun )$ is the normalized BH mass. We consider a non-spinning BH with $a=0$ for most of this work (except in \S~\ref{sec_TDE_jets} where we compute jet energies). For $a =0$, $\eta = 0.057$ and $\MdotEdd = 0.04 \Msun {\rm yr}^{-1} M_6$.   

We follow an $\alpha$-viscosity prescription with viscosity $\nu = \alpha c_s H$, and choose $\alpha =0.1 $ in this section\footnote{Later on, \S~\ref{sec:observables} highlights the disk properties sensitive to the choice of $\alpha$.}. The viscous equilibrium for a torque-free inner disk boundary $R_{\rm i}$ (corresponding to the inner-most stable circular orbit), implies $\nu \Sigma = \dot{M} f(R)/(3 \pi)$, where function $f(R) = 1 - \sqrt{R_{\rm i} / R}$. For a non-spinning BH, $R_{\rm i} = 6 r_g$, where $r_g = G M /c^2$ is the gravitational radius of the central BH. The viscous dissipation leads to a local heating rate per unit surface area of $D(R) = (9 /8) \nu \Sigma \OmgK^2$, where $\OmgK = \sqrt{ G M/R^3}$ is the Keplerian orbital frequency. This translates to a volumetric heating rate $q^+ = D(R)/H = 9 \OmgK \alpha P/8$. This dissipative heating is balanced by volumetric cooling due to radiative diffusion $\qrad = 4 \sigSB T_{\rm c}^4/(3 \kes \Sigma H)$ and radial advection $\qadv = \dot{M} P/(2 \pi R^2 \Sigma)$ \citep{frank_king_raine2002,shen_matzner2014}.  

The equations for thermal and viscous equilibrium, namely $q^{+} = q^{-} =  \qadv + \qrad$ and $\nu \Sigma = \dot{M} f(R)/(3 \pi)$, give the final algebraic equations which can be expressed only in terms of local quantities $(\rho, T_{\rm c})$:
\begin{subequations}
\begin{align} 
& \frac{3 \dot{M}}{8 \pi} \OmgK^2 f(R)  = \frac{4 \sigSB T_{\rm c}^4 \OmgK  }{3 \kes \sqrt{P(\rho,T_{\rm c}) \rho}} +  \frac{\xi \dot{M}  }{2 \pi R^2} \frac{P}{\rho}  \\[1 em]
& \frac{\alpha P^{3/2}(\rho,T_{\rm c})}{\OmgK^2 \sqrt{\rho}} = \frac{\dot{M} f(R)}{3 \pi}
\end{align}
\label{disk_visc_eqbm}
\end{subequations}
We solve these algebraic equations numerically in \texttt{Mathematica} and for this section, we choose a typical range of parameters of interest -- $M = 10^{4 -7} \Msun$ and $\dot{M} = 0.01 -10 \MdotEdd$ for $6 r_g \leq R \leq 500 r_g$. We discuss various physical properties of magnetic disks ($p_0 \neq 0$) below and compare them with their non-magnetic counterparts ($p_0 = 0$). 

\subsection{Nature of Magnetic Disks}

\begin{figure*}
\centering
\includegraphics[width=1
\textwidth]{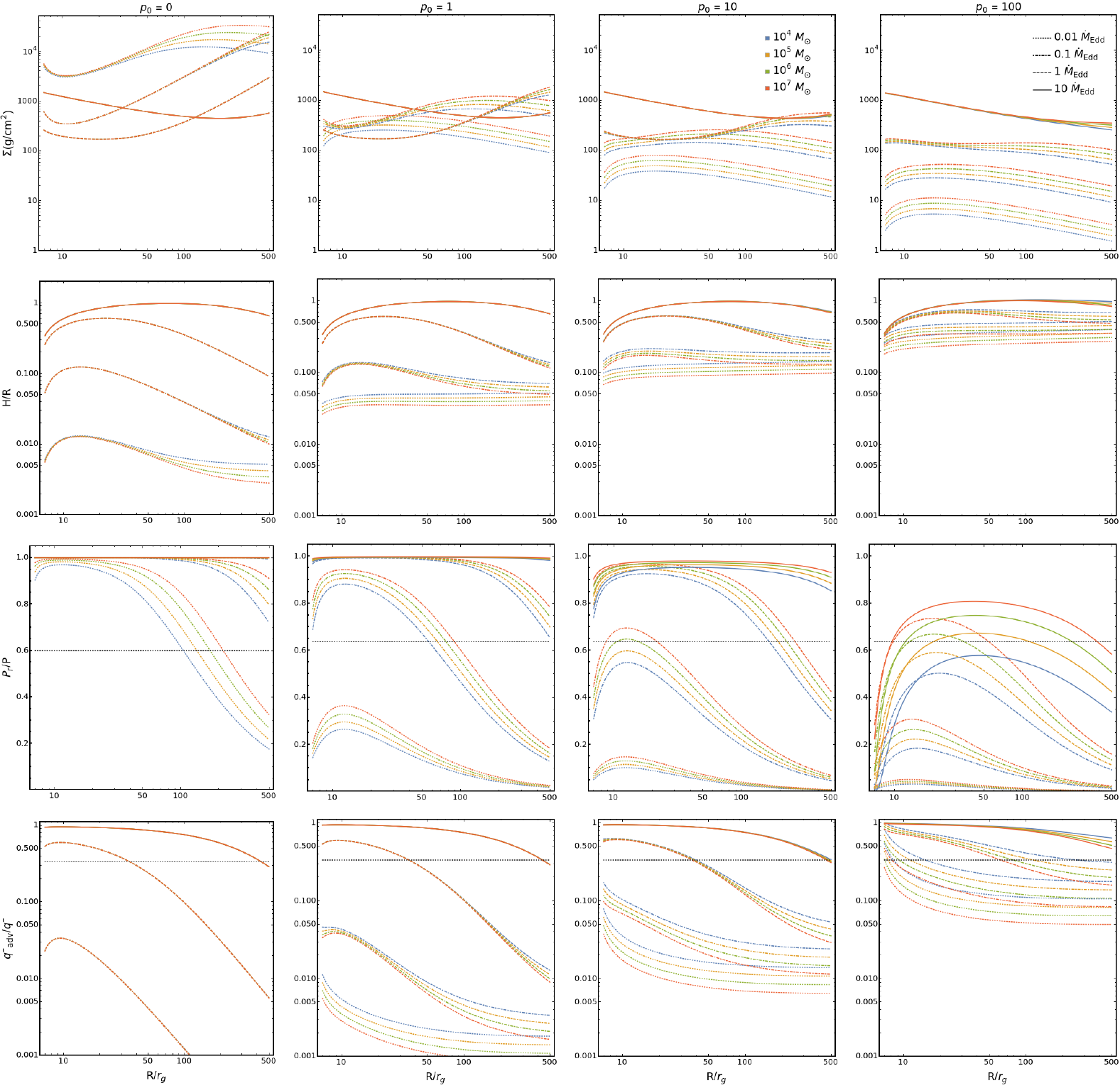}
\caption{Radial profiles of physical entities of non-magnetic ($p_0 = 0$ in the first column) and magnetic disks ($p_0=\{1,10,100 \}$ in the remaining columns with $p_0$ increasing from left to right). The first row shows gas surface densities $\Sigma$, the second row aspect ratios $H/R$, the third row the fractional contribution of radiation pressure $P_{\rm r}$ to total pressure $P$, and the fourth row the fractional contribution of advective cooling $q_{\rm adv}^{-}$ to total volumetric cooling $q^{-}$.      }
\label{fig_SS_mag_disks2}
\end{figure*}

The significance of including magnetic pressure $\Pmag$ strongly depends on the magnitude of the disk mass accretion rate $\dot{M}$. Figure~\ref{fig_SS_mag_disks2} depicts the 
physical properties of both non-magnetic ($p_0 = 0 $ in the first column) and magnetic disks (with $p_0 = \{1,10,100 \}$ in the remaining columns) for various $M$ and $\dot{M}$ of interest. For high $\dot{M} = \{ 1,10 \} \MdotEdd$ disks for which advective cooling dominates, disk properties change only a little with magnetic fields for the modest values of $p_0 = \{ 1,10 \}$. There are significant changes for high-$\dot{M}$ magnetic disks only if they have far higher \Alf speeds (i.e. the $p_0 = 100$ case) than those taken from the PP05 saturation criterion. On the contrary, low $\dot{M}$ disks (a term used here for $\dot{M} = \{0.01,0.1\} \MdotEdd$) are quite sensitive to the contribution of magnetic pressure even when considered with $p_0 = \{1 , 10 \}$.  As $p_0$ increases for these low $\dot{M}$ cases, the contribution of $\Pmag$ to the total pressure and hence its impact on disk properties increases.

The magnetic pressure acts as an additional vertical support for the disks, and leads to a decreasing fractional contribution $\Prad/P$ of radiation pressure with increasing $p_0$ (third row of figure~\ref{fig_SS_mag_disks2}). Note that the remaining non-radiative pressure support comes from $\Pgas$ for a non-magnetic disk, while it mainly comes from $\Pmag$ for magnetic disks as the contribution of $\Pgas$ is highly suppressed with $\Pgas/\Pmag = c_{\rm s,g}/(p_0 \vK) \ll 1$. For magnetic disks, $\Pgas/P  < 0.01$ in the entire parameter space considered here. Due to smaller contribution of $\Prad$, magnetic disks are relatively cooler. 

Magnetic pressure puffs up the disks, with progressively higher aspect ratios $H/R$ seen as $p_0$ increases (second row of the figure). For disks dominated by $\Pmag$, $H/R \simeq v_{\rm A}/ \vK = \sqrt{p_0 c_{\rm s,g}/ \vK}$. As a result, thicker magnetized disks can more effectively advect energy inwards. Evidently, the advective contribution $\qadv/q^{-}$ to cooling increases with increasing $p_0$ (fourth row of figure~\ref{fig_SS_mag_disks2}). Likewise, magnetic disks have smaller surface mass density $\Sigma$ compared to their non-magnetic counterparts (first row of figure~\ref{fig_SS_mag_disks2}). Smaller $\Sigma$ and higher $H$ implies smaller mass density $\rho$ for magnetic disks. 

All these trends stand out especially strongly for low-$\dot{M}$ disks 
and are much milder for higher $\dot{M}$. 
The relative contribution of advective cooling $\qadv/q^{-}$ depends sensitively on $\dot{M}$ (with only a weak dependence on $p_0$ and $M$); see the fourth row of figure~\ref{fig_SS_mag_disks2}. We can understand the above results better by constructing approximate disk solutions, presented in table~\ref{tbl_analytic_sol} of appendix~\ref{app_analytic}, with cooling only due to: (a) vertical radiative diffusion relevant for low $\dot{M}$ (and/or outer disk regions); (b). radial advection, appropriate for high $\dot{M}$ (and/or inner disk regions). We describe briefly the properties of these solutions below, but refer the interested reader to appendix \ref{app_analytic} for greater detail.

\emph{{\rm Case (a)}. Radiatively cooled disks}: Radiatively-cooled magnetically-dominated disks (case a1 of table~\ref{tbl_analytic_sol}) tend to have a lower surface density $\Sigma$, which falls for large values of $p_0$ as $p_0^{-8/9}$. In contrast to radiation-dominated disks (case a2), $\Sigma$ (and also mass density $\rho$) for these magnetic disks increases with increasing $\dot{M}$ and is a decreasing function of radial distance $R$. The magnetic disks are vertically puffier with larger disk height $H \propto R $ (approximately), with only a weak, positive dependence on $\dot{M}$. This is in contrast to radiation-dominated disks with a flat radial profile for $H \propto R^0 \dot{M}$. Magnetic disks with higher $p_0$ have higher disk heights $H \propto p_0^{4/9}$. Due to additional support from $\Pmag$ (rather than only $\Prad$), these disks are relatively cooler with mid-disk temperatures $T_{\rm c} \propto p_0^{-2/9}$. Radial profiles for $T_{\rm c} \propto R^{-8/9}$ are relatively steeper than their radiation-dominated counterparts.           

\emph{{\rm Case  (b)}. Advectively cooled disks}: It is interesting to note that the disks with only advective cooling (case b) have $P$ and $\rho$ independent of the dominant source of pressure. Consequently, the introduction of additional pressure support from magnetic fields has only a small effect on many physical properties (like $P$, $\rho$, $H$ and $\Sigma$). In the parameter space considered here, advective disks tend to retain a significant support from $\Prad$ (as evident from third row of figure~\ref{fig_SS_mag_disks2}) so that magnetically dominated advective disks (case b1) never occur (though it might be physically possible for much higher $p_0$ values). Other properties like mid-plane temperature $T_{\rm c}$ 
of advective (or high $\dot{M}$) disks are also impacted only a little with magnetic fields, even for the highest $p_0 =100$ we consider here.

\section{Disk Thermal Instability}
\label{sec:thermal}

\begin{figure*}
\begin{subfigure}{0.99\textwidth}
\centering
\includegraphics[width=0.99 \textwidth]{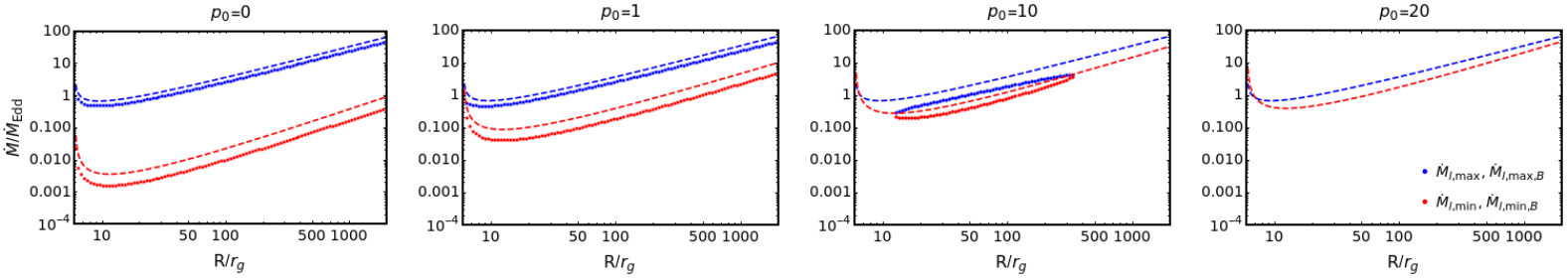}
\subcaption{$M = 10^4 \Msun$}
\end{subfigure}
\\
\begin{subfigure}{0.99\textwidth}
\centering
\includegraphics[width=0.99 \textwidth]{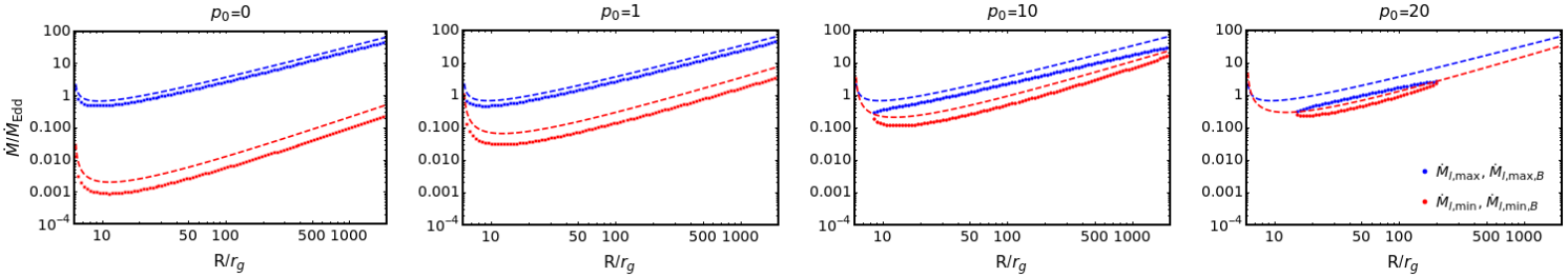}
\subcaption{$M = 10^6 \Msun$}
\end{subfigure}
\caption{The local range of instability in mass accretion rate $ \dot{M}_{\rm l, min } \leq  \dot{M} \leq \dot{M}_{\rm l, max }$ for a non-magnetic disk ($p_0 = 0$ case shown in the first column) and $\dot{M}_{\rm l, min}^{\rm B } \leq  \dot{M} \leq \dot{M}_{\rm l, max}^{\rm B }$ for magnetic disks in the remaining columns with $p_0 = \{1,10,20 \}$ increasing from left to right.  Areas below (above) the red (blue) curves are thermally stable. Analytical estimates derived in the text are shown as dashed lines; while the dots represent the results from numerical solutions of our disk model. The panels (a) and (b) represent BHs with masses $M = 10^4 \Msun$ and $10^6 \Msun$ respectively. }
\label{fig_Mdot_range_instability}
\end{figure*} 

\begin{figure*}
\begin{subfigure}{0.99\textwidth}
\centering
\includegraphics[width=0.99 \textwidth]{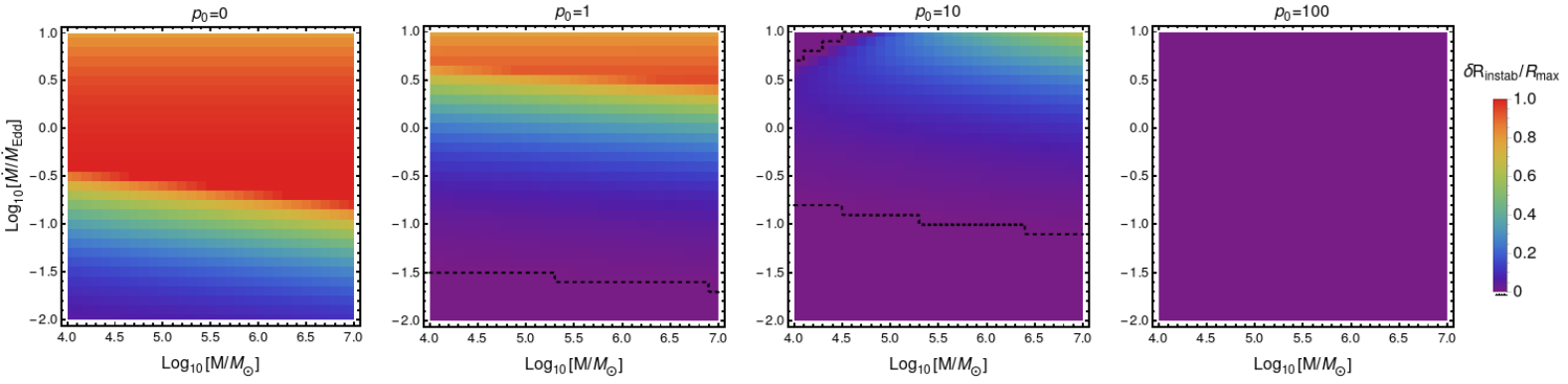}
\subcaption{$R_{\rm max} = 2000 r_g$}
\end{subfigure}
\\
\begin{subfigure}{0.99\textwidth}
\centering
\includegraphics[width=0.99 \textwidth]{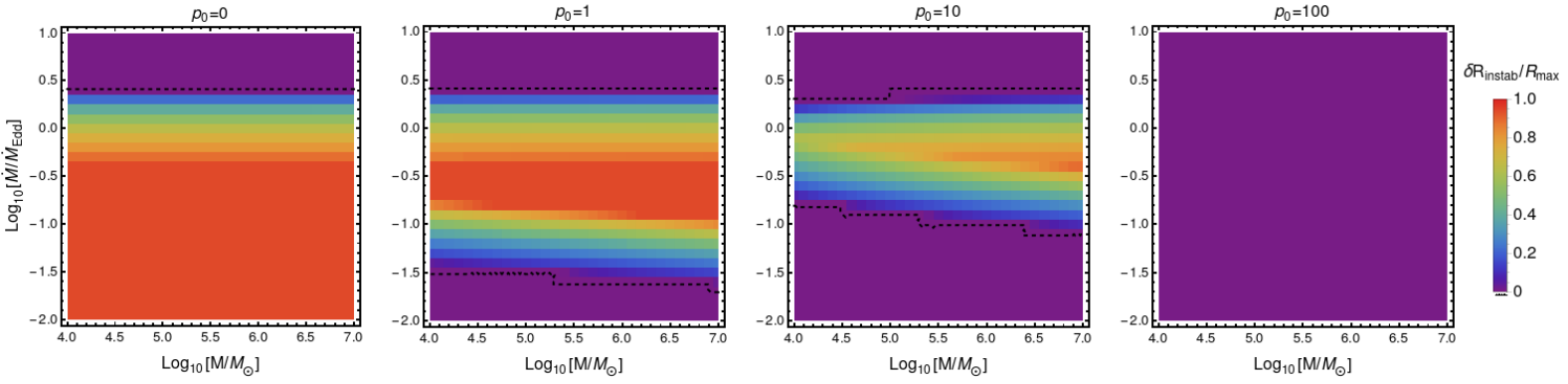}
\subcaption{$R_{\rm max} = 100 r_g$}
\end{subfigure}
\caption{The fraction of radii in a disk solution that are thermally unstable, $\delta R_{\rm instab}/R_{\rm max}$, is shown for non-magnetic ($p_0 = 0 $) and magnetic ($p_0 \neq 0$) disks. Stability checks employing relation~(\ref{thermal_instab_spl}) are done by constructing numerical disk solutions for $R \leq 2000 r_g$ (in upper panels) and $R \leq 100 r_g$ (in lower panels). Dashed black contours correspond to a marginally stable disk with $\delta R_{\rm instab} =0$. The unstable region with $\delta R_{\rm instab} \neq 0$  between the two dashed contours shrinks with increasing $p_0$. The smaller disks (lower panel) have a lower threshold $\dot{M}/\MdotEdd \simeq 2.5$ (the upper dashed curve) above which the whole disk stabilizes by advective cooling. This threshold $\dot{M}$ does not change much with increasing $p_0 \leq 10$, but the whole disk stabilizes for very high $p_0=100$.    }
\label{fig_delR_instab}
\end{figure*}

\begin{figure*}
\centering
\includegraphics[width=0.7\textwidth]{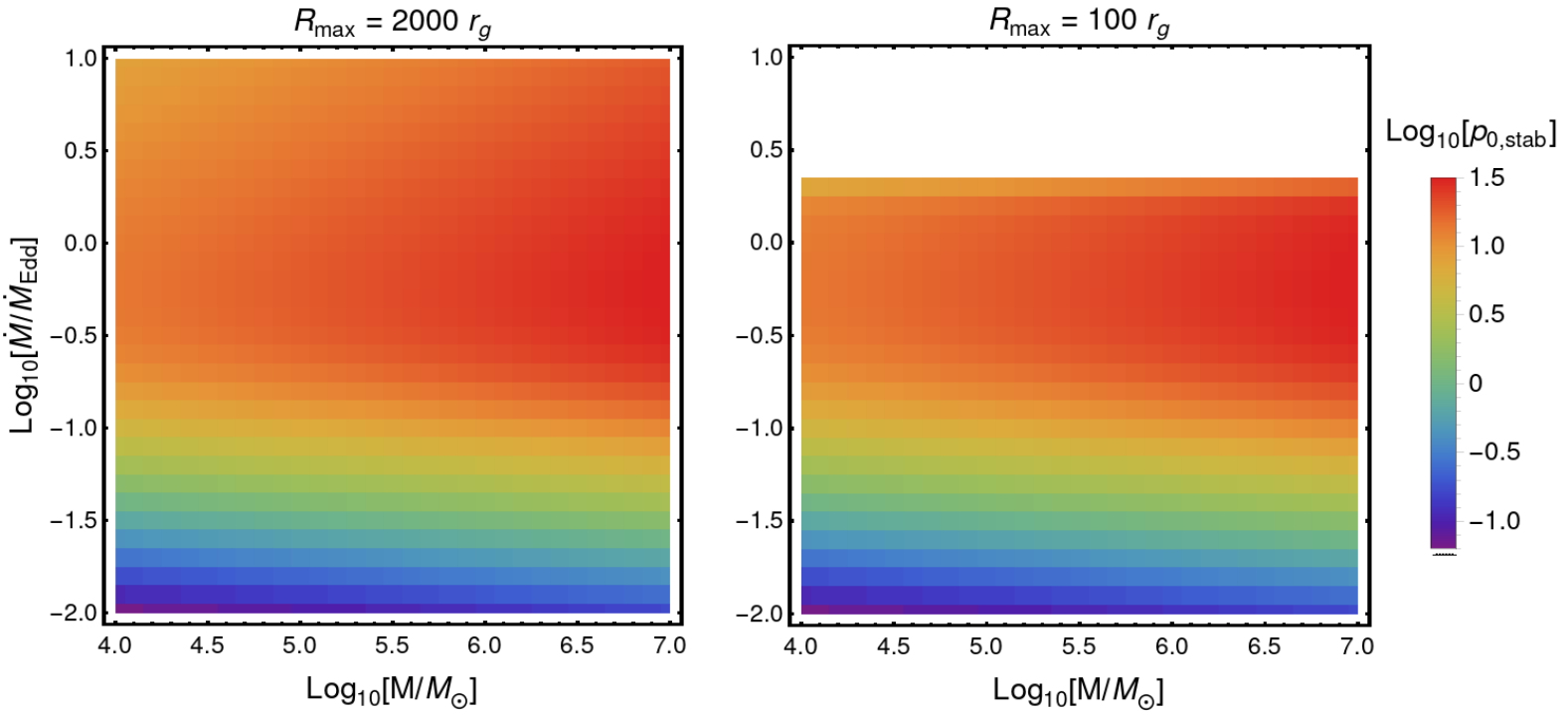}
\caption{The minimum value of $p_0 = p_{0, \rm stab}$ (in log scale) required to stabilize a disk at all radii $R \leq R_{\rm max} = \{ 2000, 100 \} r_g$ (shown, respectively, in the left and right panels), in the $\{M , \dot{M} \}$- plane. The range of $ p_{0,\rm stab } \in [0.06, 32]$ is sufficient to stabilize these disks. Blank space in the right panel (small disks) corresponds to the high $\dot{M}$ disks stabilized by advective cooling, that do not require magnetic fields for stability in the inner disk regions. } 
\label{fig_p0stab}
\end{figure*}

For high mass accretion rates, radiation pressure tends to dominate in the inner regions of a disk. This can enhance the rate of heating per unit change in the disk temperature, 
possibly resulting in thermal instability.  The following formal condition governs the occurrence of thermal instability in the linear perturbation regime \citep{frank_king_raine2002}: 
\beq 
\frac{\p q^{-}}{\p T_{\rm c}} \bigg|_{\Sigma} <  \frac{\p q^{+}}{\p T_{\rm c}} \bigg|_{\Sigma} 
\eeq 
Both derivatives are evaluated at constant $\Sigma$.  There exists a more explicit stability relation~(\ref{thermal_stab_fin}) derived in appendix~\ref{app_stability}. We rewrite it here (as a criterion for thermal instability) for the special case of interest here for which electron-scattering is the dominant opacity source ($\kappa = \kes$): 
\beq 
( \Pgas + 0.5 \Pmag + 4 \Prad ) \qadv  + ( 3 \Pgas + 3.5 \Pmag - 2 \Prad ) \qrad \leq 0
\label{thermal_instab_spl}
\eeq 
This inequality explicitly demonstrates the earlier findings about the stabilizing tendency of magnetic pressure support (BP07). This more general, but nevertheless simple, treatment of thermal stability allows us to study the stability of magnetic disks analytically. High accretion rates ($\gtrsim$ a few $\%$ of Eddington rate) may lead to dominance of $\Prad$ in the inner disk, leading to instability if the magnetic pressure support is insufficient. Above still higher accretion rates ($\gtrsim$ a few $\times 10 \%$ of Eddington rate), advective cooling can dominate, $\qadv \gg \qrad $, stabilizing the inner-most regions of the disk. For these high accretion rates, the unstable region exists only at some {\it intermediate} range of radii, with the inner-most regions stabilized by the dominance of advective cooling and outer regions stabilized by the dominance of magnetic (and/or gas) pressure. These local stability properties are emphasized in figure~\ref{fig_Mdot_range_instability}, which depicts the radial dependence of the unstable range for different Eddington ratios $\dot{M}/\MdotEdd$ (enclosed between the two curves) which satisfy the instability relation~(\ref{thermal_instab_spl}). Throughout this section, we compute the numerical disk solutions for $\alpha = 0.1$.\footnote{The question of local stability is a weak function of $\alpha$; later on, we will see that $\alpha$ does significantly effect the timescales of instability limit-cycles (\S~\ref{sec:observables}). } This demonstrates that the increasing relative contribution of magnetic pressure support (with increasing $p_0$) causes a reduction in: (1) the unstable range of Eddington ratios $\dot{M}/\MdotEdd$;  (2) the unstable range of disk radii for a given $\dot{M}/\MdotEdd$. We study below in more detail the modification of instability properties due to magnetic pressure support, discussing first the standard non-magnetic accretion disks for comparison.

\emph{Instability of non-magnetic disks}: If $\dot{M}$ is not so high, and radiative cooling is the dominant cooling mechanism ($\qrad \gg \qadv$), then the inner regions of disk dominated by $\Prad$ may become thermally unstable. For a non-magnetic disk in this regime ($P = \Prad + \Pgas$, $q^{-} = \qrad$, $\kappa = \kes$), the local thermal instability occurs for $\Prad \geq 0.6 P$ (using equation~\ref{thermal_instab_spl}). So, the minimum local mass accretion rate $\dot{M}_{\rm l,min}$ for instability to occur, is given as: 
\beq 
\frac{\dot{M}_{\rm l,min}}{\MdotEdd} = 4.7 \times 10^{-4} M_6^{-1/8} \alpha_{-1}^{-1/8} \,\frac{ (R/10 r_g)^{21/16} }{f(R)} . 
\label{Mdot_l_min}
\eeq 
Alternatively, this gives the radial distance $R_{\rm st,o}$ within which disk is unstable for a given $\dot{M}$:
\beq  
R_{\rm st,o} = 3380 r_g M_6^{2/21} \alpha_{-1}^{2/21} \bigg( \frac{ \dot{M} }{ \MdotEdd  }  \bigg)^{16/21}
\label{R_stab_o}
\eeq 
Interestingly at $R \simeq 11.4 r_g$, $\dot{M}_{\rm l,min}$  attains its lowest value $\dot{M}_{\rm min}$ and then, keeps increasing uniformly at larger radii. This global minimum $\dot{M}_{\rm min}$ is the critical accretion rate at which instability begins to occur in the disk:
\beq 
 \frac{ \dot{M}_{\rm min} }{\MdotEdd} = 2 \times 10^{-3} M_6^{-1/8} \alpha_{-1}^{-1/8} 
 \label{Mdot_min}
\eeq 

For high mass accretion rates $\dot{M}/\MdotEdd$, $\Prad$ becomes increasingly dominant as a pressure source, and the thickness of the disk increases. Under such circumstances, cooling by radial advection becomes significant. In this regime ($P = \Prad$, $q^{-} = \qrad + \qadv$, $\kappa = \kes$), the disk is locally stable if $\qadv \geq \qrad /2$. So the maximum local accretion rate $\dot{M}_{\rm l,max}$ below which instability occurs (that is, below which the advective cooling is insufficient to stabilize) is: 
\beq  
\frac{\dot{M}_{\rm l,max}}{\MdotEdd} = 0.32 \frac{(R/10 r_g)}{\sqrt{f(R)}}  
\label{Mdot_l_max}
\eeq 
$\dot{M}_{\rm l,max}$ has a minimum at $R  = 9.37 r_g$ and the corresponding $\dot{M}_{\rm max} = 0.68 \MdotEdd$ is the minimum mass accretion rate at which advective cooling begins to play a role in the disk stability. The above expression also gives the radial distance $R_{\rm st,i} = 30.9 r_g \dot{M}/\MdotEdd  $ inside which the disk is stable due to advective cooling for a given $\dot{M} \gtrsim \dot{M}_{\rm max}$.   

The analysis of above two cases implies that non-magnetic disks are unstable in the following range of radii:
\begin{itemize} 
\item  For $\dot{M}_{\rm min} \leq \dot{M} \leq \dot{M}_{\rm max}$, advective cooling is insufficient for stability throughout the inner disk and instability occurs in the inner disk for $R \leq R_{\rm st,o}$. 

\item For $\dot{M} > \dot{M}_{\rm max}$, advective cooling can stabilize the innermost regions of the disk ($R \leq R_{\rm st,i}$).  Instability due to $\Prad$ dominance will persist for $R \in [R_{\rm st,i}, R_{\rm st,o}]$ if the disk is large enough (i.e. if it extends beyond $R_{\rm st,i}$).  
\end{itemize}
It is noteworthy that $R_{\rm st,i}$ increases faster than $R_{\rm st,o}$ with $\dot{M}$, decreasing the radial range of instability. However it is not possible to completely eliminate the $\Prad$ instability for arbitrarily large disks (without any constraint on its outer radius $R_{\rm max}$) for practically feasible values of $\dot{M} \ll 3.67 \times 10^8 \MdotEdd M_6^{2/5}$. But for disks with a finite radial size $R_{\rm max}$, it is possible to determine a minimum $\dot{M} = \dot{M}_{\rm l,max} (R = R_{\rm max})$ (or equivalently an $\dot{M}$ for which $R_{\rm stab, i} = R_{\rm max}$ holds) above which the whole disk is stabilized by advective cooling. For example, a disk of size $R_{\rm max} = 100 r_g$ is stable for $\dot{M} \gtrsim 3.24 \MdotEdd$. Figure~\ref{fig_delR_instab} depicts the range of unstable disk radii $\delta R_{\rm instab} = R_{\rm st,o}-R_{\rm st,i}$ as computed from numerical disk solutions in $\{M, \dot{M}/\MdotEdd \}$-plane for two disk sizes, $R_{\rm max} = \{ 100, 2000 \} \, r_g$. This figure demonstrates that whether or not $\Prad$ instabilities exist can depend sensitively on disk size $R_{\rm max}$. 

\emph{Instability of magnetic disks}: Magnetic disks that are cooled by radiative diffusion are locally stable for $\Pmag \geq 4 \Prad / 7$. This bound is deduced by choosing $P = \Prad + \Pmag$ and $q^{-} = \qrad$ in the instability condition of equation~(\ref{thermal_instab_spl}). Here, in a simplified analysis, we completely neglect gas pressure $\Pgas = \Pmag c_{\rm s,g}/\vK \ll \Pmag$, which is a good approximation for our magnetic disk models. This gives a local minimum mass accretion rate $\dot{M}_{\rm l,min}^{\rm B}$ above which instability happens at radius $R$:  
\beq 
\frac{\dot{M}_{\rm l,min}^{\rm B}}{\MdotEdd} = 0.016 \sqrt{p_0} \frac{( R/10 r_g )^{37/32} }{f(R)} M_6^{-1/16} \alpha_{-1}^{-1/16} 
\label{Mdot_l_minB}
\eeq 
At $R = 12.3 r_g$, $\dot{M}_{\rm l,min}^{\rm B} = \dot{M}_{\rm min}^{\rm B} $ is a minimum point. So, the minimum mass accretion rate $\dot{M}_{\rm min}^{\rm B}$ for instability to happen in a disk: 
\beq 
\frac{\dot{M}_{\rm min}^{\rm B}}{\MdotEdd} = 0.066 \sqrt{p_0} M_6^{-1/16}  \alpha_{-1}^{-1/16}
\label{Mdot_minB}
\eeq 
which is more than an order of magnitude higher than $\dot{M}_{\rm min}$ for $p_0 \geq 1$. 

Equation~(\ref{Mdot_l_minB}) can be rewritten to give the disk radius $R_{\rm st,o}^{\rm B}$ within which instability occurs for a given $\dot{M}$:  
\beq 
R_{\rm st,o}^{\rm B} = 363 r_g \frac{ M_6^{2/37} \alpha_{-1}^{2/37}}{  p_0^{16/37} } \bigg(  \frac{ \dot{M}  }{ \MdotEdd  }  \bigg)^{32/37} 
\label{R_stab_oB}
\eeq 
Magnetic pressure support shrinks the outer edge of unstable radii $R_{\rm st,o}^{\rm B}$ by more than an order of magnitude for $p_0\ge 1$. $R_{\rm st,o}^{\rm B}$ gets smaller with increasing $p_0$. 

In the regime where advective cooling is significant, we find from the numerical solutions presented in figure~\ref{fig_SS_mag_disks2} that $\Prad$ still strongly dominates the local pressure at many radii  for $p_0 \leq 100$. Thus, we expect that the inclusion of $\Pmag$ will produce only small changes in the maximum $\dot{M}_{\rm l,max}$ (and $\dot{M}_{\rm max}$) above which advective cooling stabilizes the disk. Likewise, the effect of magnetic fields on $R_{\rm st,i}$ (the radius within which the disk is stable due to advective cooling) is expected to be small. This fact is illustrated clearly in figure~\ref{fig_delR_instab}.  
The unstable range of $\dot{M}$ in this figure shrinks with increasing $p_0$ as expected. However, for $R_{\rm max} = 100 r_g$ (lower panels), the upper dashed curve that encapsulates the effect of $R_{\rm st,i}$ does not show any significant variation with $p_0\leq 10$. Overall, we can summarize the range of thermally unstable disk radii in magnetic disks with the following two cases:
\begin{itemize}
    \item  For $ \dot{M}_{\rm min}^{\rm B} \leq \dot{M} \leq \dot{M}_{\rm max}$, advective cooling is ineffective in stabilizing the disk. However, due to magnetic pressure support, the thermal instability occurs over a narrower range of radii $R \in [R_{\rm i}, R_{\rm st,o}^{\rm B}]$, with a smaller outer boundary $R_{\rm st,o}^{\rm B} \ll R_{\rm st,o}$ in comparison to the unmagnetized case. 
    
    \item For $\dot{M} > \dot{M}_{\rm max}$, advective cooling stabilizes the innermost regions of the disk $R \lesssim R_{\rm st,i}$ (similar to the non-magnetic case for small $p_0$). The range of unstable radii $R \in [R_{\rm st,i}, R_{\rm st,o}^{\rm B}]$ for a magnetized disk is again narrower than that of the non-magnetized disk due to $R_{\rm st,o}^{\rm B} \ll R_{\rm st,o}$.  
\end{itemize}
We can find a maximum value of $p_0$, which we call $p_{\rm 0 m}$, needed for the local stability at a given $R$ by setting $ \dot{M}_{\rm l,min}^{\rm B} = \dot{M}_{\rm l, max}$:
\beq 
p_{\rm 0 m} = 423 M_6^{1/8} \alpha_{-1}^{1/8} f(R) \bigg( \frac{R}{10 r_g} \bigg)^{-5/16}.
\eeq 
The value of $p_{\rm 0 m}$ has a global maximum value $p_{\rm 0 g m} = 168 M_6^{1/8} \alpha_{-1}^{1/8}$ at $R = 40.6 r_g$. The predictions of this simplified analysis are higher by a factor of few as compared to numerical evaluations of $p_{\rm 0 g m}$. Figure~\ref{fig_p0stab} shows density plots in the $\{M,\dot{M} \}$-plane, with colors representing the maximum $p_0$ needed to globally stabilize a disk with size $R_{\rm max} = \{2000, 100  \} r_g$. It is evident that $p_0 \lesssim 32$ is sufficient to stabilize the magnetic disks for the whole range of parameters explored here. The mediocre performance of the simple prediction for $p_{\rm 0 g m}$ comes from neglecting the contribution of magnetic pressure $\Pmag$ in determining the smallest $\dot{M} = \dot{M}_{\rm l, max}$ above which advective cooling becomes important (especially for the high values of $p_0$). Magnetic pressure support will further facilitate stability, decreasing the limit $\dot{M}_{\rm l, max}$ meaning that significantly smaller $p_{\rm 0 m}$ (and hence $p_{\rm 0 g m}$) suffice for stability.


\section{Observational implications}
\label{sec:observables}

In this section, we investigate the observable signatures of strongly magnetized disks and the occurrence (or not) of thermal instability. In particular, we estimate (1) the timescale of instability limit-cycles for magnetized disks across a broad BH mass range, with an emphasis on the large intermediate-mass BHs and small supermassive BHs relevant for quasi-periodic eruptions, (2) the hardened black-body emission associated with magnetic disks, and the point at which they become (effectively) optically thin, and (3) the energy release in relativistic jets powered by spinning BHs in a magnetic environment.

\subsection{Implications for Quasi-Periodic Eruptions}
\label{sec_qpe}

\begin{figure*}
\begin{subfigure}{0.46\textwidth}
\centering
\includegraphics[width=0.99 \textwidth]{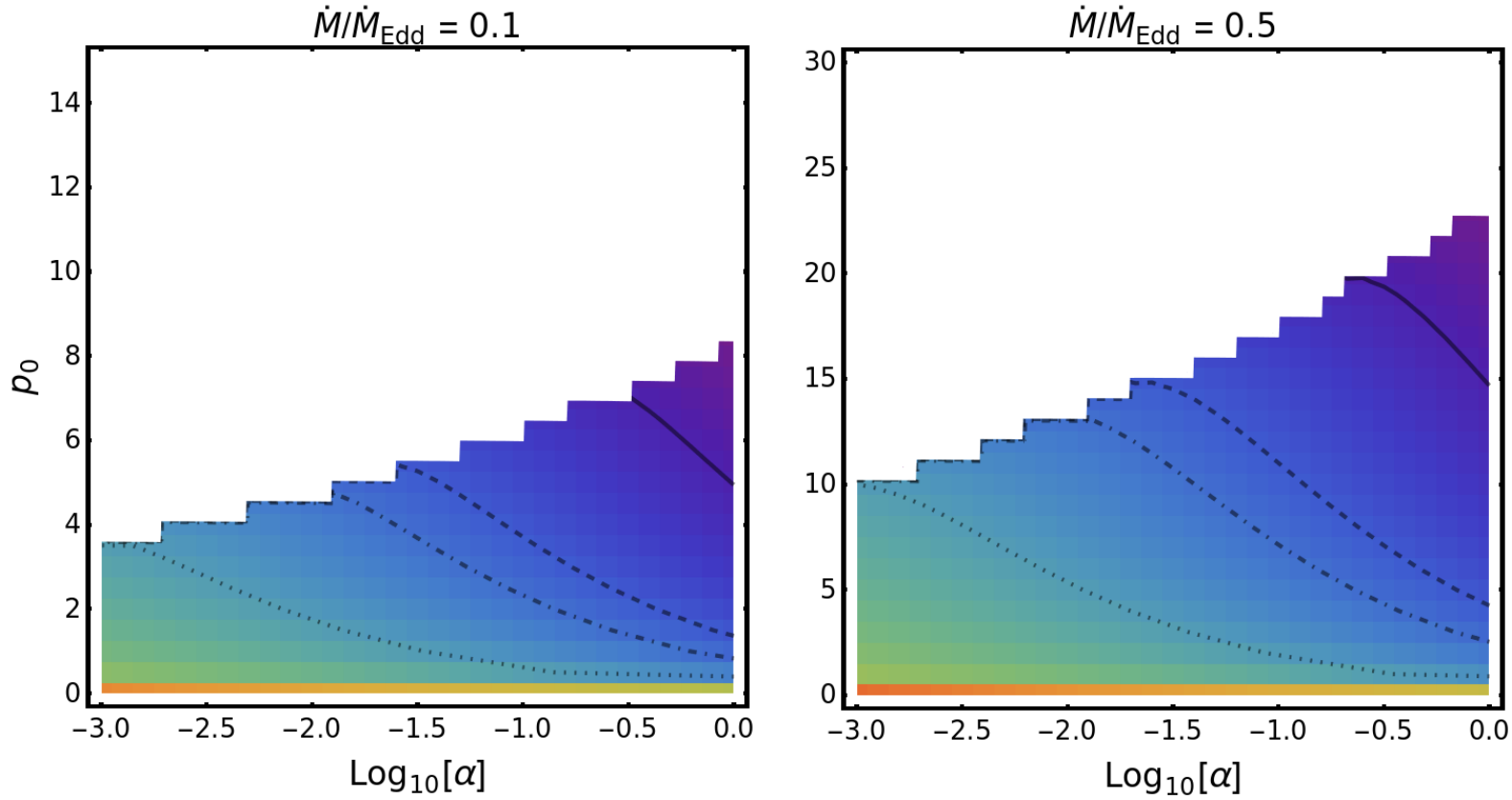}
\subcaption{$M = 10^5 \Msun$}
\end{subfigure}
\hfill 
\begin{subfigure}{0.51\textwidth}
\centering
\includegraphics[width=0.99 \textwidth]{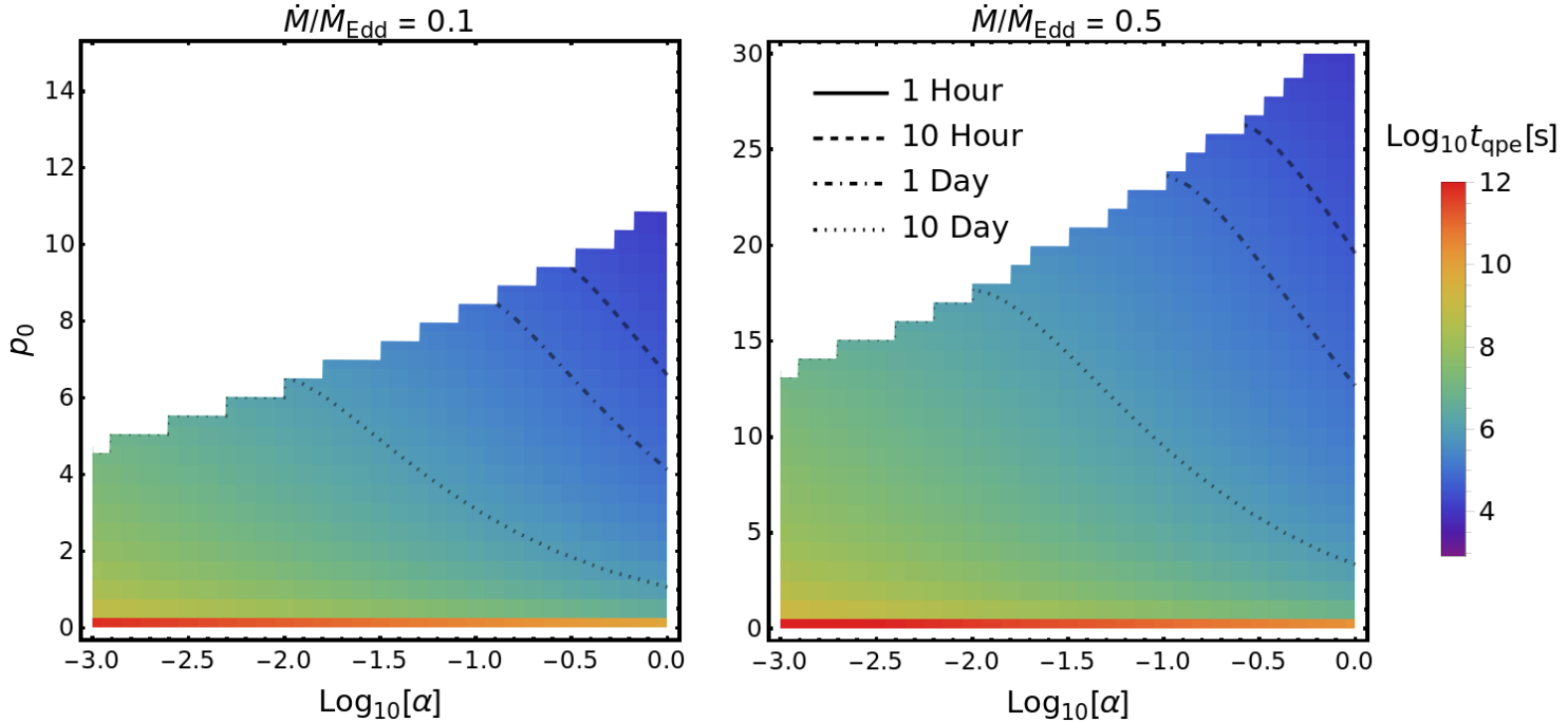}
\subcaption{$M = 10^6 \Msun$} 
\end{subfigure}
\caption{Instability periods ($t_{\rm qpe}$) are shown as a color-coded density plot in the $\{ \log_{10}\alpha,p_0  \}$-plane for a general disk with (a) $M = 10^5 \Msun$ and (b) $M=10^6 \Msun$, with two typical Eddington ratios: $\dot{M}/\MdotEdd = 0.1, 0.5$. For shorter periods consistent with QPE observations ($< 1$d), larger $\alpha$ values and moderately high $p_0$ values are needed (although $p_0$ must be low enough to avoid the region of complete disk stability, shown in white). Extremely long periods for non-magnetic disks ($p_0 = 0$) stand out vividly against the much shorter time-periods for magnetized disks ($p_0 > 0$).      
 }
\label{fig_tqpe_dp2}
\end{figure*}

\begin{figure*}
\centering
\includegraphics[width=0.99 \textwidth]{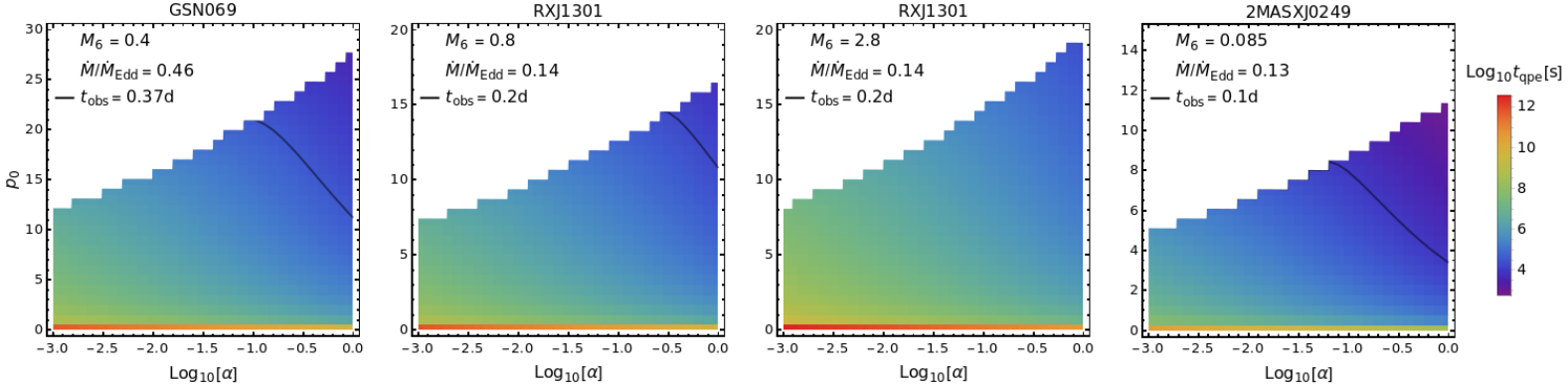}
\caption{Instability periods deduced from our disk models for three observed QPE sources in the $\{ \log_{10}\alpha,p_0  \}$-plane. RXJ1301 is presented in panels 2 and 3 for lower and upper limits on its BH mass. The solid contour represents the observed period and tracks the range of suitable $\alpha$ and $p_0$ for these sources. Larger BH masses require larger values of $\alpha$; the largest inferred BH hosting a QPE, RXJ1301, requires $\alpha \gtrsim 0.3$ (for its lower mass estimate $M = 8 \times 10^5 \Msun$; no suitable $\alpha$ and $p_0$ combination exists for its upper mass limit). Larger $p_0 > 10$ are needed for larger $M$ or $\dot{M}/\MdotEdd$ (for GSN 069 and RXJ1301); while 2MASXJ0249 needs  $p_0 \sim 4-8$.     
 }
\label{fig_tqpe_obs_dp2}
\end{figure*}

\begin{table*}
\centering
\footnotesize
\begin{tabular}{|c| c| c| c| c| c|c|}
  \hline

  Source & $M (10^6 \Msun)$ & $\dot{M} (\MdotEdd)$ & $t_{\rm obs}$(d)  & $(\alpha, p_0)_1$ & $(\alpha, p_0)_2$ &  Refs.  \\

  \hline
  
GSN 069  &  0.4 &   0.46  & 0.37 & (0.1, 21)   &   (1, 11.2)  &   \citet{Miniutti_2019} \\ 
        
\hline 

RXJ1301 & 0.8 &  0.14   & 0.2 & (0.32, 14.5) &  (1, 10.8)       & \citet{Giustini_2020} \\
               &  2.8 &         &   &   -   &  -          &     \\
               \hline
2MASXJ0249 & 0.085 &  0.13 & 0.1  & (0.063, 8.46  )  & (1, 3.4)     & \citet{Chakraborty_2021}  \\                    
      \hline 
eRO-QPE1      &    0.6 & $0.1^{*}$  &  0.77 & (0.08, 7.9 ) & (1, 3.36)  & \citet{Arcodia_2021},  \\ 
  &   &  &  &  &  &   \citet{Wevers_2022} \\
\hline
eRO-QPE2      & 0.09  & $0.1^{*}$  &  0.1 &  (0.1, 6.26) & (1, 2.9) & \citet{Arcodia_2021},  \\
 &   &  &  &  &  &   \citet{Wevers_2022} \\
\hline

\end{tabular}
\caption{Observed QPEs: this table compiles observationally measured and inferred properties of known QPE hosts.  This includes the mass of the central BH ($M$), the accretion rate in Eddington units $\dot{M}$, and the observed QPE recurrence time $t_{\rm obs}$.  The two sets of extreme parameters $(\alpha,p_0)_1$ and $(\alpha,p_0)_2$ for which the periodicity of instability limit cycles matches the observed QPE periods $t_{\rm obs}$ are also included as separate columns. These points can be interpreted as the end points of the black contours in figure~\ref{fig_tqpe_obs_dp2}: $(\alpha,p_0)_1$ has the maximum $p_0$ above which the disk is stable, and $(\alpha,p_0)_2$ gives the minimum $p_0$ (for $\alpha = 1$). ${}^{*}${We use a typical value of $\dot{M}/\MdotEdd = 0.1$ for eRO-QPE1 and eRO-QPE2.} 
}
\label{table:QPE_times}
\end{table*}

Recent X-ray observations of short period quasi-periodic eruptions (QPEs) cannot be explained by simple models of thermo-viscous instabilities in $\Prad$ dominated disks, as the viscous time is far too long \citep{Arcodia_2021}, but alternative models of accretion disk instabilities may be viable \citep{Raj_Nixon2021, Sniegowska2022, Pan_2022_viscosity_model}. There are also other a number of other hypotheses for their origin, such as repeating mass transfer from a stellar component (or a binary) orbiting the central BH \citep{King_2020,King_2022_QPE_models,Metzger_Stone_2022,Wang2022,Nixon2022,Zhao2022,Krolik2022}; interactions of an orbiting stellar component with already existing accretion disk \citep{sukova2021,Xian2021}; and gravitational lensing \citep{Ingram2021}, although this is disfavored by the achromatic evolution of QPEs. Here we focus on disk instability of magnetized disks (\S~\ref{sec:thermal}) to examine whether it may explain QPE phenomena.

Observed QPE sources generally have black holes with $M \sim 10^5 - 10^6 \Msun$, accreting roughly at an Eddington ratio of $\dot{M}/\MdotEdd \sim 0.1-0.5$, and with observed QPE recurrence times on the order of a few hours up to one day \citep{Miniutti_2019,Giustini_2020,Arcodia_2021,Chakraborty_2021}. The classical radiation pressure instability for Shakura-Sunyaev disks yields much longer limit-cycles, as the recurrence time is governed by the viscous time at the outer edge of the unstable zone. For an average mass accretion rate $\dot{M} > \dot{M}_{\rm min}$ (equation~\ref{Mdot_min}), the instability occurs within a range of radii with the outer bound $R_{\rm st,o} = 3380 r_g M_6^{2/21} \alpha_{-1}^{2/21} (\dot{M}/\MdotEdd)^{16/21}$, given in equation~\ref{R_stab_o}. The recurrence time $t_{\rm qpe}$ of radiation pressure limit-cycles should be of the same order of magnitude as the viscous timescale $t_{\nu} = R^2/\nu$ at $R_{\rm st,o}$, reflecting the timescale on which gas accumulates at the outer edge of an unstable zone \citep{Cannizzo_1996,Belloni+97}. One can estimate this timescale analytically by considering a marginally stable disk with $\Prad = 0.6 P$ (and $P = \Pgas + \Prad$, $\kappa = \kes$, $q^{-} = \qadv$ as earlier in \S~\ref{sec:thermal}):
\beq 
t_{\rm qpe} = t_{\nu}(R_{\rm st,o}) \simeq  2350~{\rm yr} \, M_6^{4/3} \alpha_{-1}^{-2/3} \bigg( \frac{\dot{M}}{\MdotEdd} \bigg)^{2/3} 
\label{t_qpe}
\eeq 
For $M = 4 \times 10^5 \Msun$ and $\dot{M} = 0.46 \MdotEdd$ (roughly matching the parameters for GSN 069, the discovery object for this class of transients), we have $t_{\rm qpe} \simeq  410 ~{\rm yr}$ (for $\alpha = 0.1$) which is orders of magnitude larger than its observed recurrence time, $\simeq 0.37~{\rm d}$ \citep{Miniutti_2019}.  

As seen earlier in \S~\ref{sec:thermal}, magnetic pressure support tends to stabilize the disk, with instability occurring only for $\dot{M} > \dot{M}_{\rm min}^{\rm B} \gg \dot{M}_{\rm min}$ (equation~\ref{Mdot_minB}), shrinking the range of unstable $\dot{M}$. Further, for a given $\dot{M}$, the range of unstable radii shrinks, with a smaller outer boundary radius $R_{\rm st,o}^{\rm B} = 360 r_g M_6^{2/37} \alpha_{-1}^{2/37} p_0^{-16/37} (\dot{M}/\MdotEdd )^{32/37} \ll R_{\rm st,o}$ (equation~\ref{R_stab_oB}). Again estimating the viscous timescale at $R_{\rm st,o}^{\rm B}$ for a marginally stable disk with $\Pmag = 4 \Prad/7$ (and $P = \Prad + \Pmag$, $\kappa = \kes$, $q^{-} = \qrad$ as in \S~\ref{sec:thermal}), gives an analytical expression for the time period $t_{\rm qpe}^{\rm B}$ of limit-cycles in a magnetic disk:
\beq 
t_{\rm qpe}^{\rm B} = t_{\nu}(R_{\rm st,o}^{\rm B}) \simeq 1.1~{\rm yr} \, \frac{ M_6^{44/37} }{ p_0^{56/37} \alpha_{-1}^{30/37} } \bigg( \frac{ \dot{M} }{\MdotEdd }  \bigg)^{38/37} 
\label{t_qpeB}
\eeq 
 For GSN 069, this expression gives $ t_{\rm qpe}^{\rm B}\simeq 0.64~{\rm d}$ for the parameters $\{ \alpha = 0.1 , p_0 = 20  \}$, which agrees at the order of magnitude level with the observed period. Choosing such moderately high values of $p_0$ gives QPE time-periods in good agreement with observations. However, a magnetized disk becomes completely stable and can not exhibit QPEs once $p_0$ exceeds a critical value, $p_0 \geq 230 M_6^{1/8} (\dot{M}/\MdotEdd )^2 \alpha_{-1}^{1/8} $ (this is obtained from the stability criterion $\dot{M} \leq \dot{M}_{\rm min}^{\rm B}$). This analytical and approximate upper bound $p_0$ is actually an overestimate. As $p_0$ (or the relative magnetic contribution to $P$) increases, the advective term, which is ignored in these analytical estimates, begins to contribute significantly to cooling due to the larger aspect ratios of magnetic disks; this exerts an additional stabilizing effect.         
 
 We numerically calculate the instability limit-cycle period, which we take to be the viscous timescale at the outer edge of the unstable region, for the general disk models of \S~\ref{sec:disks}, in which the total pressure $P = \Prad + \Pmag + \Pgas$ and the cooling rate $q^{-} = \qrad + \qadv$. The resulting  timescales are depicted in figure~\ref{fig_tqpe_dp2} in the $\{\log_{10}\alpha , p_0\}$- plane for $M = \{10^5 , 10^6  \}\Msun$ and $\dot{M}/\MdotEdd = \{0.1 ,0.5 \} $, which represent approximate parameter bounds for the observed QPE sources. Shorter recurrence timescales, which are needed to match QPE observations, tend to occur for larger $\alpha$ and $p_0$ values. The requirement that an unstable zone exist somewhere in the disk provides an upper bound on $p_0$ (the white space in figure~\ref{fig_tqpe_dp2} represents stable disks). To obtain recurrence times $\lesssim 1 $d (consistent with observations), (a) the suitable range of $\alpha$ values is mainly a function of BH mass, with $\alpha \gtrsim 0.01$ (0.1) for $M = 10^5 \Msun$ $(10^6 \Msun)$; (b) the suitable range of $p_0$ tends towards larger values for higher $M$ and $\dot{M}/\MdotEdd$. 
 It is interesting to note the extremely long timescales for non-magnetized disks ($p_0=0$, above the horizontal axis) which appears as a discontinuity in figure~\ref{fig_tqpe_dp2}. 
 
 Beyond these general considerations, we also try to find $\{\alpha,p_0\}$ combinations that reproduce observed recurrence times in the three QPE sources in figure~\ref{fig_tqpe_obs_dp2} (we select these three sources because of the availability of literature constraints on $M$ and $\dot{M}$). The solid contour in each panel is the locus of $\{\log_{10}\alpha,p_0\}$ values for which $t_{\rm qpe}^{\rm B}$ equals the observed QPE time. For GSN069 and 2MASXJ0249
 , $\alpha \gtrsim 0.1$ is required, while the higher BH mass QPE RXJ1301 needs $\alpha \gtrsim 0.3$
 . Note that if the BH mass for RXJ1301 is as large as its inferred upper limit, we are unable to produce the observed recurrence time for any value of $\{\alpha, p_0 \}$. The required values for disk magnetic field strengths, $10 \lesssim p_0 \lesssim 15-20$, are higher for systems with either higher $M$ (like RXJ1301) or $\dot{M}$ (like GSN069). We list the extreme pairs of $\{\alpha,p_0\}$ (end points of the black contours of the figure~\ref{fig_tqpe_obs_dp2}) for which the calculated limit-cycle time equals the observed QPE time in table~\ref{table:QPE_times}. 
 
 We note that \citet{Sniegowska2022, Pan_2022_viscosity_model} also introduced magnetized disk models in order to explain QPEs as limit-cycle behavior in unstable accretion disks, but unlike in this work, neither of these models considers a disk dominated by magnetic pressure support (see e.g. the equation~24 of \citealt{Sniegowska2022}, or table 1 of \citealt{Pan_2022_viscosity_model}). 
 In order to reduce viscous limit cycle times down to observed QPE values, \citet{Sniegowska2022} invoke smaller disk radii (e.g. due to TDEs), while \citet{Pan_2022_viscosity_model} invoke angular momentum loss in a magnetized wind. This is in contrast to the large reduction in viscous limit-cycle times found in our models, due primarily to the larger aspect ratios of magnetically dominated disks. 
Shorter limit-cycles in a magnetically dominated accretion disk have also been explored in some earlier works in the context of changing-look active galactic nuclei (CL AGNs) \citep{Dexter2019,Pan2021}.


\subsubsection{Instability for a broad range in BH masses }

Using the mass-scaling for the limit-cycle periods of a magnetic disk (equation~\ref{t_qpeB}) and the observationally deduced properties of QPE sources, one can  estimate approximate instability periods $t_{\rm p} \sim 1 {\rm d}~(M/10^5 \Msun)^{44/37}$ for a general $M$. This implies:
\begin{itemize}
    \item[(a)] For an X-ray binary (XRB) with a typical $M =10\Msun $, the limit-cycle period $t_{\rm p} = 1.5$s, which resembles heartbeat oscillations (on timescales $\sim 1-100$s) seen in a few XRBs \citep{Belloni+97,Altamirano_2011_XRB_obs,Bagnoli2015,Maselli2018}.  
    
    \item[(b)] For larger $M = 10^{7-8}\Msun$, the range of limit-cycle period $t_{\rm p} =0.7 - 10 $yr roughly matches the timescales for spectroscopic and photometric variability for CL AGNs \citep{Denney2014,LaMassa2015,Yang2018,Assef2018,Graham2020}.  
\end{itemize}

We calculate more exact limit-cycle timescales $t_p$ (as done earlier in this section for QPEs) for general disk models for a typical XRB source with $M = 10 \Msun$ and $\dot{M}/\MdotEdd = \{ 0.1, 0.5 , 1 \}$ and present these results in figure~\ref{fig_tp_XRBs}. The instability exists for relatively small values of $p_0\lesssim 2-6$ when $\alpha \sim 0.1$. In most of the unstable parameter space of $\{\log_{10} \alpha, p_0 \}$, $t_p \sim 1 - 100$s, in agreement with observed time-periods of heartbeat oscillations. But these instability cycles have been observed in only a few sources and are absent in the most XRBs. This 
could be explained if moderately higher $p_0$-disks (say, $p_0 \gtrsim 6$) exist in a majority of XRBs, stabilizing them against radiation pressure instabilities.  

We perform a similar calculation for AGNs with $M = 10^{7-8} \Msun$ (and $\dot{M}/\MdotEdd = \{ 0.1, 0.5 , 1 \}$) and present the resulting limit-cycle timescales in figure~\ref{fig_tp_AGNs}. Compared to XRBs, the radiation pressure instability 
can persist over a wider range of $p_0$ values. For all AGN cases, disks with $\alpha \sim 0.1$ are unstable for $p_0 \lesssim 10-40$. The majority of the unstable parameter space has instability timescales ranging from a few months to a few years, similar to observed variability timescales for CL AGNs. Instabilities in magnetized accretion disks may be a plausible cause of this variability \citep{Dexter2019,Pan2021,Sniegowska2022}, though there are other possible physical scenarios suggested in previous studies \citep{Ross2018,Noda2018,Scepi2021,Raj_Nixon2021,Raj2021,Feng2021}. The requirement of extremely long temporal baselines makes both the detection and periodicity determination of these sources hard. NGC 1566 is a repeating-CL AGN with semi-periodic optical outbursts \citep{Alloin1986,Oknyansky2019}. 
If disk instability emerges as a favourable scenario, future observations would help further to characterize the underlying magnetic accretion disks.

\subsection{Hardened thermal emission from stable TDE disks}
\label{sec_TDE_hardBB}



\begin{figure*}
\begin{subfigure}{0.99\textwidth}
\centering
\includegraphics[width=0.99 \textwidth]{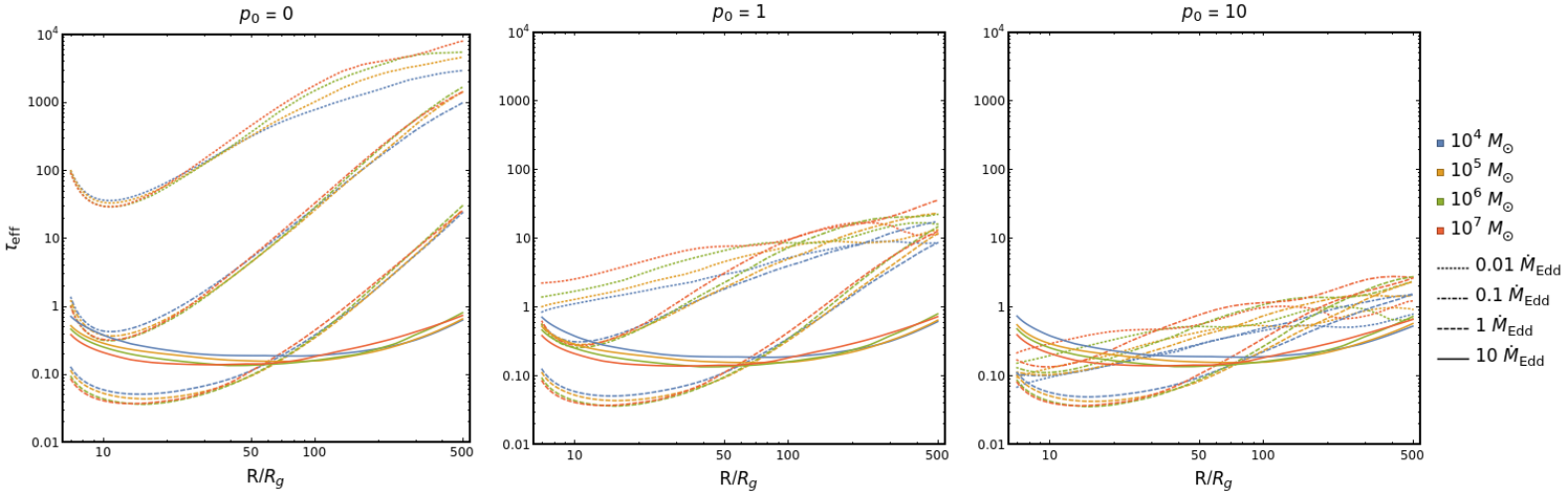}
\subcaption{$\alpha = 0.1$}
\end{subfigure}
\\
\vspace{0.5cm}
\begin{subfigure}{0.99\textwidth}
\centering
\includegraphics[width=0.99 \textwidth]{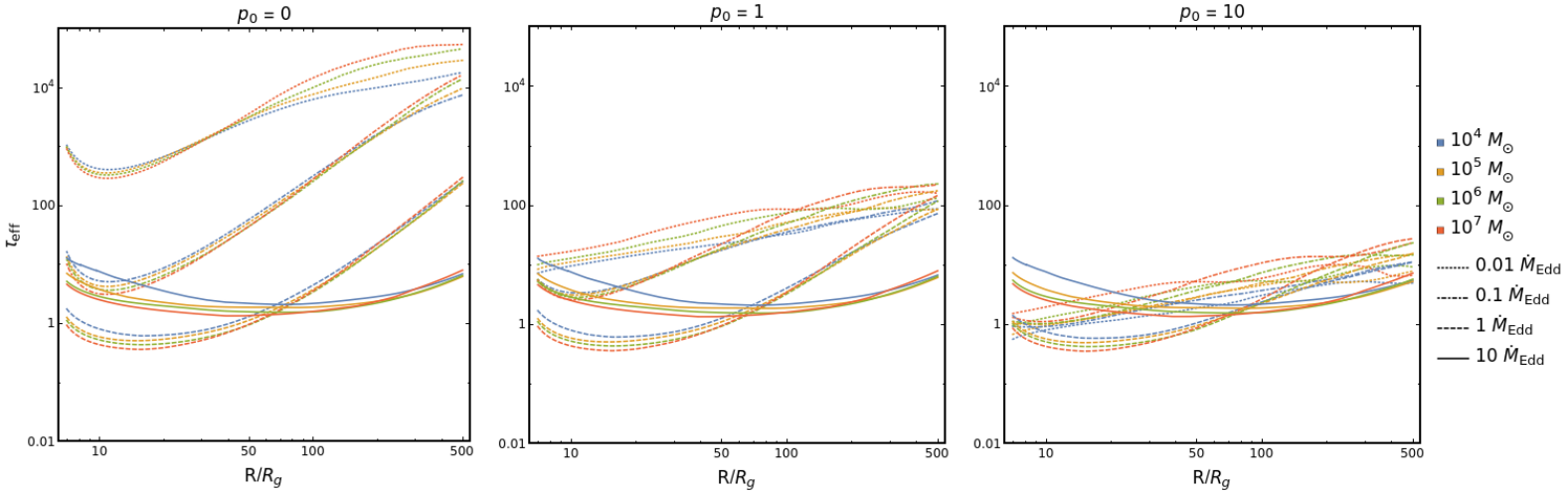}
\subcaption{$\alpha=0.01$}
\end{subfigure}
\caption{The effective optical depth $\tau_{\rm eff}$ for non-magnetic ($p_0=  0 $) and magnetic disks (for $p_0 = 1,10$) plotted as a function of dimensionless radius $R/R_{\rm g}$. As the contribution of magnetic pressure support increases with increasing $p_0$, $\tau_{\rm eff}$ falls by multiple orders of magnitude for radiatively cooled disks (i.e. lower $\dot{M}$ disks, and the outer disk regions for $\dot{M} = 1 \MdotEdd$). The properties of inner disk regions for high $\dot{M}$, where cooling by radial advection dominates, are only weakly affected by magnetic fields. For $\alpha=0.1$ (upper panel), the inner regions of a magnetized disk (for $p_0=10$) are effectively optically thin with $\tau_{\rm eff}<1$. But for smaller $\alpha = 0.01$ (lower panel), $\tau_{\rm eff}$ in the inner disk remains higher than unity for most of the parameter space.   
 }
\label{fig_tau_eff_prof}
\end{figure*}

\begin{figure*}
\centering
\includegraphics[width=0.99 \textwidth]{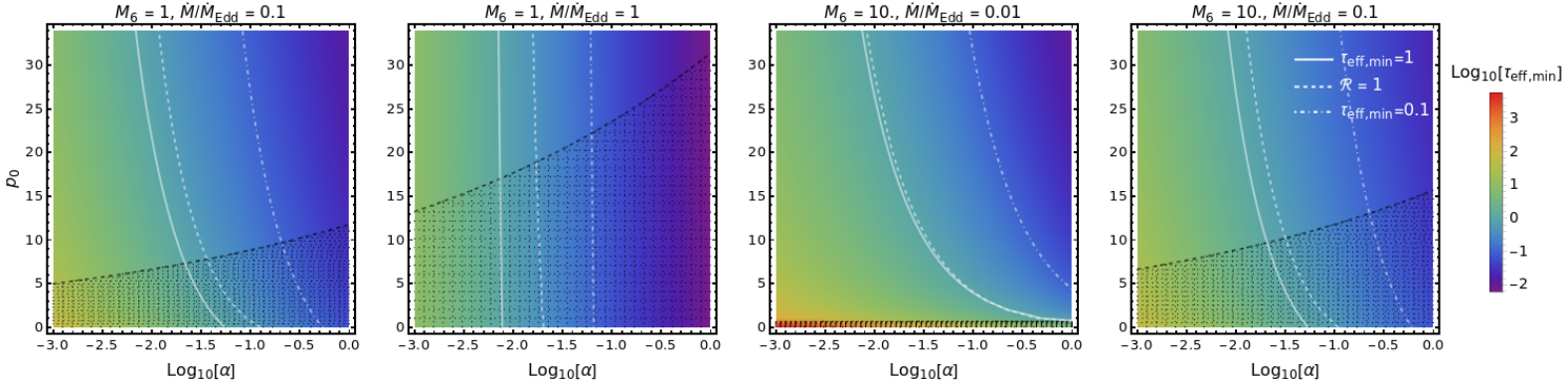}
\caption{Minimum effective optical depth $\tau_{\rm eff, min}$ in $\log_{10}[\alpha]-p_0$ plane for $M = 10^6 \Msun$ (left two panel) and $M = 10^7\Msun$ (right two panels). White contours represent different approximate criteria ($\tau_{\rm eff, min} = 1$ solid, $\mathcal{R} = 1$ dashed, $\tau_{\rm eff, min} = 0.1$ dot-dashed) for a marginally effectively optically thick disk. The dashed black lines (for $\mathcal{S}_{\rm th} = 0$) bound the unstable regions (shaded black zones are unstable).}
\label{fig_tau_eff2u}
\end{figure*} 

\begin{figure}
\centering
\includegraphics[width=0.5\textwidth, trim={0 0 0 0},clip]{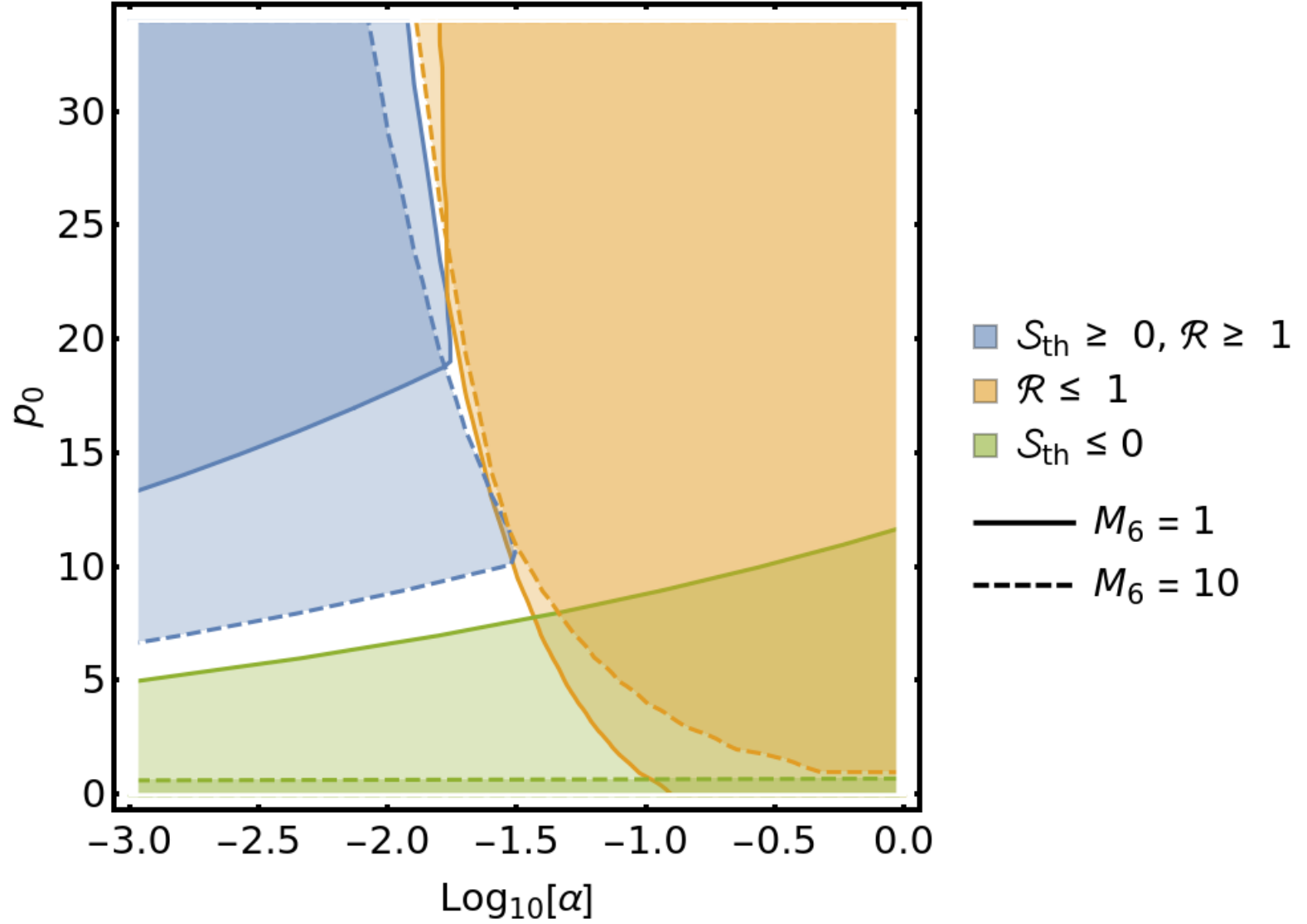}
\caption{Integrated properties concerning stability and effective optical thickness in the $\log_{10}[\alpha],p_0$-plane for $M = 10^6 \Msun$, $ 0.1 \leq \dot{M}/\MdotEdd \leq 1$ (solid lines) and $M = 10^7 \Msun$, $ 0.01 \leq \dot{M}/\MdotEdd \leq 0.1$ (dashed lines). 
The condition $\mathcal{R} \geq 1$ 
is utilized as the criterion for (possibly color-corrected or hardened) thermal emission. For all $\dot{M}$ considered here, the blue regions are always stable ($\mathcal{S}_{\rm th} \geq 0$) and effectively optically thick ($\mathcal{R} \geq 1$), the orange regions are always effectively optically thin and the green regions always have unstable zones. For $M = 10^7 \Msun$ contours, which include smaller $\dot{M}/\MdotEdd$, the always-unstable region includes only non-magnetic case of $p_0 =0$, that remains unstable for the lowest $\dot{M}/\MdotEdd = 0.01$.   } 
\label{fig_stab_opthick_summary}
\end{figure}

\begin{figure*}
\centering
\includegraphics[width=0.99 \textwidth]{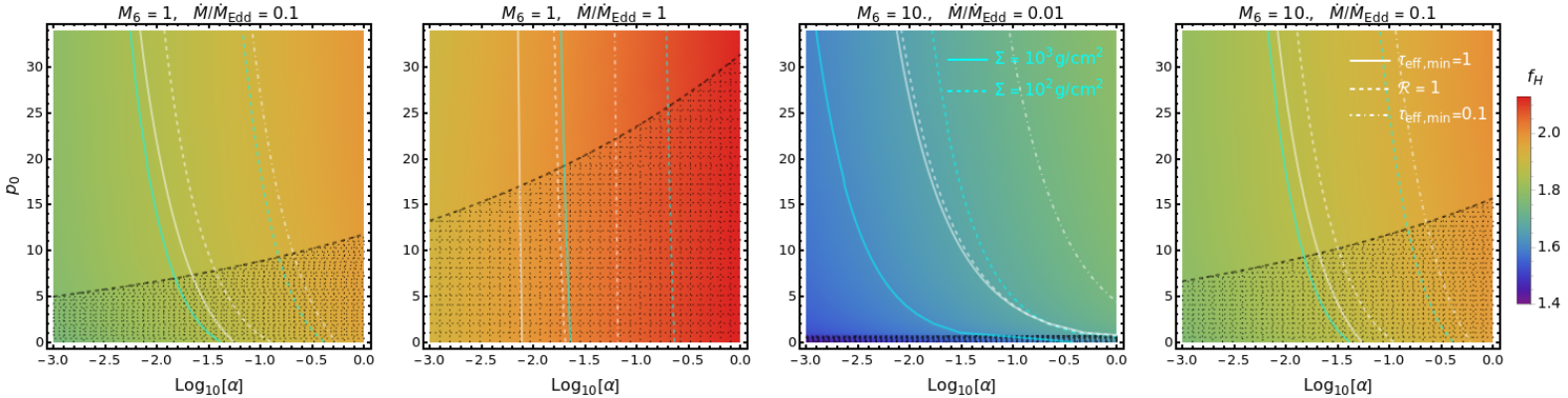}
\caption{The spectral hardening factor $f_{\rm H}$ in the $\log_{10}[\alpha]-p_0$ plane for $M = 10^6 \Msun$ (left two panels) and $M = 10^7\Msun$ (right two panels). Note that the smaller values of $\dot{M}/\MdotEdd$ are chosen for bigger $M$, as expected from TDE disk models (for the disruption of a solar mass star).  As in previous figures, white lines represent various contours for $\tau_{\rm eff, min}$ for  marginally effectively optically thick systems. Dashed black lines bound the shaded region of instability.  We also show contours for $\Sigma=10^3$ g$/{\rm cm}^3$ (solid cyan lines), which sets an approximate lower limit above which the estimates for $f_{\rm H}$ are the most trustworthy \citep{Davis2019} (see
the text for further discussion).
}
\label{fig_fH2u}
\end{figure*}

Accretion disks formed after the tidal disruption of a star span a range of initial mass accretion rates $\dot{M}_{\rm pk}/\MdotEdd$ which evolve very slowly for the first few fall-back times and which can decay at late times as $t^{-5/3}$. Recent calculations of time-dependent viscously spreading TDE disks \citep{Mummery2021arXiv,Mummery2021,Mummery_Balbus2021} predict $\dot{M}_{\rm pk}/\MdotEdd \sim 0.01-1$ for a non-spinning BH of mass $M \sim 10^6 - 10^7 \Msun$, where higher accretion rates are associated with smaller $M$ and vice versa. Direct continuum fitting of multi-epoch X-ray spectra finds a slightly higher range of near-peak accretion rates for the well-studied TDE ASASSN-14li, although later epochs descend into the $\dot{M}_{\rm pk}/\MdotEdd \sim 0.01-1$ range \citep{Wen+20}.

Instead of exhibiting any signature of thermal instability \citep{shen_matzner2014}, most TDEs whose disks have been directly observed are observed to uneventfully transit through this naively unstable range of $\dot{M}$ \citep{vanVelzen+19,Wen+20}. Magnetic pressure support is a possible way to stabilize these disks, although with the side effect of making them lighter and fluffier (i.e. smaller $\Sigma$ and $\rho$) as compared to non-magnetic Shakura-Sunyaev type disks. As a result, they can become effectively optically thin and unable to emit as a quasi-thermal, multi-colour blackbody (BB) \citep{Blaes2006,BegelmanPringle07}. This may pose a challenge for the magnetic pressure explanation of observed TDE thermal stability, as most of the TDE disks with $M \sim 10^{6-7} \Msun$ are (i) observed to emit a BB or hardened BB X-ray spectra, and (ii) do not show any prominent signs of instability \citep{Komossa2015,Miller2015, Gezari2017, Wevers2019, vanVelzen2019ZTF, vanVelzen+19, Wen+20,vanVelzen2021,Mummery_Balbus2021,Mummery2021arXiv}\footnote{There are a few TDEs which show statistically significant variability on timescales of roughly few tens of days, which might be associated with an instability. But there are no clear detection of periodicity relatable to a limit-cycle expected from radiation pressure instability. }. In this section, we study the combined stability and spectral properties of magnetized disks with parameters relevant to thermal TDEs.

We consider both of the following approximate criteria for the ability of an accretion disk to produce locally quasi-thermal (i.e. hardened or color-corrected) BB emission: 
\begin{itemize}
    \item The effective optical depth $\tau_{\rm eff} = \sqrt{3 \kappa_{\rm abs} (\kes + \kappa_{\rm abs}) } \Sigma$. If $\tau_{\rm eff} > 1$ for a disk annulus, it can be treated as a blackbody. 
    \item The heating-to-cooling ratio $\mathcal{R} = \eta_{\rm ff} H/( \sigSB T_{\rm s}^4)$; this is the ratio of energy generated by free-free emission to the expected blackbody cooling in a local disk annulus. Here, $\eta_{\rm ff}$ is the Rosseland mean of free-free emissivity. A disk annulus can emit as a BB if there is sufficient free-free emission, that is if $\mathcal{R} > 1$ \citep{Davis2019}.    
\end{itemize}
Figure~\ref{fig_tau_eff_prof} shows the radial profiles of $\tau_{\rm eff}$ for both magnetic (for $p_0 = 1,10$) and non-magnetic disks. For all the figures in this section, we evaluate $\tau_{\rm eff}$ by first generating tables of $\kappa_{\rm abs}$ using Cloudy, as detailed in appendix~\ref{app_opacity_effects}. As seen in \S~\ref{sec:disks} for other physical quantities, the sensitivity of $\tau_{\rm eff}$ on magnetic field strength (parametrized by $p_0$) depends upon $\dot{M}$. For low $\dot{M}$, and also the outer disk regions for $\dot{M} = 1 \MdotEdd$ solutions, where radiative cooling dominates, the magnitude of $\tau_{\rm eff}$ falls steeply with increasing $p_0$. High $\dot{M}$ disks dominated by advective cooling are not especially sensitive to $p_0$. For $p_0 = 10$, disks are effectively optically thin (thick) for $\alpha = 0.1$ ($0.01$) over most of the parameter space $\{ M, \dot{M} \}$ considered here. This is due to the approximate relation  $\Sigma \propto \alpha^{-1}$ (see table~\ref{tbl_analytic_sol}), which makes low-$\alpha$ disks more massive and hence more optically thick. We give some analytical estimates for $\tau_{\rm eff}$ below using the approximate disk solutions of table~\ref{tbl_analytic_sol}. For these estimates, we approximate absorption opacity by Kramer's opacity, as $\kappa_{\rm abs} \sim \kappa_{\rm Kr} = 1.9 \times 10^{24} \rho T^{-7/2}$ (which fits the more precise Cloudy absorption opacity quite well for the parameters of our interest).       

 For low $\dot{M}$, the $\Prad$-dominated non-magnetic disks (case a2 of table~\ref{tbl_analytic_sol}) have effective optical depth: 
\beq 
\tau_{\rm eff} = 37.4 \,\dot{M}_{0.01}^{-2} \alpha_{-1}^{-17/16}  M_6^{-1/16}  \bigg( \frac{R}{10 r_g} \bigg)^{93/32}
\label{tau_eff_Prad_qrad}
\eeq 
and for $\Pmag$-dominated (magnetic) disks (case a1): 
\beq 
\tau_{\rm eff} = 13 \, \dot{M}_{0.01}^{1/3} \alpha_{-1}^{-11/12} p_0^{-7/6}  M_6^{1/12}  \bigg( \frac{R}{10 r_g} \bigg)^{5/24} \,.
\label{tau_eff_Pmag_qrad}
\eeq 
Increasing the magnetic contribution to pressure ($p_0$) and the viscosity parameter ($\alpha$) tend to make disks optically thin, with $\tau_{\rm eff}$ being roughly inversely proportional to these parameters. 

As $\dot{M}$ increases, $\tau_{\rm eff}$ attains its minimum near $\dot{M} \sim \MdotEdd$, and then starts increasing for more highly super-Eddington accretion rates. For these advectively cooled disks, magnetic pressure never manages to dominate for moderate values of $p_0 \leq 32$ (which is the minimum value needed to stabilize these disks in the parameter ranges studied here; see figure~\ref{fig_p0stab}). Thus, we consider only $\Prad$ dominated case for advective disks (case b2 of table~\ref{tbl_analytic_sol}), and find approximately: 
\beq 
\tau_{\rm eff} = 0.076 \, \xi^{49/32} \dot{M}_{1}^{17/16} \alpha_{-1}^{-17/16}  M_6^{-1/16}  \bigg( \frac{R}{10 r_g} \bigg)^{-5/32}
\eeq 
This suggests that near-Eddington disks can behave as effectively optically thick only for small values of $\alpha \lesssim 0.01$, as is evident from figure~\ref{fig_tau_eff_prof} (using the optical thickness criterion $\tau_{\rm eff} \geq 1$). 

For deriving analytical expressions for $\tau_{\rm eff}$ for all these cases, we assume $\kappa_{\rm abs} \sim \kappa_{\rm Kr}$, which implies $\tau_{\rm eff} \simeq 0.8 \sqrt{\mathcal{R}}$. To evaluate $\mathcal{R}$ in this relation, we have used the Rosseland mean of free-free emissivity $\eta_{\rm ff} = 1.4 \times 10^{-27} T^{1/2} \rho^2/\mP^2$. Hence we do not expect significantly different conclusions based on the two criteria for (hardened) BB emission, though $\mathcal{R} > 1$ is a relaxed condition compared to $\tau_{\rm eff} > 1$.  


We also performed more exact calculations for $\tau_{\rm eff}$ by considering general disk models for a wide range of parameters. In figure~\ref{fig_tau_eff2u}, we plot the minimum value of $\tau_{\rm eff} = \tau_{\rm eff, min}$ achieved by a disk of size $R_{\rm max} = 100 r_g$ in the $\log_{10}\alpha - p_0$ plane for $M = 10^6\Msun$ (with $\dot{M}/\MdotEdd = \{ 1,0.1\}$) and $M = 10^7\Msun$ (with $\dot{M}/\MdotEdd =\{ 0.1, 0.01 \}$). Smaller $\dot{M}/\MdotEdd$ is chosen for higher $M$ to parallel the $\dot{M}-M$ relationships expected for TDE disks. 
As expected, $\tau_{\rm eff} \geq 1$ (left side of solid white contours) is a stricter criterion for (possibly hardened) BB emission as compared to $\mathcal{R} \geq 1$ (left side of dashed white line). 
We evaluate $\mathcal{R}$ at a radial distance $R$ where the disk surface temperature $T_{\rm s}$ is maximized and hence is expected to contribute the most to an observed X-ray spectrum. Disks with $\alpha \lesssim 0.01-0.03$ are expected to emit like a BB. Smaller $p_0 \gtrsim 10$ are sufficient for stability of larger BHs ($M = 10^7 \Msun$) as the peak $\dot{M}/\MdotEdd$ values attained are lower. On the other hand, the lower mass BHs ($M = 10^6 \Msun$) which can more easily accrete initially up to the Eddington rate require higher $p_0 \gtrsim 20$ for stability.  None of these results depend sensitively upon the chosen criterion for optical thickness ($\tau_{\rm eff,min} \geq 1$ or $\mathcal{R} \geq 1$), 

Figure~\ref{fig_stab_opthick_summary} further summarizes these results and highlights (in blue) the parameter space that is most consistent with typical TDE observations, i.e. thermally stable and (effectively) optically thick.  This region is characterized by high $p_0$ and low $\alpha$ values. This figure also delineates the low-$p_0$ region which is always thermally unstable (in green) and the high-$\alpha$ region which is always a non-thermal emitter (in orange). 
For observed TDEs with BB emission and apparent stability, these disks should satisfy these parameter ranges for low $\alpha \lesssim 0.03$ and high $p_0 \gtrsim 20$. It is interesting to note the mutually exclusive set of parameters favored by thermal and stable TDEs versus QPEs (which suggest $\alpha \gtrsim 0.1$ and $3 \lesssim p_0 \lesssim 20$ as seen earlier). This may simply reflect natural variation in magnetic field strengths and effective viscosities between different types of accretion disks.  However, there is an alternative argument that may explain the apparent stability of thermal TDE disks with $M \gtrsim 10^6 \Msun$. The instability limit-cycle period $t_{\rm p} \sim 1 {\rm d}(M/10^5 \Msun)^{44/37} \propto M^{44/37}$ increases faster with $M$, as compared to the debris fall-back timescale $\tfb \sim 36.5 {\rm d} \sqrt{M/10^6 \Msun} \propto \sqrt{M}$ on which $\dot{M}$ varies. For $M \gtrsim 3\times 10^6 \Msun$, the disk accretion rate evolves faster than anticipated instability timescales and it is seemingly impossible for these systems to display quasi-periodic limit-cycles similar to QPEs. This aspect needs to checked carefully with the time-dependent models of magnetic disks with evolving accretion rate similar to a TDE disk, but may alleviate the need for high $p_0$ values for TDEs.      



We also estimate the spectral hardening factor $f_{\rm H}$ for the color-corrected BB spectrum emitted by the inner accretion disk. We use the following fitting formula from the numerical simulations of \citet{Davis2019} (see their equation~9):
\beq
\begin{split}
f_{\rm H} &= 1.74 + 1.06 ( \log_{10}T_{\rm s} - 7  ) -0.14 (2 \log_{10}\OmgK -7)\\
& \qquad - 0.07 ( \log_{10}[\Sigma / 2] - 5)
\end{split} 
\label{fH}
\eeq 
to calculate $f_{\rm H}$ at a radial distance $R$ with the maximum disk surface temperature $T_{\rm s} = T/( \kes \Sigma )^{1/4}$. They obtained the above formula for the following region of simulated parameter space: $\Sigma \geq 10^3$g$/{\rm cm}^3$ and $ 0.01 < \dot{M}/\MdotEdd < 1 $. We calculate the hardening factor $f_{\rm H}$ for our disk models for $M = 10^6 \Msun$ (and $\dot{M}/\MdotEdd = \{ 0.1, 1 \}$) and $M = 10^7 \Msun$ (and $\dot{M}/\MdotEdd = \{ 0.01, 0.1 \}$), as suited for TDEs and plot these results in the $\log_{10}\alpha - p_0$-plane in figure~\ref{fig_fH2u}.  $f_{\rm H}$ depends strongly on $\alpha$ and $\dot{M}/\MdotEdd$ 
and appears to have a weak dependence on $M$ and $p_0$ (especially true in the advective domain of cooling for high $\dot{M}/\MdotEdd$). This fitting formula does not apply to the full parameter space we consider, but the region with $\Sigma \geq 10^3$g$/{\rm cm}^3$ 
roughly coincides with the effectively optically thick regime 
for most of these cases. 
Furthermore the expression~(\ref{fH}) assumes the solar metallicity and a non-spinning black hole; hardening is expected to increase with decreasing metallicity and increasing spin \citep{Davis2019}. 

Despite these uncertainties, the estimated  $f_{\rm H} \in [1.4,2]$  for the parameters with $\mathcal{R} \geq 1$ (avoiding the photon-starved regime), excludes the extreme hardening regime $f_{\rm H} \gtrsim 2.4$ \citep{Davis2019} and agrees well with the observations of most X-ray TDEs in soft -states for at least a few years after the peak when the Eddington ratios are not too low (check \citealt{Saxton2021} for a detailed review, see also \citealt{Wen+20}). 


\subsection{Jet Power in TDEs}
\label{sec_TDE_jets}

\begin{figure*}
\begin{subfigure}{0.99\textwidth}
\centering
\includegraphics[width=0.99 \textwidth]{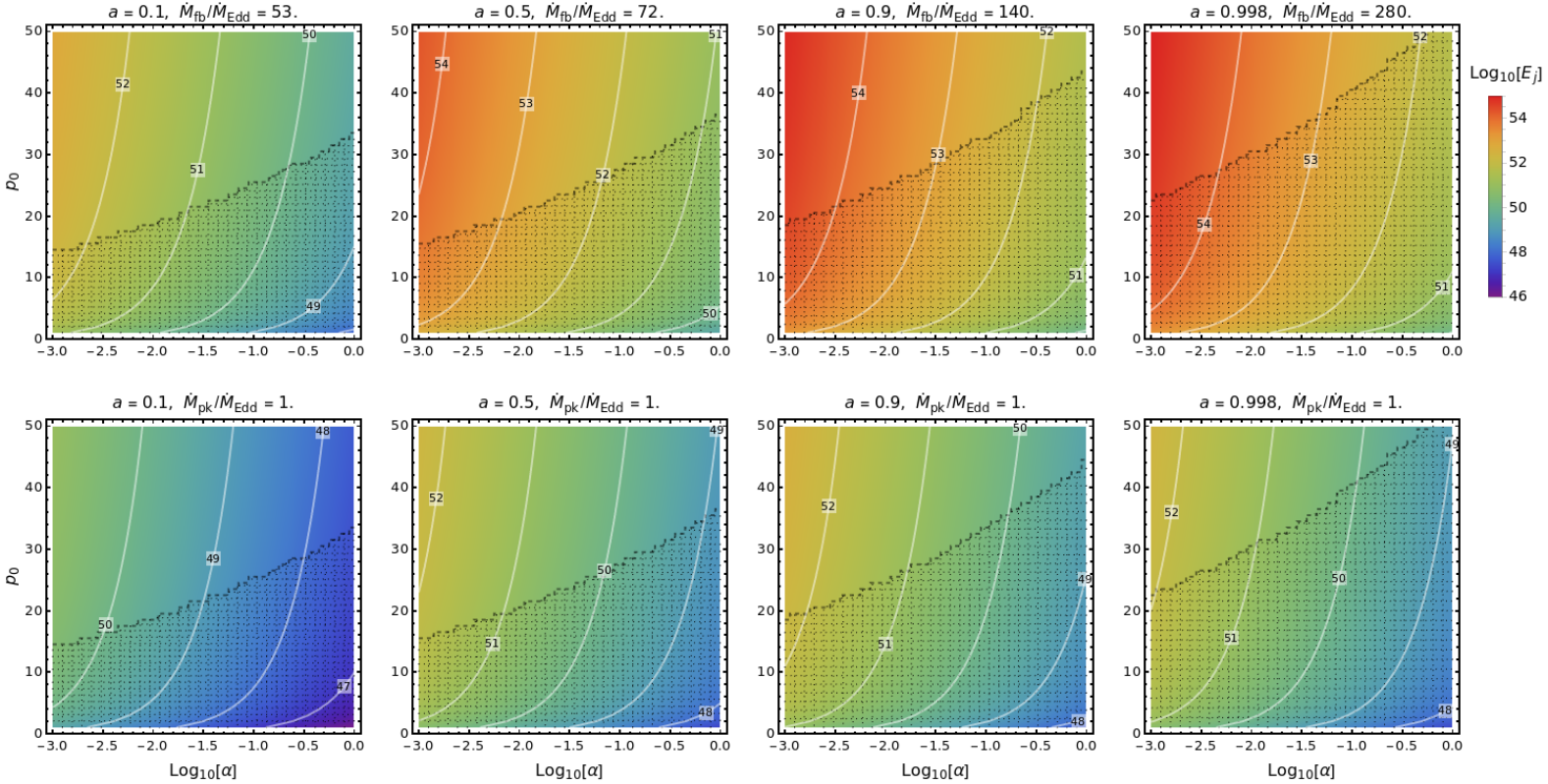}
\subcaption{$M = 10^6 \Msun$}
\end{subfigure}
\\
\begin{subfigure}{0.99\textwidth}
\centering
\includegraphics[width=0.99 \textwidth]{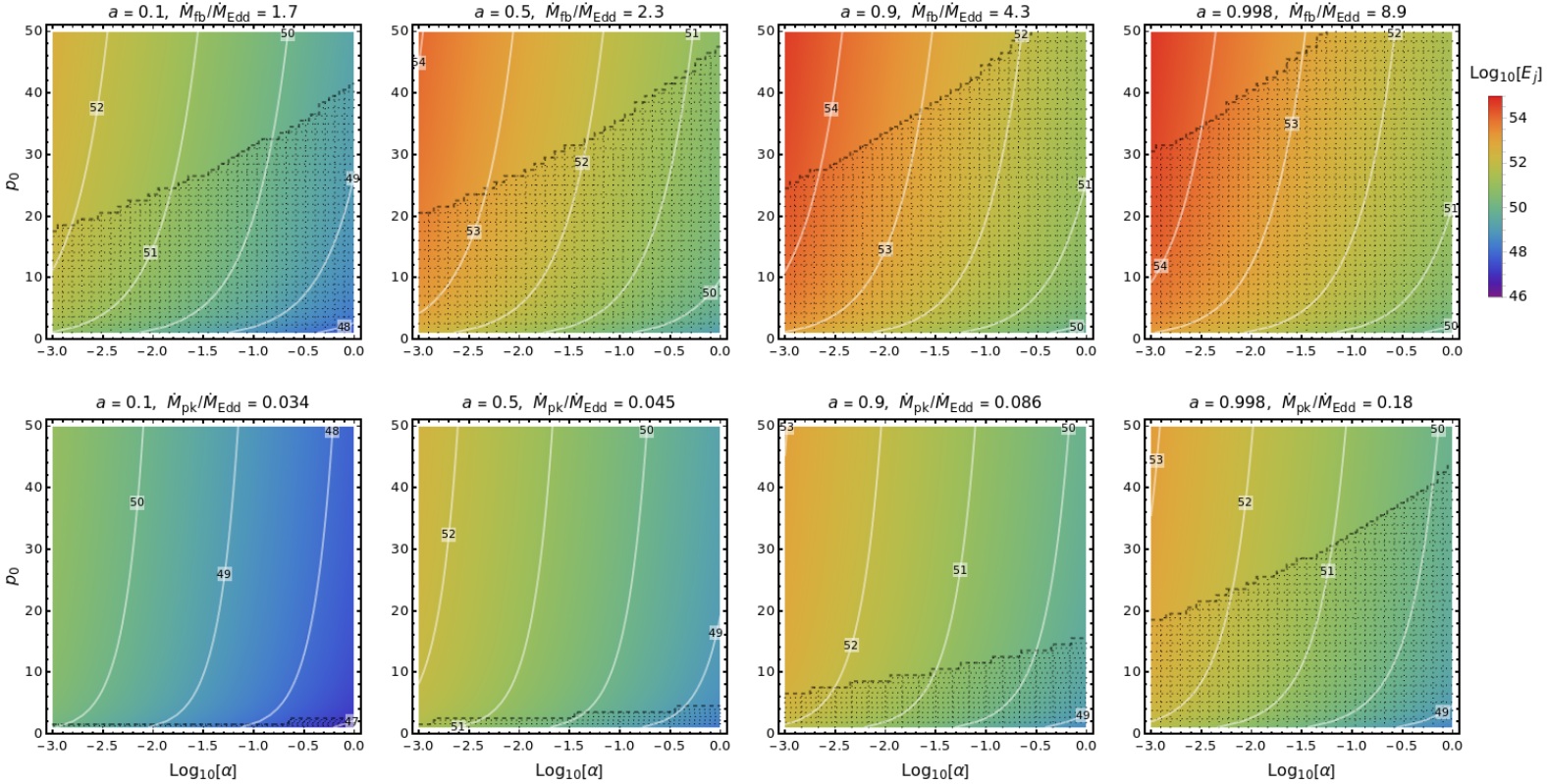}
\subcaption{$M = 10^7 \Msun$}
\end{subfigure}
\caption{Jet energy $\Ej$ (in erg) is plotted in the $\log_{10}\alpha-p_0$ plane for BH masses (a) $M = 10^6 \Msun$ and (b) $M = 10^7 \Msun$ with various spin parameters $a = 0.1 - 0.998$ (increasing from left to right) for fixed peak $\dot{M}$ (set to the classical peak fall-back rate $\dot{M}_{\rm fb}$ for the first row, and to an Eddington-capped accretion rate $\dot{M}_{\rm pk}$ for the second row of both panels a and b). The shaded region indicates the parameter range for thermally unstable disks. 
The instability region is sensitive to the choice of accretion rate model, especially for $M = 10^7 \Msun$  due to sub-Eddington $\Mdotpk$. Jet energies from the Eddington-capped accretion rates $\dot{M}_{\rm pk}$ are lower by approximately two orders of magnitude compared to the classical fall-back rates $\dot{M}_{\rm fb} $.   
}
\label{fig_Ej}
\end{figure*}

Poloidal magnetic fields around accreting BHs power highly luminous relativistic jets via the Blandford-Znajek mechanism \citep{Blandford_Znajek1977, Tchekhovskoy2010}.
Recent MHD simulations \citep{Jiang+19} indicate that poloidal magnetic fields at the BH horizon
can be roughly of the same order of magnitude as the toroidal fields in the inner disk generated by MRI. This suggests an interesting question: \emph{will the minimum disk magnetic field required to support TDE disks against thermal instability also produce observable jet energies?} This is important to check, as late-time radio observations of most TDEs only find non-detections \citep{Bower2013,vanVelzen2013}, putting upper limits on the energies associated with putative off-axis jets \citep{Generozov2017,Matsumoto2021}. In this section we approximately estimate jet power and energy emitted by a BH fed by a magnetized disk, and then specialize to the case of TDEs.      

We estimate the jet power $\Pj$ from Blandford-Znajek formula \citep{Blandford_Znajek1977} generalized by \citet{Tchekhovskoy2010} (see their equation~9) to include BHs with a general spin $ 0<a <1$ (rather than only the low-spin limit of the classical form):
\beq 
\Pj = \kj \frac{G^2 M^2}{c^3}\PhiH^2 ( \OmgH^2 + 1.38 \OmgH^4 - 9.2 \OmgH^6  ) \, .
\label{Pj_form1}
 \eeq 
 Here $\PhiH$ (in ${\rm Gauss}$) is the total (normalized) poloidal flux of magnetic field passing through the BH horizon, defined by radial coordinate $\rH = \rg (1 + \sqrt{1-a^2})$ and $\OmgH = a \rg/(2 \rH)$ is the normalized angular frequency of the BH. 
 
 To evaluate $\Pj$ for our magnetic disk models, first we must estimate the poloidal flux $\PhiH$ at the horizon. We consider force free outflow solution for the jet \citep{Narayan2007, Tchekhovskoy2010}, with the following form for (in-plane) poloidal flux profile\footnote{Any distance with a "tilde" is normalized with $\rg$ as earlier.}:
 \beq 
 \Phi_{\rm p}(R) = \int_{0}^{\Rtil} B_{\rm p} 2 \pi \Rtil' \, \rmd \Rtil' =  \PhiH \, \frac{R}{\rH}   
 \eeq 
This linear profile corresponds to choosing their outflow parameter $\nu = 1$, for which the constant pre-factor $\kj \simeq 0.044$ \citep{Tchekhovskoy2010}. Moreover, it is evident from this simulation study that magnitude of output jet power is of similar order for any $ \nu \in [0.5, 1]$ and therefore our estimates for $\Pj$ will not be significantly sensitive to this heuristic choice. Further we assume this form of poloidal flux solution to be valid until the inner edge of the disk. More specifically, we choose the disk radius $R = \Rd = 1.4 R_{\rm in}$ for this boundary, where the disk magnetic pressure $\Pmag$ is the maximum as suggested by the analytic disk solutions of table~\ref{tbl_analytic_sol}. Here $R_{\rm in}$ is the inner edge of the disk and coincides with the inner-most stable circular orbit \footnote{Check \citet{Bardeen1972} for spin $a$-dependent mathematical expressions for various quantities like the radius $R_{\rm in}$ and radiative efficiency $\eta$. }. From the above expression, we have a poloidal magnetic field with magnitude $\Bpd = \PhiH/(2 \pi \Rdtil \rHtil)$ at the inner disk radius $\Rd$. Further, the disk poloidal field $\Bpd$ can be approximately related to the toroidal magnetic field $\Btd \simeq \sqrt{8 \pi \Pmag}$ using equipartition arguments as $\Bpd \sim (H/R) \Btd$ \citep{Meier2001}. The proportionality of the poloidal field strength to disk height $H$ can be understood as turbulent eddies roughly of size $H$ giving rise to $\Bpd$. Using these arguments and expressions, one can relate the poloidal flux $\PhiH$ to the disk magnetic pressure $\Pmag$ as:
\beq
\PhiH^2 = 32 \pi^3 \bigg( \frac{H}{\Rd} \bigg)^2 \Pmag \rHtil^2 \Rdtil^2 
\label{PhiH}
 \eeq 

We evaluate jet power from equations~(\ref{Pj_form1}) and (\ref{PhiH}) by numerically evaluating $H/\Rd$ and $\Pmag$ (at the inner radius $\Rd$) for our magnetic disk models. To apply these estimates to a TDE involving a Solar mass star, we choose mass accretion rate $\dot{M}$ at initial times in the following two ways: 
\begin{itemize} 

\item[(1)] \emph{Classical} mass accretion $\dot{M} = \dot{M}_{\rm fb}$, the peak debris fall-back rate: 
\beq 
\frac{\dot{M}_{\rm fb} }{ \MdotEdd } = \frac{ 87.6 }{ M_6^{3/2} } \eta_{-1}(a)  
\label{Mdot_fb}
\eeq 
Here $\eta_{-1} =  \eta/0.1$ is the normalized radiative efficiency. For $M = 10^{6-7} \Msun$ and $a = 0.1-0.998$, the fall-back rates imply a super-Eddington accretion $\dot{M}/\MdotEdd \simeq 1.7 - 280 $, with higher Eddington ratios for lower mass $M$ and higher spin $a$.

\item[(2)] \emph{Modified} mass accretion models with much lower and Eddington-capped peak mass accretion rate $\dot{M} = {\rm Min}[ \dot{M}_{\rm pk}, \MdotEdd]$, where:
\beq 
\frac{\dot{M}_{\rm pk}}{\MdotEdd} = \frac{1.75}{M_6^{3/2}} \eta_{-1}(a)
\label{Mdot_pk}
\eeq 
This is inspired by 
observations of TDEs that rarely find highly super-Eddington disk luminosities.  The above expression implies $\dot{M}/\MdotEdd \lesssim 3 \times 10^{-2} $ for $M \gtrsim 10^7 \Msun$ and hence satisfies yet another physical inference of their model, which states that the accretion flow becomes radiatively inefficient for central BH mass above $ M \sim 10^7 \Msun$ (for $a=0$). Additionally this peak accretion rate, being dominantly controlled by viscous evolution of the outer disk, is not expected to depend significantly on BH spin $a$. 
The accretion rates are Eddington-limited with $\dot{M}/\MdotEdd = 1$ 
for $M = 10^6 \Msun$, while the rates become sub-Eddington $\dot{M}_{\rm pk}/\MdotEdd \simeq 0.03-0.2$ for $M = 10^7 \Msun$ for the entire range of BH spin $a$.   

\end{itemize}

We further assume that the accretion rate remains roughly equal to this peak value (either of the above two choices for $\dot{M}$) till the fall-back time $t_{\rm fb} \simeq 0.1 \, {\rm yr} \sqrt{M_6}$. For a BH of fixed mass $M$ and spin $a$, we thus consider a fixed peak $\dot{M}$ and evaluate the required disk properties from our magnetic disk models for given disk parameters $\alpha$ and $p_0$. The jet power $\Pj$ is computed using equations~(\ref{Pj_form1}) and (\ref{PhiH}). We then evaluate the jet energy $\Ej = \Pj \tfb$ for these cases. We restrict ourselves to only initial times $t \leq \tfb$ for calculation of $\Ej$ as most of the jet power is expected to be emitted during this time. Moreover considering an evolving jet power due to varying $\dot{M} \sim \Mdotfb (t/\tfb)^{-5/3}$ at late times (for the classical fall-back rate model), increases $\Ej$ at most by a factor $\sim 2$. We present these results in $\log_{10}\alpha -p_0$ plane in the figure~\ref{fig_Ej} for $M = \{10^6, 10^7\} \Msun$ and $a = \{ 0.1 , 0.5 , 0.9 ,0.998  \}$. Here we highlight the major features of these results. 

\begin{itemize}
    \item  Higher values for $p_0$ and smaller $\alpha$ imply higher disk magnetic pressure $\Pmag$ and hence a higher jet energy $\Ej$. This is because a smaller viscosity parameter $\alpha$ leads to higher disk surface density and hence a higher pressure is needed to support it. 
    
    
    \item For a given poloidal flux $\PhiH$, $\Pj \propto a^2$ for smaller $a$ and $\Pj \propto a^4 $ 
     for higher spin values $a$ \citep{Tchekhovskoy2010}. Further, higher spin implies: (1) a higher peak accretion rate due to higher radiative efficiency $\eta$, (2) a smaller inner disk radius $\Rdtil$ which decreases by a factor of $\sim 5$ for $a \in [0.1,0.998]$. Both these factors leads to higher $\Pmag(\Rd)$ for higher $a$, which makes $\Ej$ to increase significantly with increasing $a$. 
    
    \item Lower and Eddington-capped peak accretion rates $\dot{M}_{\rm pk}$ from the modified model result in jet energies smaller by roughly two orders of magnitude when compared with the classical fall-back rates $\Mdotfb$. 
    
    \item Significant portion of the $\{ \alpha, p_0 \}$ 
    parameter space is thermally unstable; while stability occurs for higher $p_0 \gtrsim 30-40$ for $\alpha \sim 0.1$. For sub-Eddington accretion rates $\dot{M}_{\rm pk}$ for $M = 10^7 \Msun$, stability is relatively easy to obtain, with a threshold $p_0 \sim 10$ for $a \lesssim 0.9$. The sensitivity of stability to the choice of the mass accretion model is understandable from the earlier discussions in \S~\ref{sec:thermal}. But it is feasible that this instability is unable to manifest for TDEs with higher BH mass $M \gtrsim 3\times 10^6 \Msun$ (as discussed earlier in \S~\ref{sec_TDE_hardBB}), due to faster evolution of $\dot{M}$ on timescales shorter than the anticipated periods for instability limit-cycles.

\end{itemize}

Here we present an approximate analytical study to understand better the above results. For the cases of our interest, the accretion rates are either super-Eddington or close to Eddington (especially for $M \sim 10^6 \Msun$). For both cases, there is a significant contribution of advective cooling in the inner disk. As we have seen earlier in \S~\ref{sec:disks}, radiation pressure $\Prad$ remains dominant over $\Pmag$ for this case (for modest values of $p_0$ that we consider) and analytical solution (case b2) of table~\ref{tbl_analytic_sol} should describe well the physical state of the disk. Using this solution in the definition of $\Pmag$ (equation~\ref{Pmag}) for $R = \Rd = 1.4 R_{\rm in}$ (hence $f(\Rd) = 0.155$), we have:
\beq 
\begin{split} 
& \Pmag = 1.25 \times 10^{10} {p_0 \dot{M}_1^{9/8} \xi^{25/16}     M_6^{-9/8} \Rdtil^{-37/16} \alpha_{-1}^{-9/8}      } \\
  &= 1.9 \times 10^{12} \{1, \frac{0.0065}{ \eta_{-1}^{9/8}} M_6^{27/16} , 0.01   \} \frac{ p_0  \eta_{-1}^{9/8} \xi^{25/16}}{ \alpha_{-1}^{9/8} M_6^{45/16} \Rdtil^{37/16} }              
\end{split}
\label{Pmag_jet}
\eeq 
The second equality results by using the two peak accretion rate models of equation~(\ref{Mdot_fb}) and (\ref{Mdot_pk}) with $\dot{M} = \{ \Mdotfb , \MdotEdd , \dot{M}_{\rm pk}  \} \equiv \{ {\rm Super-Eddington, \, Eddington , \, Sub-Eddington }  \}$. Using the above expression for $\Pmag$ and $H/\Rd = 0.34/\sqrt{\xi} $ (again from the analytical solution b2) in equations~(\ref{Pj_form1}) and (\ref{PhiH}) for jet power, we have: 
\beq 
\begin{split}
& \Pj = 6.3 \times 10^{45} \{1, \frac{0.0065}{ \eta_{-1}^{9/8}} M_6^{27/16} ,0.01   \} \frac{ p_0 \xi^{9/16} \mathF(a) }{   M_6^{13/16} \alpha_{-1}^{9/8} }  \\
& \mbox{where  }\\
&\mathF(a) =  \eta_{-1}^{9/8} \rHtil^2  \Rdtil^{-5/16} ( \OmgH^2 + 1.38 \OmgH^4 - 9.2 \OmgH^6)  
\end{split}
\label{Pj_form2}
\eeq 
The spin function $\mathF$ offers a magnitude range of roughly three orders with $\mathF(0.1) = 7.4 \times 10^{-4}$ and $\mathF(1) = 0.88$. Together with variation of $p_0$ and $\alpha$, this explains the wide magnitude range spanned by $\Ej$ in figure~\ref{fig_Ej}. It is then straightforward to estimate energy output of jet $\Ej$ after time $\tfb$:
\beq 
\Ej = 2 \times 10^{52} \{1, \frac{0.0065}{ \eta_{-1}^{9/8}} M_6^{27/16} , 0.01   \} \frac{ p_0 \xi^{9/16} \mathF(a)  }{  M_6^{5/16} \alpha_{-1}^{9/8} }   
\label{Ej}
\eeq 
The energy output from the two accretion rate models differ by roughly two orders of magnitude. Most TDEs are associated with sub-Eddington accretion \citep{Jonker2020}, and radio observations may offer typical upper limits as deep as $\Ej \lesssim 10^{50}$ erg (\citealt{Matsumoto2021}, although see also \citealt{Generozov2017}), which is consistent with the modified model with lower $\dot{M}/\MdotEdd$.
There are only three hitherto discovered TDEs with strong relativistic jets \citep{Burrows2011,Cenko2012, Brown2015} with energy $\Ej \sim 10^{51-52}$ erg, which is better explained by the classical super-Eddington accretion rates $\Mdotfb$. 

\section{Conclusions}
\label{sec:conclusions}

In this paper, we construct 
models of radiatively efficient, magnetically dominated accretion flows with the aim of investigating the influence of magnetic pressure support on the radiation pressure instability. The local magnetic pressure is mainly contributed by toroidal fields generated by MRI turbulence and is quantified on the basis of the MRI saturation criterion suggested by PP05 and BP07. Recent radiation-MHD simulations \citep{Jiang+19} show evidence of reasonable agreement with this criterion. 
We describe disk (and its stability) properties in terms of the parameter $p_0$, which acts as a weight calibrating the magnetic pressure $\Pmag$ term. 
We explained the occurrence (or not) of instability in different astrophysical systems on the basis of our disk models. Additionally, we investigated the impact of magnetic fields on other physical observables of disks like the thermal X-ray emission and the emitted jet energy for TDEs. We summarize our main results below.    

\begin{itemize}
	\item The minimum mass accretion rate for the onset of instability, $\dot{M}_{\rm min}/\MdotEdd \sim 2 \times 10^{-3}$, increases to a higher threshold value for magnetic disks: $\dot{M}_{\rm min}^{\rm B}/\MdotEdd \sim 0.066 \sqrt{p_0}$. The outermost unstable radius of the accretion disk, $R_{\rm st,o} \sim 3380 r_g ( \dot{M}/\MdotEdd)^{16/21}$, shrinks by more than an order of magnitude for magnetic disks, to $R_{\rm st,o}^{\rm B} \sim 363 r_g p_0^{-16/37} (\dot{M}/\MdotEdd)^{32/37}$, and continues shrinking for higher magnetic contribution for increasing $p_0$. The viscous time at this radius, which is an approximate recurrence timescale for instability limit cycles, $t_{\rm qpe} \sim 2300 {\rm yr} \, M_6^{4/3} \alpha_{-1}^{-2/3} (\dot{M}/\MdotEdd)^{2/3}$, decreases by more than three orders of magnitude in a magnetically dominated disk: $t_{\rm qpe}^{\rm B} \sim 1 {\rm yr} \, M_6^{44/37} \alpha_{-1}^{-30/37} p_0^{-56/37} (\dot{M}/\MdotEdd)^{38/37}$. In qualitative agreement with previous studies \citep{Dexter2019,Sniegowska2022}, we find higher frequency instability cycles for magnetic disks.

	\item The short repetition timescales ($\sim \,{ \rm a \, few \, hrs}$) observed for QPE sources with $M \sim 10^{5-6} \Msun$ and $\dot{M}/\MdotEdd \sim 0.1-0.5$ can be explained by our models for moderate values of $p_0\sim 3-20$ and $\alpha \gtrsim 0.1$. Earlier works on magnetic disk instability can explain these short timescales only by invoking smaller disks (due to e.g. TDEs, \citealt{Sniegowska2022}) or rapid angular momentum loss in magnetized winds \citep{Pan_2022_viscosity_model}. Notably, $p_0$ values suited to explain observed QPEs are not unique; systems with higher $M$ and $\dot{M}/\MdotEdd$ need higher values of $p_0$. 
	
	The non-occurrence of instability for most observed XRBs may be linked to the lower magnetic field strengths required for disk stability in low-$M$ systems (e.g. $p_0 \gtrsim 2-6$ for $M = 10 \Msun$ and $\alpha \sim 0.1$). It is, however, possible to have unstable systems below these small threshold $p_0$ values,
	and the resultant instability timescales $\sim 1-100$ s roughly match the heartbeat oscillations observed in a minority of XRBs.

	Further up the BH mass spectrum ($M \sim 10^{7 - 8} \Msun$) in AGN disks, our models are consistent with instability timescales $\sim$ a few months - years, matching observed timescales over which some of these systems undergo changes in their spectral or photometric states (for CL AGNs). For nominal $\dot{M}/\MdotEdd \sim 0.1$, the predicted recurrence timescales are shorter than a decade for $p_0 \sim 4-10$ (for $\alpha \sim 0.1$).

	\item For TDEs with $M \sim 10^{6-7} \Msun$, our magnetic disk models are effectively optically thick for $\alpha \lesssim 0.01$ and thermally stable for $p_0 \gtrsim 10-20$. Higher effective viscosities can only produce a thermally stable disk at the cost of a non-thermal spectrum, which is not seen in many X-ray and UV observations of TDE disks.  
	Interestingly, the domain of favourable $p_0$ and $\alpha$ is mutually exclusive for QPEs and TDEs. This apparent contradiction can be countered by these arguments: (1) even when effectively optically thin (for e.g. $\alpha \sim 0.1$), the TDE disks have only a moderate spectral hardening factor $f_{\rm H} \lesssim 2$, so their emission may still be thermal enough to match observations; (2) TDE disks around BHs with higher masses $M \gtrsim 10^6 \Msun$ might be unable to display the longer thermal instability limit-cycles due to rapidly  evolving mass accretion rates (sourced by fallback); (3) there is likely to be a range of natural variation in magnetic field strengths and effective viscosities between and within different types of accretion systems.    
	
	\item Requiring a minimum magnetic field strength for thermal stability of TDE disks also suggests a minimum efficiency for jet launching through the Blandford-Znajek mechanism.  We approximately estimate the emitted jet energies (consistent with our disk models at the inner disk boundary) for TDEs with $M \sim 10^{6-7} \Msun$, which depend sensitively on the initial disk accretion rate. For the super-Eddington accretion case (where $\dot{M}$ closely follows the mass fall-back rate of tidal debris streams), one can obtain very high jet energy $\Ej \sim 10^{51-52}$ erg, in agreement with at least two of the observed jetted TDEs. But more conservative and Eddington-limited estimates for $\dot{M}$ are consistent with $\Ej \lesssim 10^{50}$ erg, in better agreement with the upper bounds estimated from radio non-detections \citep{vanVelzen+13, Bower+13} of many thermally selected TDEs \citep{Matsumoto2021}. This confirms that the magnetic fields needed to stabilize TDE disks do not necessarily result in exorbitant jet energies. 

\end{itemize}    

It is important to note that the analysis done in this paper has been in the framework of 1D, steady-state accretion disk models with (a range of) approximate estimates for typical magnetic field strengths in accretion disk midplanes.  While our parameterization, which is adapted from PP05 and BP07, appears consistent with some state-of-the-art radiation MHD simulations \citep{Jiang+19}, the parameter space coverage of global MHD accretion simulations is limited and different saturation criteria for magnetic fields in MRI-unstable accretion flows deserve further investigation and could change our quantitative results.  Qualitatively, however, it seems that magnetic pressure dominance is a promising way to explain the observed thermal stability of accretion disks, and perhaps to shorten instability recurrence times in the inner zones of AGN disks down to the timescales of QPEs.



In the future, we hope to construct time-dependent magnetized disk models to make detailed predictions of observables like duty cycle, rise and fall times, and amplitudes of limit-cycles associated with magnetically truncated outbursts of the classic radiation pressure instability. This is important for assessing the viability of this QPE explanation, but is also crucial to predict detailed instability signatures for TDEs with $M \gtrsim 10^6 \Msun$, for which evolving mass accretion rates need to be considered. In spite of a general understanding of the stabilizing role played by the magnetic fields in an accretion disk against the classical radiation pressure instability, the detailed nature of this phenomenon remains elusive and demands more detailed analytical and numerical studies in future.

\section*{Acknowledgements}

 Authors thank Richard Saxton, Giovanni Miniutti, Margherita Giustini and Riccardo Arcodia for enlightening discussions and their useful suggestions about this work.     
Both KK and NCS gratefully acknowledge support from the Israel Science Foundation (Individual Research Grant 2565/19). 

\section*{Data Availability}

Data underlying the results of this work will be shared by the authors upon reasonable request.

\bibliographystyle{mnras}
\bibliography{disk_refs} 

\begin{thebibliography}{}
\makeatletter
\relax
\def\mn@urlcharsother{\let\do\@makeother \do\$\do\&\do\#\do\^\do\_\do\%\do\~}
\def\mn@doi{\begingroup\mn@urlcharsother \@ifnextchar [ {\mn@doi@}
  {\mn@doi@[]}}
\def\mn@doi@[#1]#2{\def\@tempa{#1}\ifx\@tempa\@empty \href
  {http://dx.doi.org/#2} {doi:#2}\else \href {http://dx.doi.org/#2} {#1}\fi
  \endgroup}
\def\mn@eprint#1#2{\mn@eprint@#1:#2::\@nil}
\def\mn@eprint@arXiv#1{\href {http://arxiv.org/abs/#1} {{\tt arXiv:#1}}}
\def\mn@eprint@dblp#1{\href {http://dblp.uni-trier.de/rec/bibtex/#1.xml}
  {dblp:#1}}
\def\mn@eprint@#1:#2:#3:#4\@nil{\def\@tempa {#1}\def\@tempb {#2}\def\@tempc
  {#3}\ifx \@tempc \@empty \let \@tempc \@tempb \let \@tempb \@tempa \fi \ifx
  \@tempb \@empty \def\@tempb {arXiv}\fi \@ifundefined
  {mn@eprint@\@tempb}{\@tempb:\@tempc}{\expandafter \expandafter \csname
  mn@eprint@\@tempb\endcsname \expandafter{\@tempc}}}

\bibitem[\protect\citeauthoryear{{Abramowicz}, {Czerny}, {Lasota}  \&
  {Szuszkiewicz}}{{Abramowicz} et~al.}{1988}]{Abramowicz+88}
{Abramowicz} M.~A.,  {Czerny} B.,  {Lasota} J.~P.,   {Szuszkiewicz} E.,  1988,
  \mn@doi [\apj] {10.1086/166683}, \href
  {https://ui.adsabs.harvard.edu/abs/1988ApJ...332..646A} {332, 646}

\bibitem[\protect\citeauthoryear{{Alloin}, {Pelat}, {Phillips}, {Fosbury}  \&
  {Freeman}}{{Alloin} et~al.}{1986}]{Alloin1986}
{Alloin} D.,  {Pelat} D.,  {Phillips} M.~M.,  {Fosbury} R.~A.~E.,   {Freeman}
  K.,  1986, \mn@doi [\apj] {10.1086/164475}, \href
  {https://ui.adsabs.harvard.edu/abs/1986ApJ...308...23A} {308, 23}

\bibitem[\protect\citeauthoryear{{Altamirano} et~al.,}{{Altamirano}
  et~al.}{2011}]{Altamirano_2011_XRB_obs}
{Altamirano} D.,  et~al., 2011, \mn@doi [\apjl] {10.1088/2041-8205/742/2/L17},
  \href {https://ui.adsabs.harvard.edu/abs/2011ApJ...742L..17A} {742, L17}

\bibitem[\protect\citeauthoryear{{Arcodia} et~al.,}{{Arcodia}
  et~al.}{2021}]{Arcodia_2021}
{Arcodia} R.,  et~al., 2021, \mn@doi [\nat] {10.1038/s41586-021-03394-6}, \href
  {https://ui.adsabs.harvard.edu/abs/2021Natur.592..704A} {592, 704}

\bibitem[\protect\citeauthoryear{{Assef}, {Stern}, {Noirot}, {Jun}, {Cutri}  \&
  {Eisenhardt}}{{Assef} et~al.}{2018}]{Assef2018}
{Assef} R.~J.,  {Stern} D.,  {Noirot} G.,  {Jun} H.~D.,  {Cutri} R.~M.,
  {Eisenhardt} P.~R.~M.,  2018, \mn@doi [\apjs] {10.3847/1538-4365/aaa00a},
  \href {https://ui.adsabs.harvard.edu/abs/2018ApJS..234...23A} {234, 23}

\bibitem[\protect\citeauthoryear{{Bagnoli} \& {in't Zand}}{{Bagnoli} \& {in't
  Zand}}{2015}]{Bagnoli2015}
{Bagnoli} T.,  {in't Zand} J.~J.~M.,  2015, \mn@doi [\mnras]
  {10.1093/mnrasl/slv045}, \href
  {https://ui.adsabs.harvard.edu/abs/2015MNRAS.450L..52B} {450, L52}

\bibitem[\protect\citeauthoryear{{Balbus} \& {Hawley}}{{Balbus} \&
  {Hawley}}{1991}]{Balbus1991}
{Balbus} S.~A.,  {Hawley} J.~F.,  1991, \mn@doi [\apj] {10.1086/170270}, \href
  {https://ui.adsabs.harvard.edu/abs/1991ApJ...376..214B} {376, 214}

\bibitem[\protect\citeauthoryear{{Bardeen}, {Press}  \& {Teukolsky}}{{Bardeen}
  et~al.}{1972}]{Bardeen1972}
{Bardeen} J.~M.,  {Press} W.~H.,   {Teukolsky} S.~A.,  1972, \mn@doi [\apj]
  {10.1086/151796}, \href
  {https://ui.adsabs.harvard.edu/abs/1972ApJ...178..347B} {178, 347}

\bibitem[\protect\citeauthoryear{{Begelman} \& {Pringle}}{{Begelman} \&
  {Pringle}}{2007}]{BegelmanPringle07}
{Begelman} M.~C.,  {Pringle} J.~E.,  2007, \mn@doi [\mnras]
  {10.1111/j.1365-2966.2006.11372.x}, \href
  {https://ui.adsabs.harvard.edu/abs/2007MNRAS.375.1070B} {375, 1070}

\bibitem[\protect\citeauthoryear{{Belloni}, {M{\'e}ndez}, {King}, {van der
  Klis}  \& {van Paradijs}}{{Belloni} et~al.}{1997}]{Belloni+97}
{Belloni} T.,  {M{\'e}ndez} M.,  {King} A.~R.,  {van der Klis} M.,   {van
  Paradijs} J.,  1997, \mn@doi [\apjl] {10.1086/310595}, \href
  {https://ui.adsabs.harvard.edu/abs/1997ApJ...479L.145B} {479, L145}

\bibitem[\protect\citeauthoryear{{Blaes}, {Davis}, {Hirose}, {Krolik}  \&
  {Stone}}{{Blaes} et~al.}{2006}]{Blaes2006}
{Blaes} O.~M.,  {Davis} S.~W.,  {Hirose} S.,  {Krolik} J.~H.,   {Stone} J.~M.,
  2006, \mn@doi [\apj] {10.1086/503741}, \href
  {https://ui.adsabs.harvard.edu/abs/2006ApJ...645.1402B} {645, 1402}

\bibitem[\protect\citeauthoryear{{Blandford} \& {Znajek}}{{Blandford} \&
  {Znajek}}{1977}]{Blandford_Znajek1977}
{Blandford} R.~D.,  {Znajek} R.~L.,  1977, \mn@doi [\mnras]
  {10.1093/mnras/179.3.433}, \href
  {https://ui.adsabs.harvard.edu/abs/1977MNRAS.179..433B} {179, 433}

\bibitem[\protect\citeauthoryear{{Bonnerot} \& {Stone}}{{Bonnerot} \&
  {Stone}}{2021}]{BonnerotStone21}
{Bonnerot} C.,  {Stone} N.~C.,  2021, \mn@doi [\ssr]
  {10.1007/s11214-020-00789-1}, \href
  {https://ui.adsabs.harvard.edu/abs/2021SSRv..217...16B} {217, 16}

\bibitem[\protect\citeauthoryear{{Bower}, {Metzger}, {Cenko}, {Silverman}  \&
  {Bloom}}{{Bower} et~al.}{2013a}]{Bower2013}
{Bower} G.~C.,  {Metzger} B.~D.,  {Cenko} S.~B.,  {Silverman} J.~M.,   {Bloom}
  J.~S.,  2013a, \mn@doi [\apj] {10.1088/0004-637X/763/2/84}, \href
  {https://ui.adsabs.harvard.edu/abs/2013ApJ...763...84B} {763, 84}

\bibitem[\protect\citeauthoryear{{Bower}, {Metzger}, {Cenko}, {Silverman}  \&
  {Bloom}}{{Bower} et~al.}{2013b}]{Bower+13}
{Bower} G.~C.,  {Metzger} B.~D.,  {Cenko} S.~B.,  {Silverman} J.~M.,   {Bloom}
  J.~S.,  2013b, \mn@doi [\apj] {10.1088/0004-637X/763/2/84}, \href
  {https://ui.adsabs.harvard.edu/abs/2013ApJ...763...84B} {763, 84}

\bibitem[\protect\citeauthoryear{{Burrows} et~al.,}{{Burrows}
  et~al.}{2011}]{Burrows2011}
{Burrows} D.~N.,  et~al., 2011, \mn@doi [\nat] {10.1038/nature10374}, \href
  {https://ui.adsabs.harvard.edu/abs/2011Natur.476..421B} {476, 421}

\bibitem[\protect\citeauthoryear{{Cannizzo}}{{Cannizzo}}{1993}]{Cannizzo93}
{Cannizzo} J.~K.,  1993, \mn@doi [\apj] {10.1086/173486}, \href
  {https://ui.adsabs.harvard.edu/abs/1993ApJ...419..318C} {419, 318}

\bibitem[\protect\citeauthoryear{{Cannizzo}}{{Cannizzo}}{1996}]{Cannizzo_1996}
{Cannizzo} J.~K.,  1996, \mn@doi [\apjl] {10.1086/310167}, \href
  {https://ui.adsabs.harvard.edu/abs/1996ApJ...466L..31C} {466, L31}

\bibitem[\protect\citeauthoryear{{Cenko} et~al.,}{{Cenko}
  et~al.}{2012}]{Cenko2012}
{Cenko} S.~B.,  et~al., 2012, \mn@doi [\apj] {10.1088/0004-637X/753/1/77},
  \href {https://ui.adsabs.harvard.edu/abs/2012ApJ...753...77C} {753, 77}

\bibitem[\protect\citeauthoryear{{Chakraborty}, {Kara}, {Masterson},
  {Giustini}, {Miniutti}  \& {Saxton}}{{Chakraborty}
  et~al.}{2021}]{Chakraborty_2021}
{Chakraborty} J.,  {Kara} E.,  {Masterson} M.,  {Giustini} M.,  {Miniutti} G.,
   {Saxton} R.,  2021, \mn@doi [\apjl] {10.3847/2041-8213/ac313b}, \href
  {https://ui.adsabs.harvard.edu/abs/2021ApJ...921L..40C} {921, L40}

\bibitem[\protect\citeauthoryear{{Davis} \& {El-Abd}}{{Davis} \&
  {El-Abd}}{2019}]{Davis2019}
{Davis} S.~W.,  {El-Abd} S.,  2019, \mn@doi [\apj] {10.3847/1538-4357/ab05c5},
  \href {https://ui.adsabs.harvard.edu/abs/2019ApJ...874...23D} {874, 23}

\bibitem[\protect\citeauthoryear{{Denney} et~al.,}{{Denney}
  et~al.}{2014}]{Denney2014}
{Denney} K.~D.,  et~al., 2014, \mn@doi [\apj] {10.1088/0004-637X/796/2/134},
  \href {https://ui.adsabs.harvard.edu/abs/2014ApJ...796..134D} {796, 134}

\bibitem[\protect\citeauthoryear{{Dexter} \& {Begelman}}{{Dexter} \&
  {Begelman}}{2019}]{Dexter2019}
{Dexter} J.,  {Begelman} M.~C.,  2019, \mn@doi [\mnras]
  {10.1093/mnrasl/sly213}, \href
  {https://ui.adsabs.harvard.edu/abs/2019MNRAS.483L..17D} {483, L17}

\bibitem[\protect\citeauthoryear{{Done}, {Gierli{\'n}ski}  \& {Kubota}}{{Done}
  et~al.}{2007}]{Done+07}
{Done} C.,  {Gierli{\'n}ski} M.,   {Kubota} A.,  2007, \mn@doi [\aapr]
  {10.1007/s00159-007-0006-1}, \href
  {https://ui.adsabs.harvard.edu/abs/2007A&ARv..15....1D} {15, 1}

\bibitem[\protect\citeauthoryear{{Evans} \& {Kochanek}}{{Evans} \&
  {Kochanek}}{1989}]{EvansKochanek89}
{Evans} C.~R.,  {Kochanek} C.~S.,  1989, \mn@doi [\apjl] {10.1086/185567},
  \href {https://ui.adsabs.harvard.edu/abs/1989ApJ...346L..13E} {346, L13}

\bibitem[\protect\citeauthoryear{{Feng}, {Cao}, {Li}  \& {Gu}}{{Feng}
  et~al.}{2021}]{Feng2021}
{Feng} J.,  {Cao} X.,  {Li} J.-w.,   {Gu} W.-M.,  2021, \mn@doi [\apj]
  {10.3847/1538-4357/ac07a6}, \href
  {https://ui.adsabs.harvard.edu/abs/2021ApJ...916...61F} {916, 61}

\bibitem[\protect\citeauthoryear{{Ferland} et~al.,}{{Ferland}
  et~al.}{2013}]{Cloudy2013}
{Ferland} G.~J.,  et~al., 2013, \rmxaa, \href
  {https://ui.adsabs.harvard.edu/abs/2013RMxAA..49..137F} {49, 137}

\bibitem[\protect\citeauthoryear{{Ferland} et~al.,}{{Ferland}
  et~al.}{2017}]{Cloudy2017}
{Ferland} G.~J.,  et~al., 2017, \rmxaa, \href
  {https://ui.adsabs.harvard.edu/abs/2017RMxAA..53..385F} {53, 385}

\bibitem[\protect\citeauthoryear{{Frank}, {King}  \& {Raine}}{{Frank}
  et~al.}{2002}]{frank_king_raine2002}
{Frank} J.,  {King} A.,   {Raine} D.~J.,  2002, {Accretion Power in
  Astrophysics: Third Edition}

\bibitem[\protect\citeauthoryear{{Generozov}, {Mimica}, {Metzger}, {Stone},
  {Giannios}  \& {Aloy}}{{Generozov} et~al.}{2017}]{Generozov2017}
{Generozov} A.,  {Mimica} P.,  {Metzger} B.~D.,  {Stone} N.~C.,  {Giannios} D.,
    {Aloy} M.~A.,  2017, \mn@doi [\mnras] {10.1093/mnras/stw2439}, \href
  {https://ui.adsabs.harvard.edu/abs/2017MNRAS.464.2481G} {464, 2481}

\bibitem[\protect\citeauthoryear{{Gezari}, {Chornock}, {Lawrence}, {Rest},
  {Jones}, {Berger}, {Challis}  \& {Narayan}}{{Gezari}
  et~al.}{2015}]{Gezari+15}
{Gezari} S.,  {Chornock} R.,  {Lawrence} A.,  {Rest} A.,  {Jones} D.~O.,
  {Berger} E.,  {Challis} P.~M.,   {Narayan} G.,  2015, \mn@doi [\apjl]
  {10.1088/2041-8205/815/1/L5}, \href
  {https://ui.adsabs.harvard.edu/abs/2015ApJ...815L...5G} {815, L5}

\bibitem[\protect\citeauthoryear{{Gezari}, {Cenko}  \& {Arcavi}}{{Gezari}
  et~al.}{2017}]{Gezari2017}
{Gezari} S.,  {Cenko} S.~B.,   {Arcavi} I.,  2017, \mn@doi [\apjl]
  {10.3847/2041-8213/aaa0c2}, \href
  {https://ui.adsabs.harvard.edu/abs/2017ApJ...851L..47G} {851, L47}

\bibitem[\protect\citeauthoryear{{Giustini}, {Miniutti}  \&
  {Saxton}}{{Giustini} et~al.}{2020}]{Giustini_2020}
{Giustini} M.,  {Miniutti} G.,   {Saxton} R.~D.,  2020, \mn@doi [\aap]
  {10.1051/0004-6361/202037610}, \href
  {https://ui.adsabs.harvard.edu/abs/2020A&A...636L...2G} {636, L2}

\bibitem[\protect\citeauthoryear{{Graham} et~al.,}{{Graham}
  et~al.}{2020}]{Graham2020}
{Graham} M.~J.,  et~al., 2020, \mn@doi [\mnras] {10.1093/mnras/stz3244}, \href
  {https://ui.adsabs.harvard.edu/abs/2020MNRAS.491.4925G} {491, 4925}

\bibitem[\protect\citeauthoryear{{Grz{\c{e}}dzielski}, {Janiuk}  \&
  {Czerny}}{{Grz{\c{e}}dzielski} et~al.}{2017}]{Grzdzielski+17}
{Grz{\c{e}}dzielski} M.,  {Janiuk} A.,   {Czerny} B.,  2017, \mn@doi [\apj]
  {10.3847/1538-4357/aa7dd9}, \href
  {https://ui.adsabs.harvard.edu/abs/2017ApJ...845...20G} {845, 20}

\bibitem[\protect\citeauthoryear{{Guillochon} \& {Ramirez-Ruiz}}{{Guillochon}
  \& {Ramirez-Ruiz}}{2013}]{GuillochonRamirezRuiz13}
{Guillochon} J.,  {Ramirez-Ruiz} E.,  2013, \mn@doi [\apj]
  {10.1088/0004-637X/767/1/25}, \href
  {https://ui.adsabs.harvard.edu/abs/2013ApJ...767...25G} {767, 25}

\bibitem[\protect\citeauthoryear{{Guillochon}, {Manukian}  \&
  {Ramirez-Ruiz}}{{Guillochon} et~al.}{2014}]{Guillochon+14}
{Guillochon} J.,  {Manukian} H.,   {Ramirez-Ruiz} E.,  2014, \mn@doi [\apj]
  {10.1088/0004-637X/783/1/23}, \href
  {https://ui.adsabs.harvard.edu/abs/2014ApJ...783...23G} {783, 23}

\bibitem[\protect\citeauthoryear{{Hills}}{{Hills}}{1975}]{Hills75}
{Hills} J.~G.,  1975, \mn@doi [\nat] {10.1038/254295a0}, \href
  {https://ui.adsabs.harvard.edu/abs/1975Natur.254..295H} {254, 295}

\bibitem[\protect\citeauthoryear{{Hirose}, {Krolik}  \& {Blaes}}{{Hirose}
  et~al.}{2009}]{Hirose+09}
{Hirose} S.,  {Krolik} J.~H.,   {Blaes} O.,  2009, \mn@doi [\apj]
  {10.1088/0004-637X/691/1/16}, \href
  {https://ui.adsabs.harvard.edu/abs/2009ApJ...691...16H} {691, 16}

\bibitem[\protect\citeauthoryear{{Honma}, {Matsumoto}  \& {Kato}}{{Honma}
  et~al.}{1991}]{Honma+91}
{Honma} F.,  {Matsumoto} R.,   {Kato} S.,  1991, \pasj, \href
  {https://ui.adsabs.harvard.edu/abs/1991PASJ...43..147H} {43, 147}

\bibitem[\protect\citeauthoryear{{Ingram}, {Motta}, {Aigrain}  \&
  {Karastergiou}}{{Ingram} et~al.}{2021}]{Ingram2021}
{Ingram} A.,  {Motta} S.~E.,  {Aigrain} S.,   {Karastergiou} A.,  2021, \mn@doi
  [\mnras] {10.1093/mnras/stab609}, \href
  {https://ui.adsabs.harvard.edu/abs/2021MNRAS.503.1703I} {503, 1703}

\bibitem[\protect\citeauthoryear{{Janiuk} \& {Misra}}{{Janiuk} \&
  {Misra}}{2012}]{JaniukMisra12}
{Janiuk} A.,  {Misra} R.,  2012, \mn@doi [\aap] {10.1051/0004-6361/201118765},
  \href {https://ui.adsabs.harvard.edu/abs/2012A&A...540A.114J} {540, A114}

\bibitem[\protect\citeauthoryear{{Janiuk}, {Czerny}  \&
  {Siemiginowska}}{{Janiuk} et~al.}{2002}]{Janiuk+02}
{Janiuk} A.,  {Czerny} B.,   {Siemiginowska} A.,  2002, \mn@doi [\apj]
  {10.1086/341804}, \href
  {https://ui.adsabs.harvard.edu/abs/2002ApJ...576..908J} {576, 908}

\bibitem[\protect\citeauthoryear{{Jiang}, {Davis}  \& {Stone}}{{Jiang}
  et~al.}{2016}]{Jiang+16}
{Jiang} Y.-F.,  {Davis} S.~W.,   {Stone} J.~M.,  2016, \mn@doi [\apj]
  {10.3847/0004-637X/827/1/10}, \href
  {https://ui.adsabs.harvard.edu/abs/2016ApJ...827...10J} {827, 10}

\bibitem[\protect\citeauthoryear{{Jiang}, {Stone}  \& {Davis}}{{Jiang}
  et~al.}{2019a}]{Jiang2019superEdd}
{Jiang} Y.-F.,  {Stone} J.~M.,   {Davis} S.~W.,  2019a, \mn@doi [\apj]
  {10.3847/1538-4357/ab29ff}, \href
  {https://ui.adsabs.harvard.edu/abs/2019ApJ...880...67J} {880, 67}

\bibitem[\protect\citeauthoryear{{Jiang}, {Blaes}, {Stone}  \& {Davis}}{{Jiang}
  et~al.}{2019b}]{Jiang+19}
{Jiang} Y.-F.,  {Blaes} O.,  {Stone} J.~M.,   {Davis} S.~W.,  2019b, \mn@doi
  [\apj] {10.3847/1538-4357/ab4a00}, \href
  {https://ui.adsabs.harvard.edu/abs/2019ApJ...885..144J} {885, 144}

\bibitem[\protect\citeauthoryear{{Jonker}, {Stone}, {Generozov}, {van Velzen}
  \& {Metzger}}{{Jonker} et~al.}{2020}]{Jonker2020}
{Jonker} P.~G.,  {Stone} N.~C.,  {Generozov} A.,  {van Velzen} S.,   {Metzger}
  B.,  2020, \mn@doi [\apj] {10.3847/1538-4357/ab659c}, \href
  {https://ui.adsabs.harvard.edu/abs/2020ApJ...889..166J} {889, 166}

\bibitem[\protect\citeauthoryear{{King}}{{King}}{2020}]{King_2020}
{King} A.,  2020, \mn@doi [\mnras] {10.1093/mnrasl/slaa020}, \href
  {https://ui.adsabs.harvard.edu/abs/2020MNRAS.493L.120K} {493, L120}

\bibitem[\protect\citeauthoryear{{King}}{{King}}{2022}]{King_2022_QPE_models}
{King} A.,  2022, \mn@doi [\mnras] {10.1093/mnras/stac1641}, \href
  {https://ui.adsabs.harvard.edu/abs/2022MNRAS.tmp.1590K} {}

\bibitem[\protect\citeauthoryear{{Komossa}}{{Komossa}}{2015}]{Komossa2015}
{Komossa} S.,  2015, \mn@doi [Journal of High Energy Astrophysics]
  {10.1016/j.jheap.2015.04.006}, \href
  {https://ui.adsabs.harvard.edu/abs/2015JHEAp...7..148K} {7, 148}

\bibitem[\protect\citeauthoryear{{Krolik} \& {Linial}}{{Krolik} \&
  {Linial}}{2022}]{Krolik2022}
{Krolik} J.~H.,  {Linial} I.,  2022, arXiv e-prints, \href
  {https://ui.adsabs.harvard.edu/abs/2022arXiv220902786K} {p. arXiv:2209.02786}

\bibitem[\protect\citeauthoryear{{LaMassa} et~al.,}{{LaMassa}
  et~al.}{2015}]{LaMassa2015}
{LaMassa} S.~M.,  et~al., 2015, \mn@doi [\apj] {10.1088/0004-637X/800/2/144},
  \href {https://ui.adsabs.harvard.edu/abs/2015ApJ...800..144L} {800, 144}

\bibitem[\protect\citeauthoryear{{Lasota}}{{Lasota}}{2001}]{Lasota01}
{Lasota} J.-P.,  2001, \mn@doi [\nar] {10.1016/S1387-6473(01)00112-9}, \href
  {https://ui.adsabs.harvard.edu/abs/2001NewAR..45..449L} {45, 449}

\bibitem[\protect\citeauthoryear{{Lightman} \& {Eardley}}{{Lightman} \&
  {Eardley}}{1974}]{LightmanEardley74}
{Lightman} A.~P.,  {Eardley} D.~M.,  1974, \mn@doi [\apjl] {10.1086/181377},
  \href {https://ui.adsabs.harvard.edu/abs/1974ApJ...187L...1L} {187, L1}

\bibitem[\protect\citeauthoryear{{Lu} \& {Quataert}}{{Lu} \&
  {Quataert}}{2022}]{LuQuataert22}
{Lu} W.,  {Quataert} E.,  2022, arXiv e-prints, \href
  {https://ui.adsabs.harvard.edu/abs/2022arXiv221008023L} {p. arXiv:2210.08023}

\bibitem[\protect\citeauthoryear{{Maselli}, {Capitanio}, {Feroci}, {Massa},
  {Massaro}  \& {Mineo}}{{Maselli} et~al.}{2018}]{Maselli2018}
{Maselli} A.,  {Capitanio} F.,  {Feroci} M.,  {Massa} F.,  {Massaro} E.,
  {Mineo} T.,  2018, \mn@doi [\aap] {10.1051/0004-6361/201732097}, \href
  {https://ui.adsabs.harvard.edu/abs/2018A&A...612A..33M} {612, A33}

\bibitem[\protect\citeauthoryear{{Massaro}, {Ventura}, {Massa}, {Feroci},
  {Mineo}, {Cusumano}, {Casella}  \& {Belloni}}{{Massaro}
  et~al.}{2010}]{Massaro2010}
{Massaro} E.,  {Ventura} G.,  {Massa} F.,  {Feroci} M.,  {Mineo} T.,
  {Cusumano} G.,  {Casella} P.,   {Belloni} T.,  2010, \mn@doi [\aap]
  {10.1051/0004-6361/200912908}, \href
  {https://ui.adsabs.harvard.edu/abs/2010A&A...513A..21M} {513, A21}

\bibitem[\protect\citeauthoryear{{Matsumoto} \& {Piran}}{{Matsumoto} \&
  {Piran}}{2021}]{Matsumoto2021}
{Matsumoto} T.,  {Piran} T.,  2021, \mn@doi [\mnras] {10.1093/mnras/stab2418},
  \href {https://ui.adsabs.harvard.edu/abs/2021MNRAS.507.4196M} {507, 4196}

\bibitem[\protect\citeauthoryear{{Meier}}{{Meier}}{2001}]{Meier2001}
{Meier} D.~L.,  2001, \mn@doi [\apjl] {10.1086/318921}, \href
  {https://ui.adsabs.harvard.edu/abs/2001ApJ...548L...9M} {548, L9}

\bibitem[\protect\citeauthoryear{{Metzger}, {Stone}  \& {Gilbaum}}{{Metzger}
  et~al.}{2022}]{Metzger_Stone_2022}
{Metzger} B.~D.,  {Stone} N.~C.,   {Gilbaum} S.,  2022, \mn@doi [\apj]
  {10.3847/1538-4357/ac3ee1}, \href
  {https://ui.adsabs.harvard.edu/abs/2022ApJ...926..101M} {926, 101}

\bibitem[\protect\citeauthoryear{{Meyer} \& {Meyer-Hofmeister}}{{Meyer} \&
  {Meyer-Hofmeister}}{1981}]{MeyerMeyerHofmeister81}
{Meyer} F.,  {Meyer-Hofmeister} E.,  1981, \aap, \href
  {https://ui.adsabs.harvard.edu/abs/1981A&A...104L..10M} {104, L10}

\bibitem[\protect\citeauthoryear{{Miller} et~al.,}{{Miller}
  et~al.}{2015}]{Miller2015}
{Miller} J.~M.,  et~al., 2015, \mn@doi [\nat] {10.1038/nature15708}, \href
  {https://ui.adsabs.harvard.edu/abs/2015Natur.526..542M} {526, 542}

\bibitem[\protect\citeauthoryear{{Miniutti} et~al.,}{{Miniutti}
  et~al.}{2019}]{Miniutti_2019}
{Miniutti} G.,  et~al., 2019, \mn@doi [\nat] {10.1038/s41586-019-1556-x}, \href
  {https://ui.adsabs.harvard.edu/abs/2019Natur.573..381M} {573, 381}

\bibitem[\protect\citeauthoryear{{Mishra}, {Fragile}, {Anderson},
  {Blankenship}, {Li}  \& {Nalewajko}}{{Mishra} et~al.}{2022}]{Mishra2022}
{Mishra} B.,  {Fragile} P.~C.,  {Anderson} J.,  {Blankenship} A.,  {Li} H.,
  {Nalewajko} K.,  2022, arXiv e-prints, \href
  {https://ui.adsabs.harvard.edu/abs/2022arXiv220903317M} {p. arXiv:2209.03317}

\bibitem[\protect\citeauthoryear{{Mummery}}{{Mummery}}{2021a}]{Mummery2021arXiv}
{Mummery} A.,  2021a, arXiv e-prints, \href
  {https://ui.adsabs.harvard.edu/abs/2021arXiv210406212M} {p. arXiv:2104.06212}

\bibitem[\protect\citeauthoryear{{Mummery}}{{Mummery}}{2021b}]{Mummery2021}
{Mummery} A.,  2021b, \mn@doi [\mnras] {10.1093/mnras/stab1187}, \href
  {https://ui.adsabs.harvard.edu/abs/2021MNRAS.504.5144M} {504, 5144}

\bibitem[\protect\citeauthoryear{{Mummery} \& {Balbus}}{{Mummery} \&
  {Balbus}}{2021}]{Mummery_Balbus2021}
{Mummery} A.,  {Balbus} S.~A.,  2021, \mn@doi [\mnras]
  {10.1093/mnras/stab1141}, \href
  {https://ui.adsabs.harvard.edu/abs/2021MNRAS.505.1629M} {505, 1629}

\bibitem[\protect\citeauthoryear{{Narayan} \& {Yi}}{{Narayan} \&
  {Yi}}{1995}]{narayan_yi1995}
{Narayan} R.,  {Yi} I.,  1995, \mn@doi [\apj] {10.1086/176343}, \href
  {https://ui.adsabs.harvard.edu/abs/1995ApJ...452..710N} {452, 710}

\bibitem[\protect\citeauthoryear{{Narayan}, {McKinney}  \& {Farmer}}{{Narayan}
  et~al.}{2007}]{Narayan2007}
{Narayan} R.,  {McKinney} J.~C.,   {Farmer} A.~J.,  2007, \mn@doi [\mnras]
  {10.1111/j.1365-2966.2006.11272.x}, \href
  {https://ui.adsabs.harvard.edu/abs/2007MNRAS.375..548N} {375, 548}

\bibitem[\protect\citeauthoryear{{Nixon} \& {Coughlin}}{{Nixon} \&
  {Coughlin}}{2022}]{Nixon2022}
{Nixon} C.~J.,  {Coughlin} E.~R.,  2022, \mn@doi [\apjl]
  {10.3847/2041-8213/ac5118}, \href
  {https://ui.adsabs.harvard.edu/abs/2022ApJ...927L..25N} {927, L25}

\bibitem[\protect\citeauthoryear{{Noda} \& {Done}}{{Noda} \&
  {Done}}{2018}]{Noda2018}
{Noda} H.,  {Done} C.,  2018, \mn@doi [\mnras] {10.1093/mnras/sty2032}, \href
  {https://ui.adsabs.harvard.edu/abs/2018MNRAS.480.3898N} {480, 3898}

\bibitem[\protect\citeauthoryear{{Oda}, {Machida}, {Nakamura}  \&
  {Matsumoto}}{{Oda} et~al.}{2009}]{Oda+09}
{Oda} H.,  {Machida} M.,  {Nakamura} K.~E.,   {Matsumoto} R.,  2009, \mn@doi
  [\apj] {10.1088/0004-637X/697/1/16}, \href
  {https://ui.adsabs.harvard.edu/abs/2009ApJ...697...16O} {697, 16}

\bibitem[\protect\citeauthoryear{{Ohsuga}}{{Ohsuga}}{2006}]{Ohsuga06}
{Ohsuga} K.,  2006, \mn@doi [\apj] {10.1086/500184}, \href
  {https://ui.adsabs.harvard.edu/abs/2006ApJ...640..923O} {640, 923}

\bibitem[\protect\citeauthoryear{{Oknyansky}, {Winkler}, {Tsygankov},
  {Lipunov}, {Gorbovskoy}, {van Wyk}, {Buckley}  \& {Tyurina}}{{Oknyansky}
  et~al.}{2019}]{Oknyansky2019}
{Oknyansky} V.~L.,  {Winkler} H.,  {Tsygankov} S.~S.,  {Lipunov} V.~M.,
  {Gorbovskoy} E.~S.,  {van Wyk} F.,  {Buckley} D.~A.~H.,   {Tyurina} N.~V.,
  2019, \mn@doi [\mnras] {10.1093/mnras/sty3133}, \href
  {https://ui.adsabs.harvard.edu/abs/2019MNRAS.483..558O} {483, 558}

\bibitem[\protect\citeauthoryear{{Pan}, {Li}  \& {Cao}}{{Pan}
  et~al.}{2021}]{Pan2021}
{Pan} X.,  {Li} S.-L.,   {Cao} X.,  2021, \mn@doi [\apj]
  {10.3847/1538-4357/abe766}, \href
  {https://ui.adsabs.harvard.edu/abs/2021ApJ...910...97P} {910, 97}

\bibitem[\protect\citeauthoryear{{Pan}, {Li}, {Cao}, {Miniutti}  \& {Gu}}{{Pan}
  et~al.}{2022}]{Pan_2022_viscosity_model}
{Pan} X.,  {Li} S.-L.,  {Cao} X.,  {Miniutti} G.,   {Gu} M.,  2022, \mn@doi
  [\apjl] {10.3847/2041-8213/ac5faf}, \href
  {https://ui.adsabs.harvard.edu/abs/2022ApJ...928L..18P} {928, L18}

\bibitem[\protect\citeauthoryear{{Pessah} \& {Psaltis}}{{Pessah} \&
  {Psaltis}}{2005}]{pessah_psaltis2005}
{Pessah} M.~E.,  {Psaltis} D.,  2005, \mn@doi [\apj] {10.1086/430940}, \href
  {https://ui.adsabs.harvard.edu/abs/2005ApJ...628..879P} {628, 879}

\bibitem[\protect\citeauthoryear{{Piran}}{{Piran}}{1978}]{Piran78}
{Piran} T.,  1978, \mn@doi [\apj] {10.1086/156069}, \href
  {https://ui.adsabs.harvard.edu/abs/1978ApJ...221..652P} {221, 652}

\bibitem[\protect\citeauthoryear{{Piran}, {Svirski}, {Krolik}, {Cheng}  \&
  {Shiokawa}}{{Piran} et~al.}{2015}]{Piran+15}
{Piran} T.,  {Svirski} G.,  {Krolik} J.,  {Cheng} R.~M.,   {Shiokawa} H.,
  2015, \mn@doi [\apj] {10.1088/0004-637X/806/2/164}, \href
  {https://ui.adsabs.harvard.edu/abs/2015ApJ...806..164P} {806, 164}

\bibitem[\protect\citeauthoryear{{Pringle}, {Rees}  \& {Pacholczyk}}{{Pringle}
  et~al.}{1973}]{Pringle+73}
{Pringle} J.~E.,  {Rees} M.~J.,   {Pacholczyk} A.~G.,  1973, \aap, \href
  {https://ui.adsabs.harvard.edu/abs/1973A&A....29..179P} {29, 179}

\bibitem[\protect\citeauthoryear{{Raj} \& {Nixon}}{{Raj} \&
  {Nixon}}{2021}]{Raj_Nixon2021}
{Raj} A.,  {Nixon} C.~J.,  2021, \mn@doi [\apj] {10.3847/1538-4357/abdc25},
  \href {https://ui.adsabs.harvard.edu/abs/2021ApJ...909...82R} {909, 82}

\bibitem[\protect\citeauthoryear{{Raj}, {Nixon}  \& {Do{\u{g}}an}}{{Raj}
  et~al.}{2021}]{Raj2021}
{Raj} A.,  {Nixon} C.~J.,   {Do{\u{g}}an} S.,  2021, \mn@doi [\apj]
  {10.3847/1538-4357/abdc24}, \href
  {https://ui.adsabs.harvard.edu/abs/2021ApJ...909...81R} {909, 81}

\bibitem[\protect\citeauthoryear{{Rees}}{{Rees}}{1988}]{Rees88}
{Rees} M.~J.,  1988, \mn@doi [\nat] {10.1038/333523a0}, \href
  {https://ui.adsabs.harvard.edu/abs/1988Natur.333..523R} {333, 523}

\bibitem[\protect\citeauthoryear{{Ross}, {Latter}  \& {Tehranchi}}{{Ross}
  et~al.}{2017}]{Ross+17}
{Ross} J.,  {Latter} H.~N.,   {Tehranchi} M.,  2017, \mn@doi [\mnras]
  {10.1093/mnras/stx564}, \href
  {https://ui.adsabs.harvard.edu/abs/2017MNRAS.468.2401R} {468, 2401}

\bibitem[\protect\citeauthoryear{{Ross} et~al.,}{{Ross}
  et~al.}{2018}]{Ross2018}
{Ross} N.~P.,  et~al., 2018, \mn@doi [\mnras] {10.1093/mnras/sty2002}, \href
  {https://ui.adsabs.harvard.edu/abs/2018MNRAS.480.4468R} {480, 4468}

\bibitem[\protect\citeauthoryear{{Saxton}, {Komossa}, {Auchettl}  \&
  {Jonker}}{{Saxton} et~al.}{2021}]{Saxton2021}
{Saxton} R.,  {Komossa} S.,  {Auchettl} K.,   {Jonker} P.~G.,  2021,
  {Correction to: X-Ray Properties of TDEs}, Space Science Reviews, Volume 217,
  Issue 1, article id.18 (\mn@eprint {arXiv} {2103.15442}),
  \mn@doi{10.1007/s11214-020-00759-7}

\bibitem[\protect\citeauthoryear{{Scepi}, {Begelman}  \& {Dexter}}{{Scepi}
  et~al.}{2021}]{Scepi2021}
{Scepi} N.,  {Begelman} M.~C.,   {Dexter} J.,  2021, \mn@doi [\mnras]
  {10.1093/mnrasl/slab002}, \href
  {https://ui.adsabs.harvard.edu/abs/2021MNRAS.502L..50S} {502, L50}

\bibitem[\protect\citeauthoryear{{Shakura} \& {Sunyaev}}{{Shakura} \&
  {Sunyaev}}{1973}]{ShakuraSunyaev73}
{Shakura} N.~I.,  {Sunyaev} R.~A.,  1973, \aap, \href
  {https://ui.adsabs.harvard.edu/abs/1973A&A....24..337S} {500, 33}

\bibitem[\protect\citeauthoryear{{Shakura} \& {Sunyaev}}{{Shakura} \&
  {Sunyaev}}{1976}]{ShakuraSunyaev76}
{Shakura} N.~I.,  {Sunyaev} R.~A.,  1976, \mn@doi [\mnras]
  {10.1093/mnras/175.3.613}, \href
  {https://ui.adsabs.harvard.edu/abs/1976MNRAS.175..613S} {175, 613}

\bibitem[\protect\citeauthoryear{{Shen} \& {Matzner}}{{Shen} \&
  {Matzner}}{2014}]{shen_matzner2014}
{Shen} R.-F.,  {Matzner} C.~D.,  2014, \mn@doi [\apj]
  {10.1088/0004-637X/784/2/87}, \href
  {https://ui.adsabs.harvard.edu/abs/2014ApJ...784...87S} {784, 87}

\bibitem[\protect\citeauthoryear{{Shiokawa}, {Krolik}, {Cheng}, {Piran}  \&
  {Noble}}{{Shiokawa} et~al.}{2015}]{Shiokawa+15}
{Shiokawa} H.,  {Krolik} J.~H.,  {Cheng} R.~M.,  {Piran} T.,   {Noble} S.~C.,
  2015, \mn@doi [\apj] {10.1088/0004-637X/804/2/85}, \href
  {https://ui.adsabs.harvard.edu/abs/2015ApJ...804...85S} {804, 85}

\bibitem[\protect\citeauthoryear{{S{\k{a}}dowski}}{{S{\k{a}}dowski}}{2016}]{Sadowski16}
{S{\k{a}}dowski} A.,  2016, \mn@doi [\mnras] {10.1093/mnras/stw913}, \href
  {https://ui.adsabs.harvard.edu/abs/2016MNRAS.459.4397S} {459, 4397}

\bibitem[\protect\citeauthoryear{{{\'S}niegowska}, {Grz{\k{e}}dzielski},
  {Czerny}  \& {Janiuk}}{{{\'S}niegowska} et~al.}{2022}]{Sniegowska2022}
{{\'S}niegowska} M.,  {Grz{\k{e}}dzielski} M.,  {Czerny} B.,   {Janiuk} A.,
  2022, arXiv e-prints, \href
  {https://ui.adsabs.harvard.edu/abs/2022arXiv220410067S} {p. arXiv:2204.10067}

\bibitem[\protect\citeauthoryear{{Stone}, {Sari}  \& {Loeb}}{{Stone}
  et~al.}{2013}]{Stone+13}
{Stone} N.,  {Sari} R.,   {Loeb} A.,  2013, \mn@doi [\mnras]
  {10.1093/mnras/stt1270}, \href
  {https://ui.adsabs.harvard.edu/abs/2013MNRAS.435.1809S} {435, 1809}

\bibitem[\protect\citeauthoryear{{Sukov{\'a}}, {Zaja{\v{c}}ek}, {Witzany}  \&
  {Karas}}{{Sukov{\'a}} et~al.}{2021}]{sukova2021}
{Sukov{\'a}} P.,  {Zaja{\v{c}}ek} M.,  {Witzany} V.,   {Karas} V.,  2021,
  \mn@doi [\apj] {10.3847/1538-4357/ac05c6}, \href
  {https://ui.adsabs.harvard.edu/abs/2021ApJ...917...43S} {917, 43}

\bibitem[\protect\citeauthoryear{{Szuszkiewicz} \& {Miller}}{{Szuszkiewicz} \&
  {Miller}}{1998}]{SzuszkiewiczMiller98}
{Szuszkiewicz} E.,  {Miller} J.~C.,  1998, \mn@doi [\mnras]
  {10.1046/j.1365-8711.1998.01668.x}, \href
  {https://ui.adsabs.harvard.edu/abs/1998MNRAS.298..888S} {298, 888}

\bibitem[\protect\citeauthoryear{{Taam}, {Chen}  \& {Swank}}{{Taam}
  et~al.}{1997}]{Taam1997}
{Taam} R.~E.,  {Chen} X.,   {Swank} J.~H.,  1997, \mn@doi [\apjl]
  {10.1086/310812}, \href
  {https://ui.adsabs.harvard.edu/abs/1997ApJ...485L..83T} {485, L83}

\bibitem[\protect\citeauthoryear{{Tchekhovskoy}, {Narayan}  \&
  {McKinney}}{{Tchekhovskoy} et~al.}{2010}]{Tchekhovskoy2010}
{Tchekhovskoy} A.,  {Narayan} R.,   {McKinney} J.~C.,  2010, \mn@doi [\apj]
  {10.1088/0004-637X/711/1/50}, \href
  {https://ui.adsabs.harvard.edu/abs/2010ApJ...711...50T} {711, 50}

\bibitem[\protect\citeauthoryear{{Wang}, {Yin}, {Ma}  \& {Wu}}{{Wang}
  et~al.}{2022}]{Wang2022}
{Wang} M.,  {Yin} J.,  {Ma} Y.,   {Wu} Q.,  2022, \mn@doi [\apj]
  {10.3847/1538-4357/ac75e6}, \href
  {https://ui.adsabs.harvard.edu/abs/2022ApJ...933..225W} {933, 225}

\bibitem[\protect\citeauthoryear{{Wen}, {Jonker}, {Stone}, {Zabludoff}  \&
  {Psaltis}}{{Wen} et~al.}{2020}]{Wen+20}
{Wen} S.,  {Jonker} P.~G.,  {Stone} N.~C.,  {Zabludoff} A.~I.,   {Psaltis} D.,
  2020, \mn@doi [\apj] {10.3847/1538-4357/ab9817}, \href
  {https://ui.adsabs.harvard.edu/abs/2020ApJ...897...80W} {897, 80}

\bibitem[\protect\citeauthoryear{{Wen}, {Jonker}, {Stone}  \&
  {Zabludoff}}{{Wen} et~al.}{2021}]{Wen+21}
{Wen} S.,  {Jonker} P.~G.,  {Stone} N.~C.,   {Zabludoff} A.~I.,  2021, \mn@doi
  [\apj] {10.3847/1538-4357/ac00b5}, \href
  {https://ui.adsabs.harvard.edu/abs/2021ApJ...918...46W} {918, 46}

\bibitem[\protect\citeauthoryear{{Wevers} et~al.,}{{Wevers}
  et~al.}{2019}]{Wevers2019}
{Wevers} T.,  et~al., 2019, \mn@doi [\mnras] {10.1093/mnras/stz1602}, \href
  {https://ui.adsabs.harvard.edu/abs/2019MNRAS.487.4136W} {487, 4136}

\bibitem[\protect\citeauthoryear{{Wevers}, {Pasham}, {Jalan}, {Rakshit}  \&
  {Arcodia}}{{Wevers} et~al.}{2022}]{Wevers_2022}
{Wevers} T.,  {Pasham} D.~R.,  {Jalan} P.,  {Rakshit} S.,   {Arcodia} R.,
  2022, \mn@doi [\aap] {10.1051/0004-6361/202243143}, \href
  {https://ui.adsabs.harvard.edu/abs/2022A&A...659L...2W} {659, L2}

\bibitem[\protect\citeauthoryear{{Xian}, {Zhang}, {Dou}, {He}  \& {Shu}}{{Xian}
  et~al.}{2021}]{Xian2021}
{Xian} J.,  {Zhang} F.,  {Dou} L.,  {He} J.,   {Shu} X.,  2021, \mn@doi [\apjl]
  {10.3847/2041-8213/ac31aa}, \href
  {https://ui.adsabs.harvard.edu/abs/2021ApJ...921L..32X} {921, L32}

\bibitem[\protect\citeauthoryear{{Yang} et~al.,}{{Yang}
  et~al.}{2018}]{Yang2018}
{Yang} Q.,  et~al., 2018, \mn@doi [\apj] {10.3847/1538-4357/aaca3a}, \href
  {https://ui.adsabs.harvard.edu/abs/2018ApJ...862..109Y} {862, 109}

\bibitem[\protect\citeauthoryear{{Zhao}, {Wang}, {Zou}, {Wang}  \&
  {Dai}}{{Zhao} et~al.}{2022}]{Zhao2022}
{Zhao} Z.~Y.,  {Wang} Y.~Y.,  {Zou} Y.~C.,  {Wang} F.~Y.,   {Dai} Z.~G.,  2022,
  \mn@doi [\aap] {10.1051/0004-6361/202142519}, \href
  {https://ui.adsabs.harvard.edu/abs/2022A&A...661A..55Z} {661, A55}

\bibitem[\protect\citeauthoryear{{van Velzen}, {Frail}, {K{\"o}rding}  \&
  {Falcke}}{{van Velzen} et~al.}{2013a}]{vanVelzen2013}
{van Velzen} S.,  {Frail} D.~A.,  {K{\"o}rding} E.,   {Falcke} H.,  2013a,
  \mn@doi [\aap] {10.1051/0004-6361/201220426}, \href
  {https://ui.adsabs.harvard.edu/abs/2013A&A...552A...5V} {552, A5}

\bibitem[\protect\citeauthoryear{{van Velzen}, {Frail}, {K{\"o}rding}  \&
  {Falcke}}{{van Velzen} et~al.}{2013b}]{vanVelzen+13}
{van Velzen} S.,  {Frail} D.~A.,  {K{\"o}rding} E.,   {Falcke} H.,  2013b,
  \mn@doi [\aap] {10.1051/0004-6361/201220426}, \href
  {https://ui.adsabs.harvard.edu/abs/2013A&A...552A...5V} {552, A5}

\bibitem[\protect\citeauthoryear{{van Velzen} et~al.,}{{van Velzen}
  et~al.}{2019a}]{vanVelzen2019ZTF}
{van Velzen} S.,  et~al., 2019a, \mn@doi [\apj] {10.3847/1538-4357/aafe0c},
  \href {https://ui.adsabs.harvard.edu/abs/2019ApJ...872..198V} {872, 198}

\bibitem[\protect\citeauthoryear{{van Velzen}, {Stone}, {Metzger}, {Gezari},
  {Brown}  \& {Fruchter}}{{van Velzen} et~al.}{2019b}]{vanVelzen+19}
{van Velzen} S.,  {Stone} N.~C.,  {Metzger} B.~D.,  {Gezari} S.,  {Brown}
  T.~M.,   {Fruchter} A.~S.,  2019b, \mn@doi [\apj] {10.3847/1538-4357/ab1844},
  \href {https://ui.adsabs.harvard.edu/abs/2019ApJ...878...82V} {878, 82}

\bibitem[\protect\citeauthoryear{{van Velzen} et~al.,}{{van Velzen}
  et~al.}{2021}]{vanVelzen2021}
{van Velzen} S.,  et~al., 2021, \mn@doi [\apj] {10.3847/1538-4357/abc258},
  \href {https://ui.adsabs.harvard.edu/abs/2021ApJ...908....4V} {908, 4}

\makeatother
\end{thebibliography}

\appendix

\onecolumn

\FloatBarrier

\section{Stability criterion} 
\label{app_stability}

Here we derive thermal stability condition for a general thin disk, supported by pressure, composed of three components:
\beq 
P = \Pgas + \Prad + \Pmag = \frac{\rho \kB T_{\rm c}}{\mu \mP} + \frac{4 \sigSB T_{\rm c}^4}{3 c} + p_0 \vK \rho \sqrt{\frac{\kB T_{\rm c}}{\mu \mP}} \,.
\label{Pdef}
\eeq   
Volumetric heating rate due to viscous dissipation of these $\alpha$-disks is given by the usual expression: 
\beq  
q^{+} = \frac{D(R)}{H} = \frac{9}{8} \OmgK \alpha P \,. 
\label{q_heating_def}
\eeq 
Volumetric cooling rate $q^{-} = \qrad + \qadv$ is due to radiative diffusion in vertical direction and advection in radial direction, 
\begin{subequations}
\begin{align} 
& \qrad = \frac{4 \sigSB T_{\rm c}^4 \OmgK^2  }{3 \kappa(\rho,T_{\rm c}) P(\rho,T_{\rm c})} \\
& \qadv = \frac{\xi \dot{M}  }{2 \pi R^2} \frac{P}{\Sigma}  = \xi \frac{\nu P}{R^2} = \frac{ \xi \alpha  }{R^2 \OmgK} \frac{P^2(\rho,T_{\rm c})}{\rho}    \, .                       
\end{align} 
\label{q_cooling_def}
\end{subequations} 


Linear thermal stability of these thin disk equilibria demands that rate of volumetric cooling per unit change in temperature should exceed its heating analogue on thermal timescales (much shorter than viscous time so that local surface density $\Sigma$ does not change) and this translates to the following stability condition \citep{frank_king_raine2002}:
\beq  
\frac{\p q^{-}}{\p T_{\rm c}} \bigg|_{\Sigma} >  \frac{\p q^{+}}{\p T_{\rm c}} \bigg|_{\Sigma}  \, .
\label{stab_rel_formal}
\eeq 
We define a stability parameter $\Sth$ which measures the strength of dependence of cooling terms on $T_{\rm c}$ relative to heating:  
\beq 
 \Sth = \frac{\qadv}{q^+} \frac{\rmd \ln \qadv  }{\rmd \ln T_{\rm c}} \bigg|_\Sigma + \frac{\qrad}{q^+} \frac{\rmd \ln \qrad  }{\rmd \ln T_{\rm c}} \bigg|_\Sigma -  \frac{\rmd \ln q^+  }{\rmd \ln T_{\rm c}} \bigg|_\Sigma 
 \label{S_def}
 \eeq 
 Note that $\Sth >0$ for thermal stability.  

Partial derivatives of a general quantity $Q(\rho,T_{\rm c})$ with $T_{\rm c}$ at fixed $\Sigma$ can be evaluated using the chain rule: 
\beq  
\frac{\p Q}{\p T_{\rm c}} \bigg|_{\Sigma} = \frac{\p Q}{\p T_{\rm c}} \bigg|_{\rho} + \frac{\p Q}{\p \rho} \bigg|_{T_{\rm c}} \frac{\p \rho}{\p T_{\rm c}} \bigg|_{\Sigma}  \,. 
\label{partial_Q}
\eeq  
To assist the further analysis, it is good to note that hydrostatic equilibrium ($H = c_s/\OmgK$) implies $\rho P(\rho,T_{\rm c}) = \Sigma^2 \OmgK^2$, which gives $\rho(\Sigma,T_{\rm c})$ as an explicit function. So, it is straightforward to deduce:
\beq 
\begin{split} 
& \frac{\p \rho}{\p T_{\rm c}} \bigg|_{\Sigma} = - \lamP \frac{\rho}{T_{\rm c}} \\
& \mbox{where   } \lamP =  \frac{ \Pgas + 0.5 \Pmag + 4 \Prad  }{ 2 \Pgas + 2 \Pmag + \Prad } \,.
\end{split}
\label{partial_rho}
\eeq 
Further using equations~(\ref{Pdef}) and (\ref{partial_Q}), it can be readily checked that,
\beq 
\frac{\p P}{\p T_{\rm c}} \bigg|_{\Sigma} = \lamP \frac{P}{T_{\rm c}} \, .
\label{partial_P}
\eeq 
So the rate of increase of heating rate $q^{+}$ per unit change in $T_{\rm c}$, 
\beq  
\frac{\p q^{+} }{\p T_{\rm c}} \bigg|_{\Sigma} = \lamP \frac{q^{+}}{T_{\rm c}} \,. 
\label{partial_q+}
\eeq
From equation~(\ref{q_cooling_def}), the rate of change of radiative cooling with $T_{\rm c}$ is,
\beq 
\begin{split} 
& \frac{\p \qrad }{\p T_{\rm c}} \bigg|_{\Sigma} = \frac{\qrad}{T_{\rm c}} 
\bigg[ 4 - \frac{T_{\rm c}}{P} \frac{\p P}{\p T_{\rm c}}\bigg|_{\Sigma} - \frac{T_{\rm c}}{\kappa} \frac{\p \kappa}{\p T_{\rm c}}\bigg|_{\Sigma}    \bigg] = \Crad \frac{\qrad}{T_{\rm c}} \\
& \mbox{where  }\;\; \Crad =  4 - \lamP - \alT  + \lamP \alrho   \, \\
& \mbox{and   }\;\; \alT = \frac{\p \log \kappa}{\p \log T_{\rm c}} \bigg|_{\rho} \quad, \quad \alrho = \frac{\p \log \kappa}{\p \log \rho} \bigg|_{T_{\rm c}} . 
\end{split} 
\label{partial_qrad} 
\eeq 
Similarly, advective cooling rate \qadv changes with $T_{\rm c}$ as, 
\beq  
\frac{\p \qadv }{\p T_{\rm c}} \bigg|_{\Sigma} = \frac{\qadv}{T_{\rm c}} \bigg[ \frac{2 T_{\rm c}}{P} \frac{\p P }{\p T_{\rm c}} \bigg|_{\Sigma} -  \frac{ T_{\rm c}}{\rho} \frac{\p \rho }{\p T_{\rm c}} \bigg|_{\Sigma}  \bigg] = 3 \lamP \frac{\qadv}{T_{\rm c}} \,. 
\label{partial_qadv} 
\eeq 
Using equations~(\ref{partial_q+})- (\ref{partial_qadv}) in (\ref{S_def}), gives the condition of thermal stability of general thin disk equilibrium:
\beq
 \Sth = 2 \lamP \frac{\qadv}{q^+} + ( \Crad - \lamP ) \frac{\qrad}{q^+} >0  
\label{thermal_stab_fin}
\eeq


\section{Effect of Realistic Opacity}
\label{app_opacity_effects}

\begin{figure}
\centering
\includegraphics[width=0.5 
\textwidth]{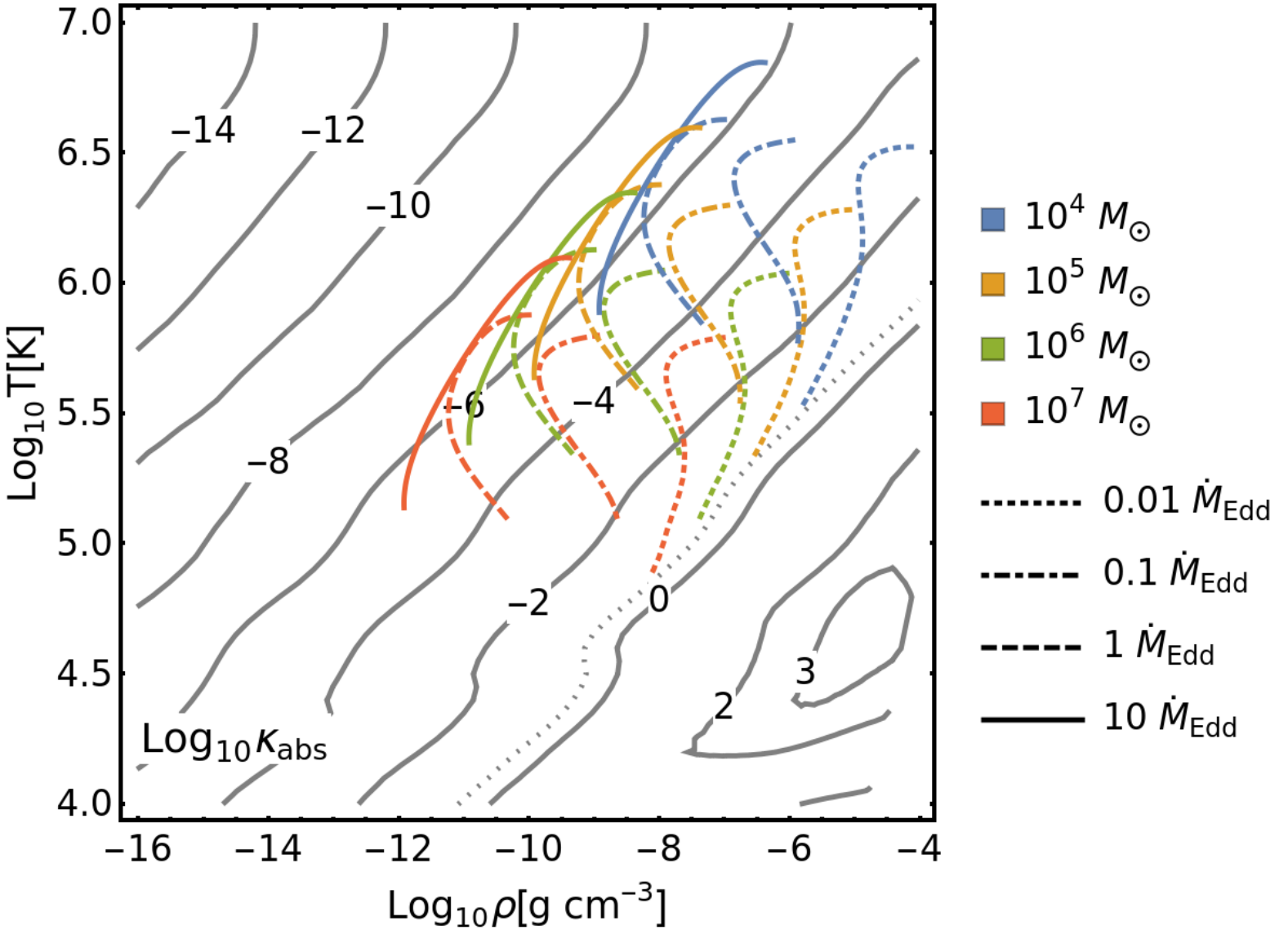}
\caption{Absorption opacity $\kappa_{\rm abs}$ evaluated in Cloudy, is shown in solid, black contours (values in log scale). A dotted black contour corresponds to $\kappa_{\rm abs} = \kes$. The disk solutions for non-magnetic models are plotted for various blackhole masses $M$ and accretion rates $\dot{M}$. For most of these models, $\kappa_{\rm abs} \ll \kes$ with notable exceptions only for the outer disk regions for lowest $\dot{M} = 0.01 \MdotEdd$ where $\kappa_{\rm abs}$ becomes comparable to $\kes$.  }
\label{fig_kappa_abs}
\end{figure}

\begin{figure}
\centering
\includegraphics[width=0.99 
\textwidth]{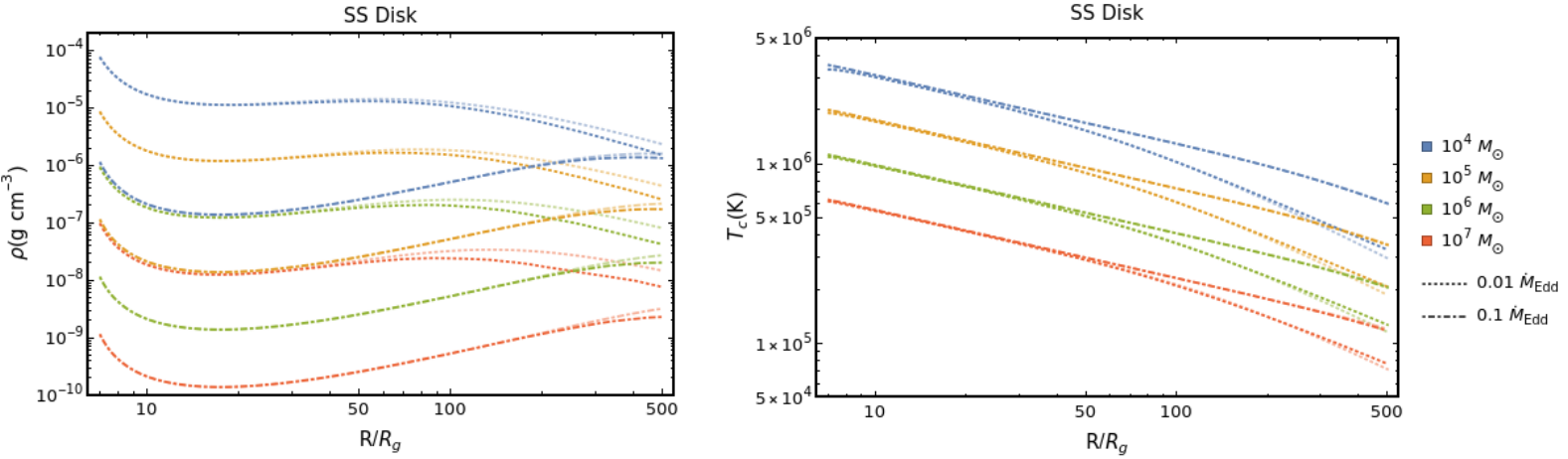}
\caption{Non-magnetic disk solutions - mid-plane disk density $\rho$ (left panel) and temperature $T_{\rm c}$ (right panel) - for Rosseland mean opacity $\kappa$ (in deep shades) and electron-scattering opacity $\kes$ (in light shades) for low Eddington ratios $\dot{M}/\MdotEdd = \{0.01,0.1 \}$ and various BH masses $M$ in different colours. The solutions are approximately equal for larger $\dot{M}/\MdotEdd$ and are not shown here.     }
\label{fig_sol_diff_kappa}
\end{figure}

\begin{figure}
\centering
\includegraphics[width=0.5 
\textwidth]{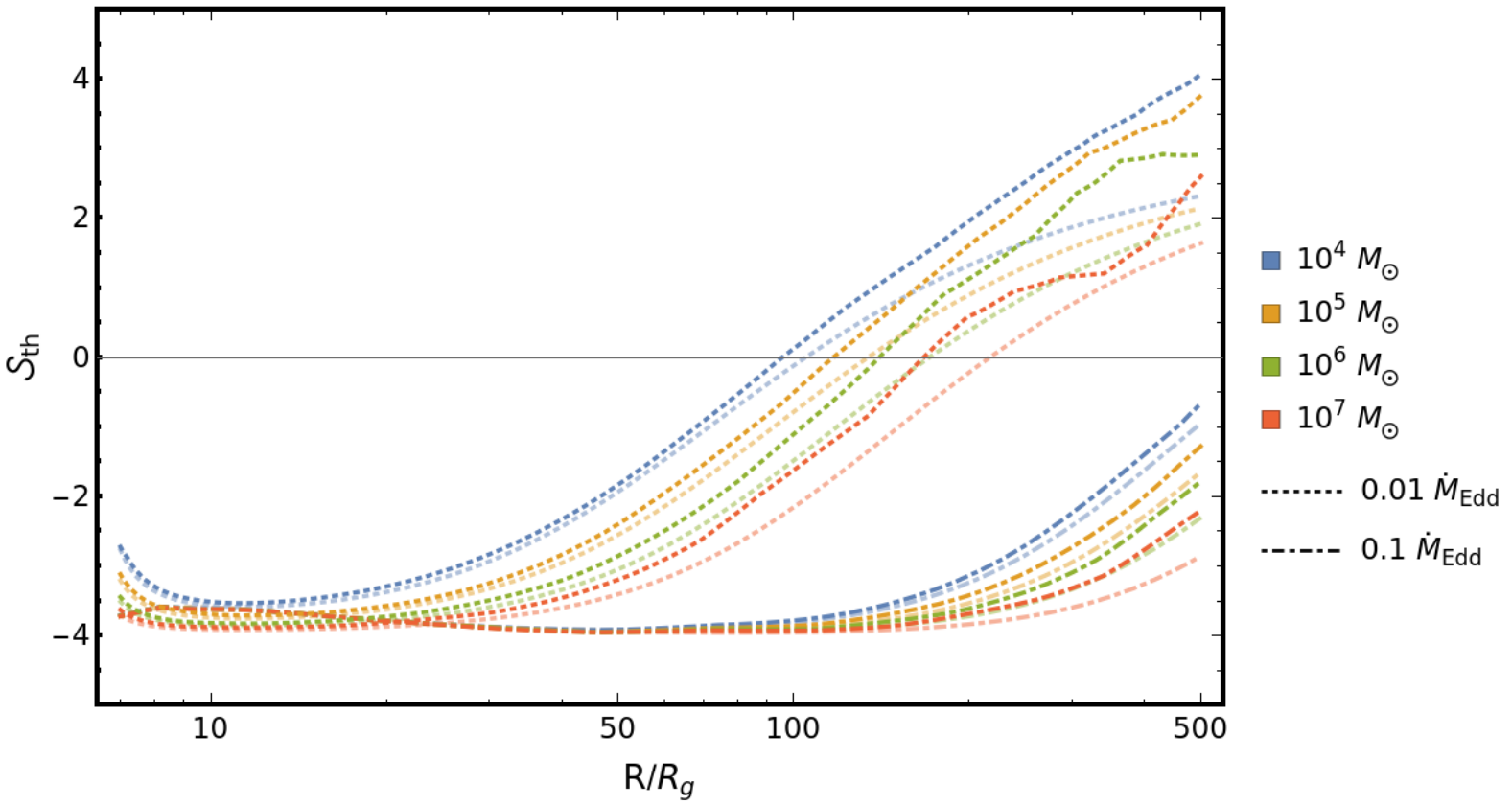}
\caption{Thermal stability parameter $\mathcal{S}_{\rm th}$ for non-magnetic disk solution with real $\kappa$ (deep shades) and $\kes$ (light shades) for low Eddington ratios. The radial range with $\mathcal{S}_{\rm th}>0$ implies thermal stability and vice versa. Consideration of real opacity $\kappa$ increases slightly the radial range of stability in the outer disk region for small $\dot{M}$; while the inner disk remains unstable. There are not any noticeable changes in stability trends for higher $\dot{M}$ (not shown here).        }
\label{fig_SSstab_diffK}
\end{figure}

We generated tables of Rosseland mean opacity $\kappa$ and absorption opacity $\kappa_{\rm abs}$ for the broad range of densities $\rho$ and temperatures $T_{\rm c}$ for solar metallicity in the photoionization code Cloudy \citep{Cloudy2013,Cloudy2017}. For a given $\{ \rho ,T_{\rm c} \}$, we choose background radiation field as an isotropic blackbody at $T_{\rm c}$. Figure~\ref{fig_kappa_abs} shows a contour map of $\kappa_{\rm abs}$ in $\{\rho, T_{\rm c} \}$ plane, along with solutions of non-magnetic disk model for various black hole masses $M$ and mass accretion rates $\dot{M}$. For the parameters of our interest, $\kes \gg \kappa_{\rm abs}$ except only for outer regions of disks with $\dot{M} = 0.01 \MdotEdd$. Even for this case, the disk solutions $\{ \rho , T_{\rm c} \}$ differ from $\kes$-only solutions by less than a few $10\%$. These differences are even smaller for fluffy magnetic disks, characterized by smaller densities (and hence smaller $\kappa_{\rm abs}$). Figure~\ref{fig_sol_diff_kappa} compares the non-magnetic disk solutions for tabulated Rosseland mean opacity $\kappa$ and constant electron-scattering opacity $\kes$. The maximum differences (for the smallest $\dot{M} = 0.01 \MdotEdd$) in the disk density $\rho$ and temperature $T_{\rm c}$ are less than a few times $10 \%$ and $1\%$ respectively. These differences are even smaller for low-density disks with higher $\dot{M}$ and/or magnetic pressure support and are not shown in the figure. We also investigate the impact of real $\kappa$ on the thermal stability trends.  Figure~\ref{fig_SSstab_diffK} showcases the radial profiles for stability parameter $\mathcal{S}_{\rm th}$ of equation~(\ref{thermal_stab_fin}) for non-magnetic disks with both cases of opacity -- $\kappa$ and $\kes$-- for low Eddington ratios $\dot{M}/\MdotEdd = \{0.01, 0.1 \}$. Employing real $\kappa$ increases the range of stability in the outer disk only slightly for low $\dot{M}$ while the inner disk remains unstable. This demonstrates the inability of contribution of absorption opacity to stabilize the disk, and motivates the consideration of magnetic pressure support for stability.                  

\section{Analytical steady state disk solutions} 
\label{app_analytic} 

Here we present approximate steady state disk solutions for special cases for dominant cooling phenomenon and pressure support in table~\ref{tbl_analytic_sol}. 
\begin{table}

\begin{subtable}{0.45\textwidth} 
\begin{center}
\begin{tabular}{|c|}

  \hline 
  
  \\[0.1em]
  
Case (a). Radiative Cooling ($q^{-} = \qrad$)   \\[1em]
  
  \hline
  
  \\[0.1em]
  
a1). Magnetically dominated Disks ($P = \Pmag$)   \\[1em] 

$\rho(T_{\rm c}) = \displaystyle{ \frac{\dot{M} f(R)}{ 3 \pi \alpha p_0^{3/2} }  \bigg( \frac{\mu \mP}{ \kB T_{\rm c} } \bigg)^{3/4} \frac{ \OmgK^{1/2}}{ R^{3/2}}  }   $ \quad ; \quad $\rho = 2.56 \times 10^{-5} \dot{M}_1^{2/3} M_6^{-5/6} p_0^{-4/3} \widetilde{R}^{-19/12} \alpha_{-1}^{-5/6}     $  \\[1em]

$P(T_{\rm c}) = \displaystyle{ \frac{\dot{M} f(R)}{ 3 \pi \alpha p_0^{1/2} } \bigg( \frac{\mu \mP}{ \kB T_{\rm c} } \bigg)^{1/4}   \frac{ \OmgK^{3/2}}{ R^{1/2}}         } $ \quad ; \quad $ P = 6.76 \times 10^{13} \dot{M}_1^{8/9} M_6^{-17/18} p_0^{-4/9} \widetilde{R}^{-91/36} \alpha_{-1}^{-17/18} $ \\[1em]

$ T_{\rm c}^{9/2}  = \displaystyle{ \frac{3 \kes (\dot{M} f(R))^2  }{ 32 \pi^2 \sigSB \alpha p_0 } \sqrt{ \frac{\mu \mP}{\kB} }  \frac{\OmgK^2}{R}           } $ \quad ; \quad $T_{\rm c} = 5.8 \times 10^7 \dot{M}_1^{4/9} M_6^{-2/9} p_0^{-2/9} \widetilde{R}^{-8/9} \alpha_{-1}^{-2/9} $ \\[1em]

$H(T_{\rm c}) = \displaystyle{ \sqrt{\frac{p_0 R}{\OmgK}} \bigg(  \frac{\kB T_{\rm c}}{ \mu \mP} \bigg)^{1/4}       }$  \quad ; \quad $\displaystyle{\frac{H}{R} = 0.054  \dot{M}_1^{1/9} M_6^{-1/18} p_0^{4/9} \widetilde{R}^{1/36} \alpha_{-1}^{-1/18}      }$ \\[1em]

$ \Sigma(T_{\rm c}) = \displaystyle{ \frac{\dot{M} f(R)}{3 \pi \alpha p_0 R} \bigg( \frac{\mu \mP  }{ \kB T_{\rm c}  }  \bigg)^{1/2}  }$ \quad ; \quad $\Sigma = 2.05 \times 10^5 \dot{M}_1^{7/9} M_6^{1/9} p_0^{-8/9} \widetilde{R}^{-5/9} \alpha_{-1}^{-8/9} $ \\[1em]

\hline 

\\[0.1em]
 
a2). Radiation dominated disks ($P = \Prad$)  \\[1em]

$\rho = \displaystyle{\frac{ 8^3 \pi^2 c^3 }{ 9^2 \alpha \kes^3  (\dot{M} f(R))^2 \OmgK}} $ \quad ; \quad $\rho  = 3.28 \times 10^{-13} \dot{M}_1^{-2} M_6^{-1}  \widetilde{R}^{3/2} \alpha_{-1}^{-1} $    \\[1em]

 $P = \displaystyle{ \frac{8  c \OmgK}{ 9 \alpha \kes}  } $  \quad ; \quad $P = 1.58 \times 10^{11} M_6^{-1}  \widetilde{R}^{-3/2} \alpha_{-1}^{-1} $   \\[1em] 
 
 $T_{\rm c}^4 = \displaystyle{ \frac{2 c^2 \OmgK}{ 3 \sigSB \alpha \kes} }$ \quad ; \quad $T_{\rm c} = 2.8 \times 10^6  M_6^{-1/4}  \widetilde{R}^{-3/8} \alpha_{-1}^{-1/4}  $  \\[1em]
 
 $ H = \displaystyle{ \frac{ 3 \kes \dot{M} f(R)   }{8 \pi c  }  }$ \quad ; \quad $\displaystyle{\frac{H}{R}} = 23.2 \dot{M}_1 \widetilde{R}^{-1}  $ \\[1em]
 
 $\Sigma = \displaystyle{ \frac{8^2 \pi c^2}{3^3 \alpha \kes^2 \dot{M} f(R) \OmgK} }$  \quad ; \quad $\Sigma = 1.1 \dot{M}_1^{-1}  \widetilde{R}^{3/2} \alpha_{-1}^{-1} $ \\[1em]

\hline
 
\end{tabular}
\end{center}
\end{subtable}
\\
\hspace{-4cm}
\begin{subtable}{0.45\textwidth} 
\begin{center}
\begin{tabular}{|c |}

  \hline
  \\[0.1em]
  
Case (b). Advective cooling  ($q^{-} = \qadv$)   \\[1em]
  
  \hline
  
  \\
  
 \hspace{1.62cm}  $\rho = \displaystyle{ \frac{8 \xi^{3/2} \dot{M}}{3^{5/2} \pi \alpha \sqrt{f(R)} \OmgK  R^3} }$ \quad ; \quad $\rho = 6.29 \times 10^{-9} \dot{M}_1 M_6^{-1}  \widetilde{R}^{-3/2} \alpha_{-1}^{-1} \xi^{3/2}   $ \hspace{1.62cm}  \\[1em] 
   
        $P = \displaystyle{ \frac{2 \dot{M} \sqrt{\xi f(R)} \OmgK   }{3^{3/2}  \pi \alpha R  } }$ \quad ; \quad  $P = 4.24 \times 10^{12} \dot{M}_1 M_6^{-1}  \widetilde{R}^{-5/2} \alpha_{-1}^{-1} \xi^{1/2}   $ \\[1em] 
        
        $H = \displaystyle{\frac{ \sqrt{3 f(R)} R }{ 2 \sqrt{\xi}  }  }$ \quad ; \quad $ \displaystyle{\frac{H}{R}} = 0.87 \xi^{-1/2}  $ \\[1em]
        
        $\Sigma = \displaystyle{\frac{4 \xi \dot{M}}{ 9 \pi \alpha \OmgK R^2 }}  $ \quad ; \quad $\Sigma = 804  \dot{M}_1 \widetilde{R}^{-1/2} \alpha_{-1}^{-1} \xi     $ \\[1.5em]
        
        \hdashline
        \\[0.1em]
  
b1). Magnetically dominated Disks ($P = \Pmag$)   \\[1em] 

$ \displaystyle{\sqrt{ \frac{ \kB T_{\rm c} }{ \mu \mP  } }  } = \displaystyle{ \frac{ 3 f(R) R \OmgK  }{ 4 \xi p_0  }} \quad ; \quad T_{\rm c} = 3.77 \times 10^{12} p_0^{-2} \widetilde{R}^{-1} \xi^{-2}  $ \\[1em]

\hline 
\\[0.1em]
 
b2). Radiation dominated disks ($P = \Prad$)  \\[1em] 

 $T_{\rm c}^4 = \displaystyle{ \frac{ \dot{M} c \sqrt{\xi f(R)}  \OmgK   }{ 2 \sqrt{3} \pi \alpha \sigSB R  } }$ \quad ; \quad $T_{\rm c} = 6.4 \times 10^6 \dot{M}_1^{1/4} M_6^{-1/4}  \widetilde{R}^{-5/8} \alpha_{-1}^{-1/4} \xi^{1/8}  $ \\[1em]

\hline

\end{tabular}
\end{center}
\end{subtable}
 \caption{Approximate disk solutions. Here $M_6 \equiv M/10^6 \Msun$, $\dot{M}_1 \equiv \dot{M}/\MdotEdd$, $\widetilde{R} \equiv R/r_g$, $\alpha_{-1} \equiv \alpha/0.1 $. }
\label{tbl_analytic_sol}
\end{table}

\FloatBarrier

\section{Additional Results}
\label{app_Additional}

\vspace{-10cm}

\begin{figure*}
\centering
\includegraphics[width=0.8 \textwidth]{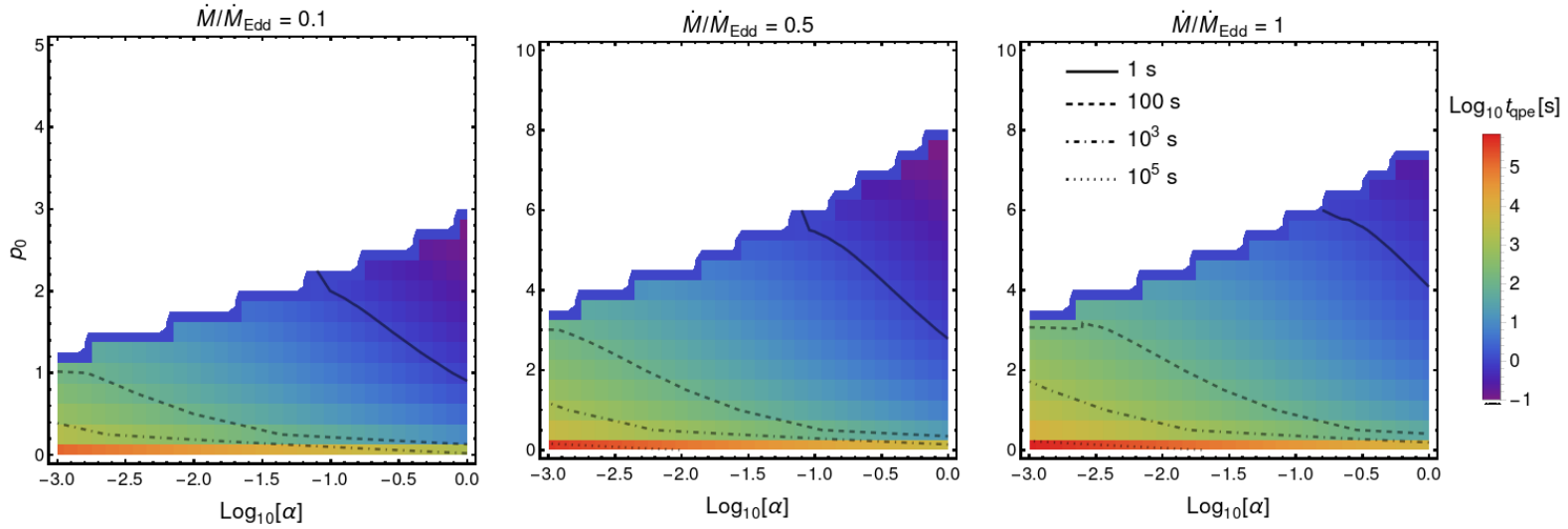}
\caption{Limit-cycle time-periods from our disk models for a typical XRB source with $M = 10 \Msun$ (for three tentative values of Eddington ratios) in $\{ \log_{10}\alpha,p_0  \}$-plane. Disks stabilize even for moderate values for $p_0$; say $p_0 \gtrsim 6$ for $\alpha \sim 0.1$ and $\dot{M}/\MdotEdd \sim 0.5-1$. This qualitatively explains the stability of majority of observed XRBs. Only a handful of sources display limit-cycles with time-periods $\sim 1-100$s, which occupies the bulk of unstable region in these figures (area enclosed between solid $t_{\rm qpe}=1$s and dashed $t_{\rm qpe}=100$s contours). It is notable that the timescales $\lesssim 1$d for roughly entire parameter space.  
 }
\label{fig_tp_XRBs}
\end{figure*}
\begin{figure*}
\begin{subfigure}{0.9\textwidth}
\centering
\includegraphics[width=0.9 \textwidth]{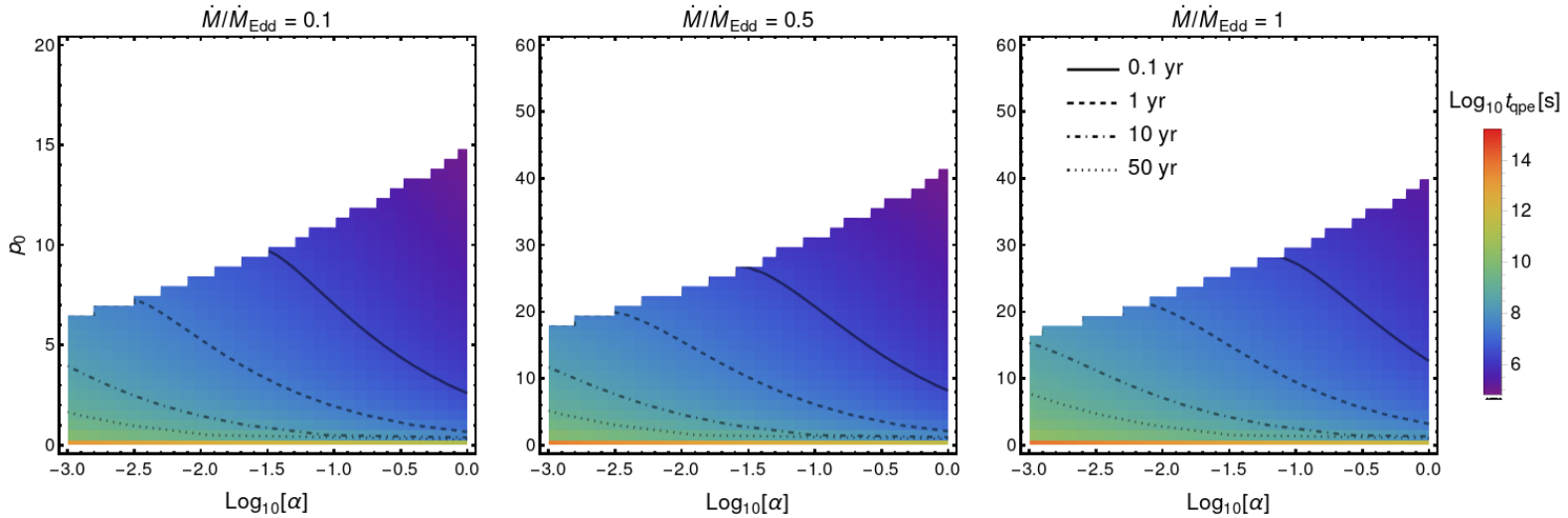}
\subcaption{$M = 10^7 \Msun$}
\end{subfigure}
\\

\begin{subfigure}{0.9\textwidth}
\centering
\includegraphics[width=0.9 \textwidth]{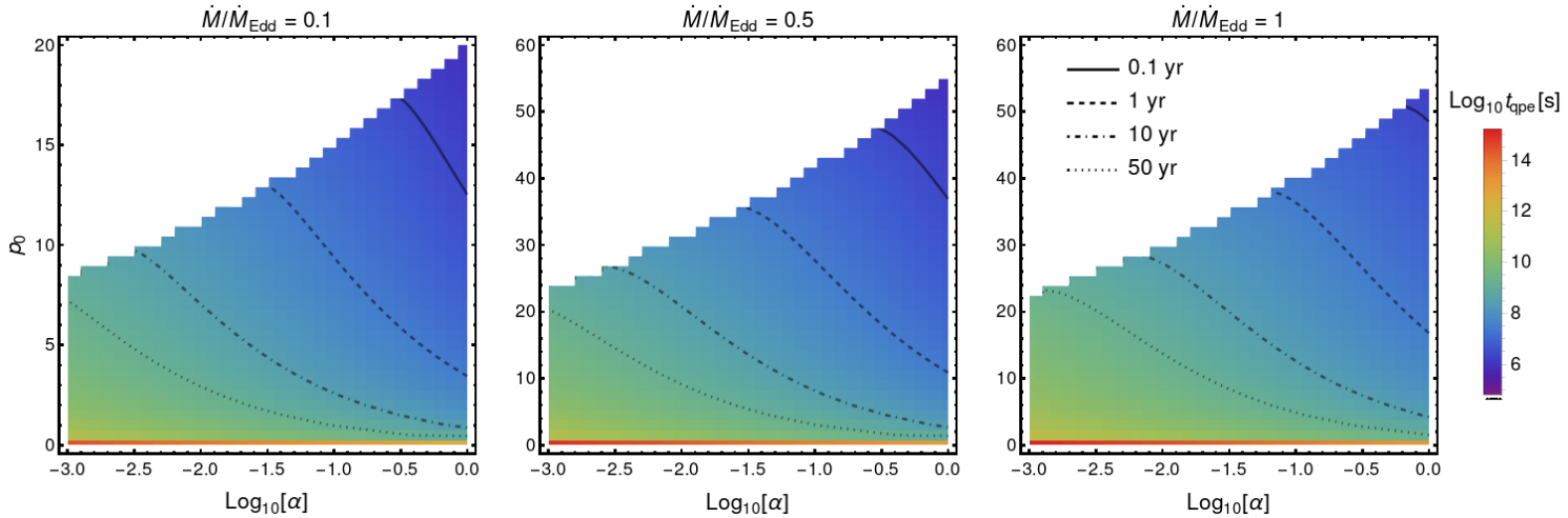}
\subcaption{$M = 10^8 \Msun$} 
\end{subfigure}
\caption{Limit-cycle time-periods for typical AGN sources with (a) $M = 10^7 \Msun$ [upper panel] (b) $M = 10^8 \Msun$ [lower panel] with three Eddington ratios in different columns. Instability occurs easily till moderate magnitudes of $p_0$ (eg. with $\alpha \simeq 0.1$ and $\dot{M}/\MdotEdd \in [0.5-1]$, disks are unstable for $p_0 \lesssim 30$(40) for case a (b)), and ceases to exist only for higher $p_0$. A major fraction of the unstable parameter space is occupied by timescales ranging from a few months to a few years (area bounded by solid and dot-dashed contours) consistent with the variability of CL AGNs and quasars. 
 }
\label{fig_tp_AGNs}
\end{figure*}

 \end{document}